\documentclass[useAMS,usenatbib]{mn2e}
\usepackage{epsf}
\usepackage{amssymb}
\usepackage[usenames]{color}

\newcommand {\bc}{\begin {center}}
\newcommand {\ec}{\end {center}}
\newcommand {\be}{\begin {equation}}
\newcommand {\ee}{\end {equation}}

\def\plotone#1{\centering \leavevmode
\epsfxsize=\columnwidth \epsfbox{#1}}

\def\plottwo#1#2{\centering \leavevmode
\epsfxsize=1.\columnwidth \epsfbox{#1}\hfil 
\epsfxsize=1.\columnwidth \epsfbox{#2}}
\def\plottwosm#1#2{\centering \leavevmode
\epsfxsize=.8\columnwidth \epsfbox{#1}\hfil 
\epsfxsize=.8\columnwidth \epsfbox{#2}}

\def\disp {\displaystyle}
\def\e {{\rm e}}
\def\D {{\rm D}}
\def\p {{\rm p}}
\def\turb {{\rm turb}}
\def\s {{\rm s}}
\def\i {{\rm i}}
\def\d {{\rm d}}
\def\c {{\rm c}}

\def\line {{\rm line}}
\def\tot {{\rm tot}}
\def\cool {{\rm cool}}
\def\r {{\rm r}}
\def\vir {{\rm vir}}
\def\min {{\rm min}}
\def\max {{\rm max}}
\def\v {{\rm v}}
\def\esc {{\rm esc}}
\def\next {{\rm next}}

\def\rms {{\rm rms}}
\def\line {{\rm line}}
\def\tot {{\rm tot}}
\def\deg{$^{\circ}$}
\def\central{{\rm central}}
\def\outer{{\rm outer}}

\title[Polarization of X-ray lines from galaxy clusters]{Polarization of X-ray lines from galaxy clusters and elliptical
  galaxies -- a way to measure tangential component of gas velocity}
\author[Zhuravleva et
  al.]{I.V.Zhuravleva$^{1}$\thanks{izhur@mpa-garching.mpg.de},
  E.M.Churazov$^{1,2}$, S.Yu.Sazonov$^{2,1}$, R.A.Sunyaev$^{1,2}$,
  \newauthor W.Forman$^{3}$, K.Dolag$^{1}$\\ \\
$^{1}$MPI for Astrophysik, Karl-Schwarzschild str. 1, Garching, 85741, Germany\\
$^{2}$Space Research Institute, Profsoyuznaya str. 84/32, Moscow,
  117997, Russia\\
$^{3}$Harvard-Smithsonian Center for Astrophysics, 60 Garden St.,
Cambridge, MA 02138, USA}

\begin{document}

\date{Accepted .... Received ...}

\pagerange{\pageref{firstpage}--\pageref{lastpage}} \pubyear{2009}

\maketitle

\label{firstpage}

\begin{abstract}
  We study the impact of gas motions on the polarization of bright
  X-ray emission lines from the hot intercluster medium (ICM). The
  polarization naturally arises from resonant scattering of 
  emission lines owing to a quadrupole component in the radiation
  field produced by a centrally peaked gas density distribution. If
  differential gas motions are present then a photon emitted in one 
region of the cluster will be scattered in another region only if their 
relative velocities are small enough and the Doppler shift of the photon 
energy does not exceed the line width. This affects both the degree and 
the direction of polarization. The changes in the polarization signal
  are in particular sensitive to the gas motions {\sl perpendicular}
  to the line of sight. \\ We calculate the expected degree of
  polarization for several patterns of gas motions, including a slow
  inflow expected in a simple cooling flow model and a
  fast outflow in an expanding spherical shock wave. In both cases,
  the effect of non-zero gas velocities is found to be minor. We also
  calculate the polarization signal for a set of clusters, taken from
  large-scale structure simulations and evaluate the impact of the
  gas bulk motions on the polarization signal. \\ We argue that
  the expected degree of polarization is within reach of the next
  generation of space X-ray polarimeters.
\end{abstract}

\section{Introduction} 
\label{sec:intro}
Galaxy clusters are the largest gravitationally bound structures in
the Universe with  cluster masses of $10^{14}- 10^{15}$
M$_\odot$. About $80 \%$ of this mass is due to dark matter, $15 \%$
due to hot gas and only a few $\%$ of the  mass corresponds to
stars. Hot gas is therefore the dominant baryonic component of 
clusters and the largest mass constituent which can be observed
directly.

In relaxed clusters, during periods of time without strong mergers,
the hot gas is in approximate hydrostatic equilibrium,
when characteristic gas velocities are small compared to the sound
speed. However, during rich cluster mergers, the gas
velocities can be as high as $\sim$ 4000 km/s as expected from numerical
simulations and suggested by recent Chandra and XMM-Newton
measurements \citep{Mark04}. But even in seemingly relaxed clusters,
bulk motions with velocities of hundreds of km/s should exist. 
The existence of gas bulk motions in clusters of galaxies is
supported by several independent theoretical and numerical studies
\citep[e.g.][]{Sun03,Dol05,Vaz09}. Recent grid-based, adaptive mesh refinement (AMR)
simulations have reached unprecedented spatial resolution and are now
providing very detailed information on the velocity of the ICM on
small scales \citep[][]{Iap08,Vaz09}. In these
simulations velocities as large as few hundred km/s are found all the
way to the cluster center.

The most direct way to detect gas bulk velocities along the line of
sight is through the measurement of the Doppler shifts of X-ray
spectral lines \citep{Ino03}. However, robust detection of intracluster gas
velocities has to await  a new generation of X-ray spectrometers
with high spectral resolution, given that the expected shift, of
e.g. 6.7 keV line of He-like iron, is $\sim 20$ eV for a
line-of-sight velocity of 1000 km/s. Measuring the line broadening by 
microturbulence or bulk motions also
requires high spectral resolution, which will become possible after the
launch of the ${\it NEXT}$ (New exploration X-ray telescope) mission with X-Ray microcalorimeter on-board. This
microcalorimeter  will have an energy resolution of 4 eV at the 6.7 keV
line \citep{Mit08, Mit09}.

Even more difficult is to measure the gas motion perpendicular to the
line of sight through the usual spectroscopic techniques. Indeed, the
shift of the line centroid due to this velocity component is proportional
to the $\disp \frac{1}{2}\left(\frac{{\it v_\perp}}{{\it c}} \right )^2 \ll 1$
(transverse Doppler effect) and amounts to $\sim 0.3$ eV for 6.7 keV line
if ${\it v_\perp}=3000~{\rm km~s^{-1}}$.  We discuss below a unique possibility
to determine such velocities using the polarization, due to resonant
scattering, of X-ray lines from galaxy clusters.

\begin{figure}
\plottwosm{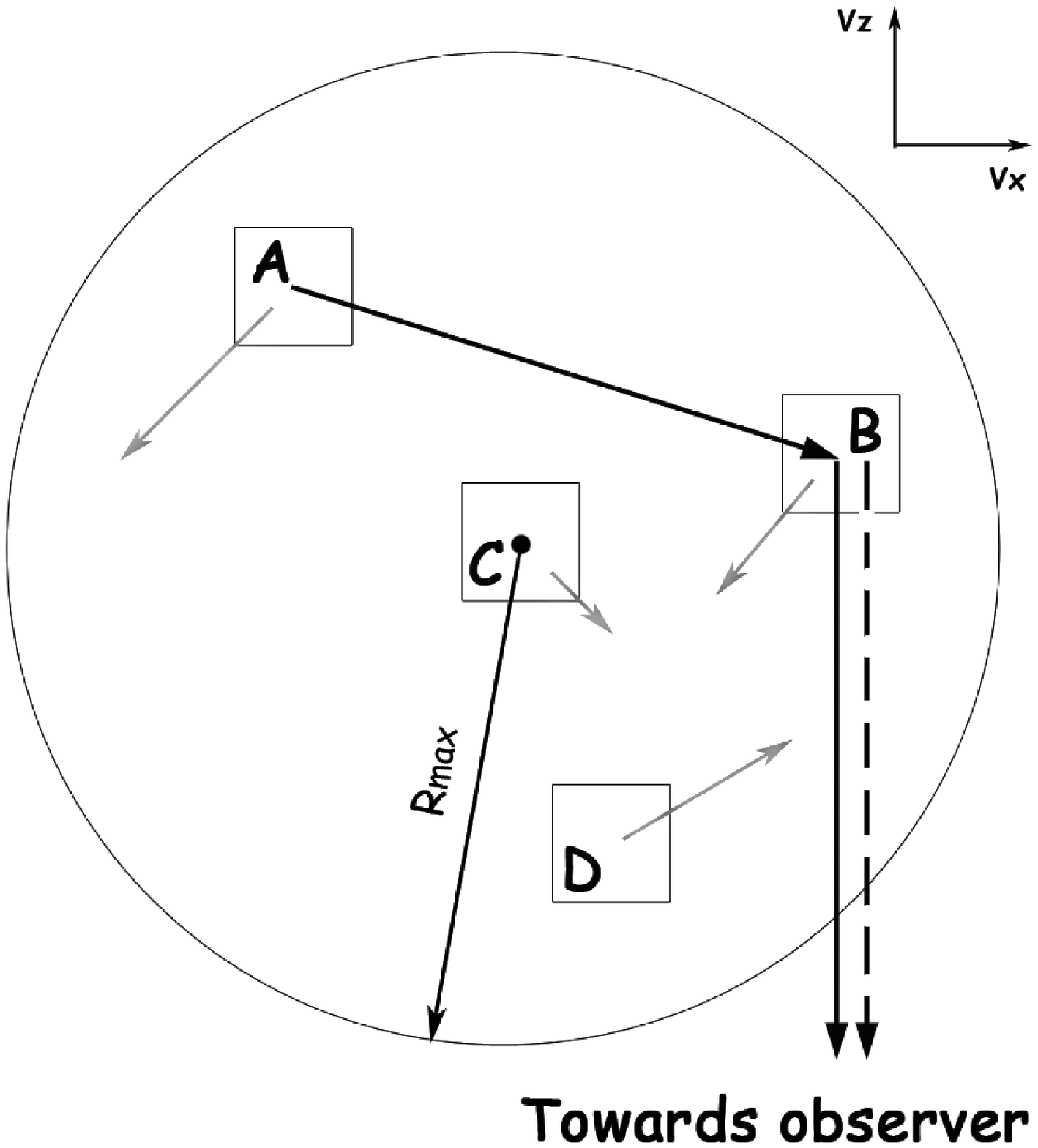}{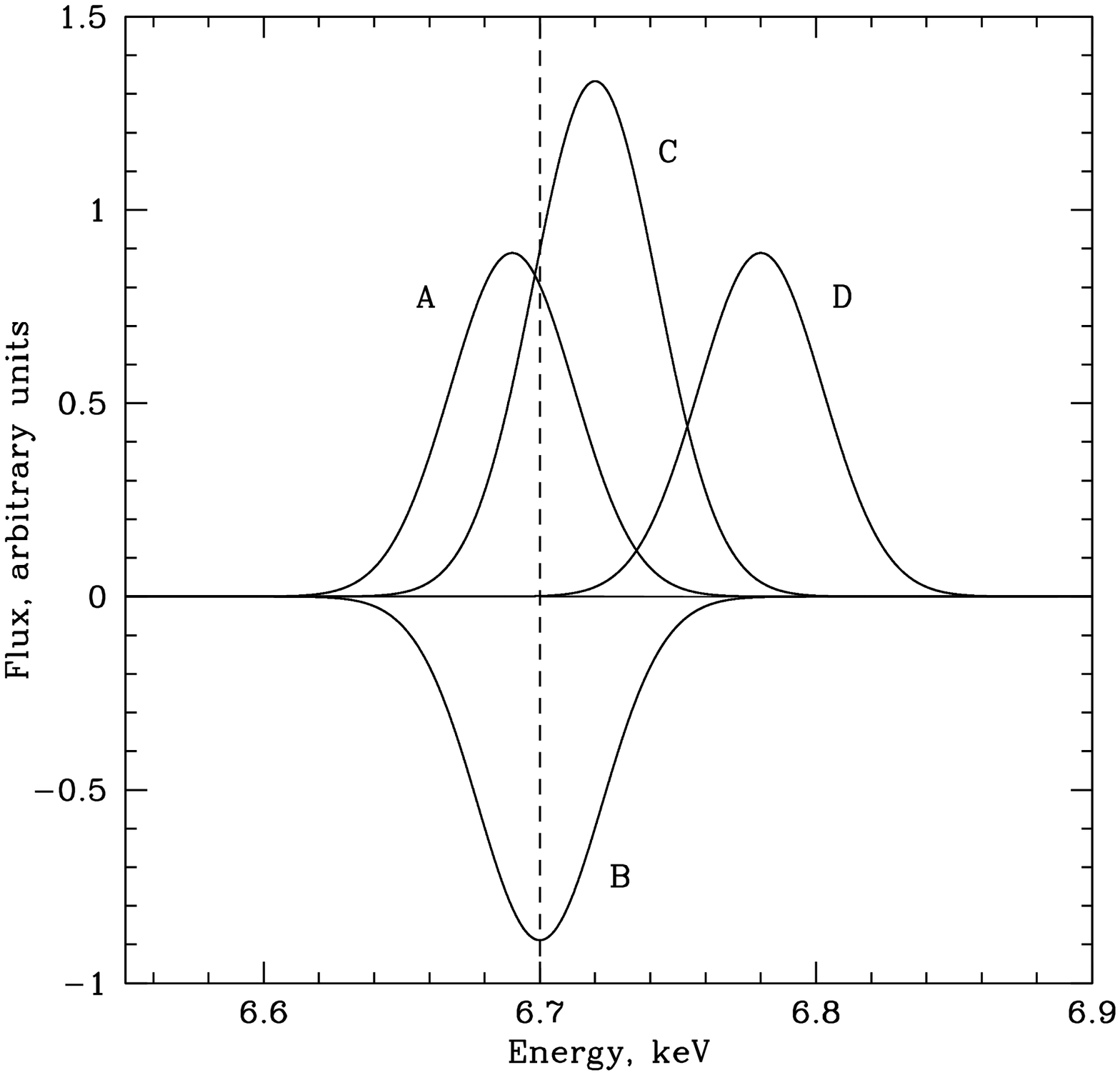}
\caption{Geometry of the problem. {\bf Top panel:} squares show
  different gas lumps located in one $(x,z)$ plane. Line of sight is
  along $z$-axis. Grey arrows show
  the projection of gas lump velocities onto this plane. The dashed
  arrow shows a photon emitted towards the observed from a gas lump
  B. Solid black arrows show a photon which was emitted in lump A and
  subsequently scattered towards the observer in lump B. The
  efficiency of photon scattering depends on the gas density and
  temperature, on the line broadening and on the relative velocities
  of the lumps along the direction between the lumps.  The motion of a
  given gas lump along the line of sight (${\it v}_Z$) affects the
  observed energy of the lines , while the motion perpendicular to the
  line of sight (i.e. ${\it v}_X$ and/or ${\it v}_Y$) affects both
  direction and degree of polarization (see also
  Fig.\ \ref{fig:intro}). The ${\it v}_Z$ component also affects the
  poalrization degree by means of transverse Doppler effect, but much
  less than the other two components, therefore, we neglect this. {\bf Bottom panel:} Line
  profiles in the reference system of lump B. The rest energy of the
  line is 6.7 keV.  The energy dependent scattering cross section is
  schematically shown as a negative Gaussian and is marked with
  ``B''. Emission spectra coming from lumps A, C and D are marked as
  ``A'', ``C'', and ``D'' respectively. The quadrupole asymmetry
  needed for polarization arises both due to variations of the total
  line flux coming to lump B from different directions and due to the
  Doppler shift of the lines caused by the differential bulk motions
  of the lumps.
\label{fig:geometry}
}
\end{figure}

The continuum emission from hot gas in galaxy clusters is optically
thin. However at the energies of strong resonant transitions, the gas
can be optically thick \citep{Gil87}. Oscillator strengths of resonant
lines are large and if ion concentrations along the line of sight are
sufficient, photons with energies of the resonant lines will be
scattered. The geometry of the problem is sketched in
Fig.\ref{fig:geometry} (top panel) - from any gas lump we observe both direct
(thick dashed line) and scattered (thick solid black line) emission.

If there is an asymmetry in the radiation field (in particular, a
quadrupole moment) then the scattering emission will be
polarized. This asymmetry can be i) due to the centrally concentrated
gas distribution and ii) owing to differential gas motions (see
Fig.\ \ref{fig:geometry} and 
\S\ref{sec:resscat}).

In clusters, polarization arises rather naturally. It is well known
\citep{Sun82} that the scattering of radiation from a bright central
source leads to a very high (up to 66\%) degree of linear polarization of the
scattered radiation (for a King density distribution).

\begin{figure*}
\plottwosm{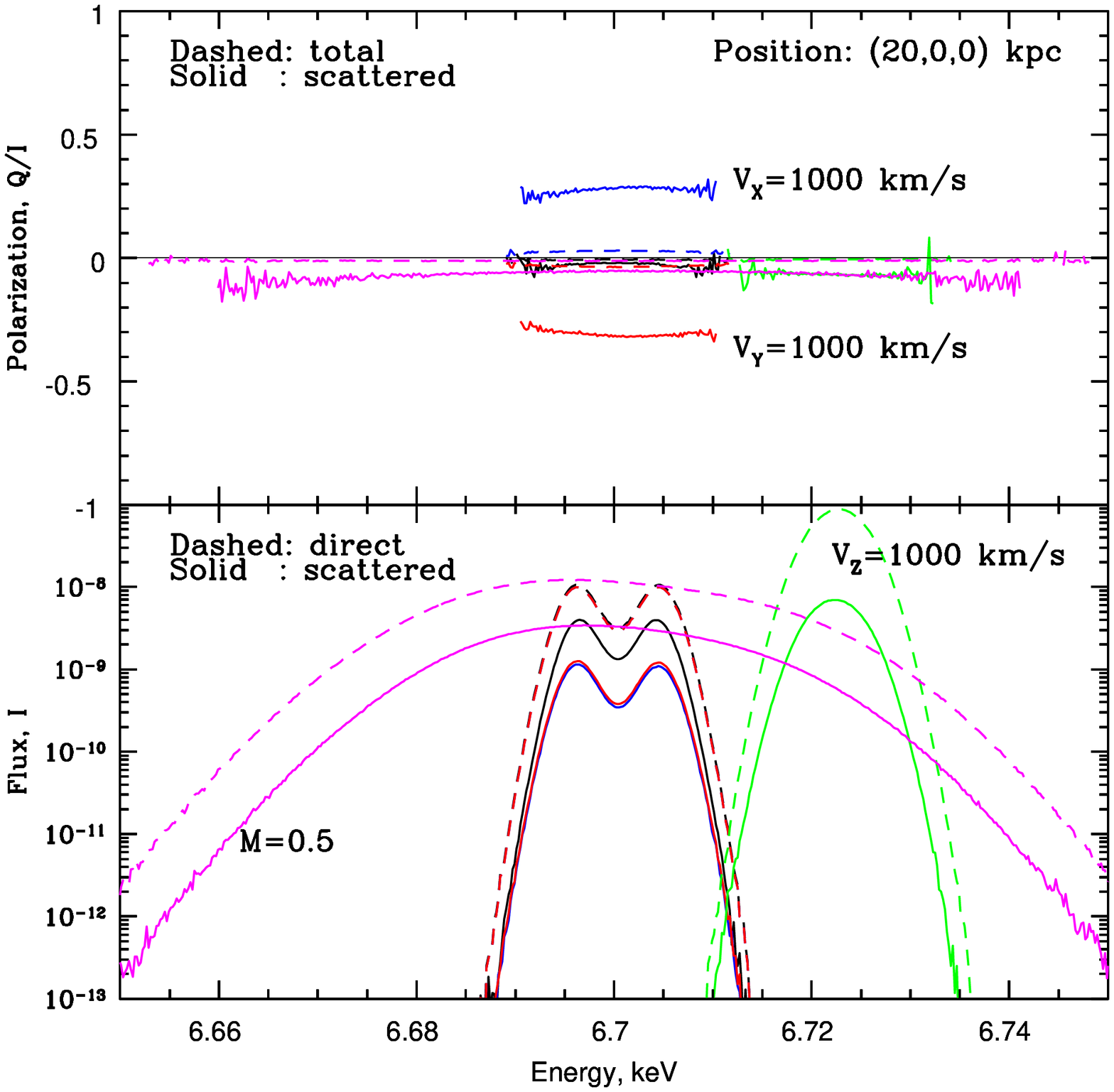}{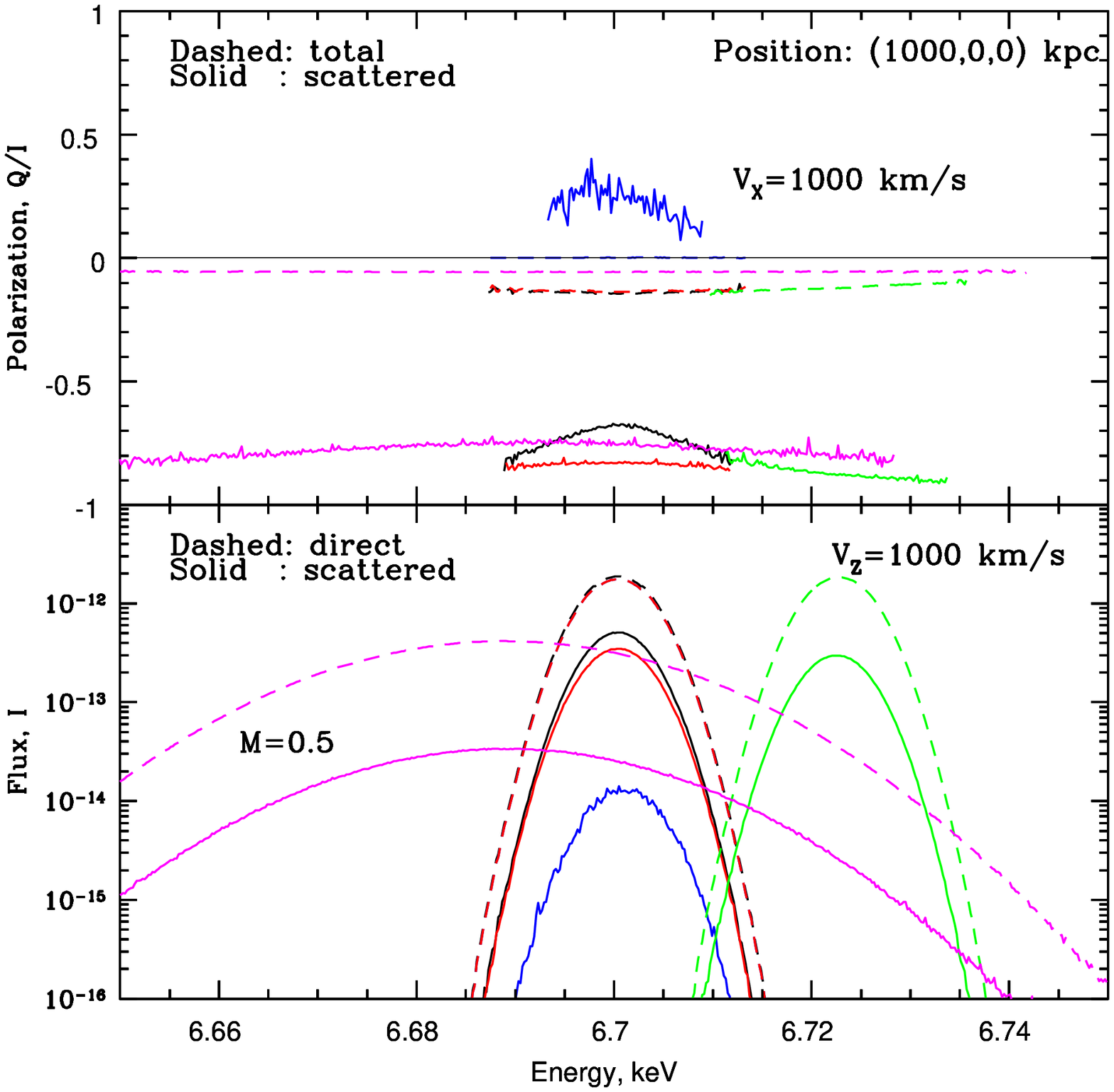}
\caption{Line profile and polarization calculated assuming a single
  scattering for two small regions of a simulated galaxy cluster.
  {\bf Left:} A region close to the cluster center with coordinates
  (20,0,0) kpc.  $X$ and $Y$ are the coordinates in the picture plane
  along horizontal and vertical directions respectively, $Z$ is along
  the line of sight. In the bottom panel dashed lines show the line
  photons which are emitted within a given region and directly reached
  an observer without scatterings $I_{dir}$ (see dashed line in
  Fig.\ref{fig:geometry}). Solid lines show the emission scattered in
  a given region $I_{scat}$ (see solid thick black line in
  Fig.\ref{fig:geometry}). In the top panel the solid line is the
  polarization of the scattered flux only (i.e. $Q_{scat}/I_{scat}$),
  while the dashed line shows the polarization of the total emission,
  coming from a given region (i.e.
  $Q_{scat}/(I_{scat}+I_{dir})$).\newline {\bf Color coding:}\newline
  {\it Black} -- lines are broadened by thermal ion velocities only,
  all bulk velocities are set to zero. The characteristic depression
  at the line center is caused by the large depth at the center of the
  line. The degree of polarization is small since this region is
  exposed to essentially isotropic radiation. \newline {\it Blue} -
  the same as the black curves, but $X$ component of the gas velocity
  $v_X$ in a given region is set to 1000 km/s. Scattered flux
  decreases since along the direction of motion the line emission left
  the resonance. Polarization of the scattered flux is strong
  ($\sim$30\%) in the direction perpendicular to the direction of
  motion.  \newline {\it Red} - the same as the black curves, but $Y$
  component of the gas velocity $v_Y$ in the region is set to 1000
  km/s. Polarization of scattered flux is strong ($\sim$30\%) with the
  polarization plane rotated by 90\deg compared to blue
  curves.  \newline {\it Green} - the same as the black curves, but
  $Z$ component (along the line of sight) of the gas velocity $v_Z$ is
  set to 1000 km/s. The line energy is strongly shifted, while the
  polarization characteristics are similar to the case without gas
  motions (black curves). The larger flux of the line (compared to the
  case without gas motions) is due to a smaller optical depth towards
  the observer.  \newline {\it Magenta} - the same as the black
  curves, but the gas bulk velocities are not set to zero, but are
  taken from simulations. The lines are broadened by micro-turbulence
  with the parameter $M=0.5$.\newline {\bf Right:} the same as in the
  left panel for a gas lump located at (1000,0,0) kpc relative to the
  cluster center. An appreciable polarization signal is present even in
  the case of the gas with zero velocities.
\label{fig:intro}
}
\end{figure*}

X-ray line emission is strongly concentrated toward the cluster
center due to the collisional nature of the emission process and the
decrease of the plasma density with distance from the cluster
center. \citet{Gil87} have shown that resonance scattering will
decrease the line surface brightness in cluster cores and increase the
line brightness in the periphery of clusters. \citet{Saz02}
have shown that this process should lead to a high degree of
polarization in the resonance lines in high spectral resolution maps
of galaxy clusters. For the richest regular clusters and clusters
whose X-ray emission is dominated by a central cooling flow,
the expected polarization degree is about $15 \%$ \citep{Saz02}.

In this paper we are interested primarily in the second possibility,
i.e. in the modification of the polarization signal due to large scale
gas motions. Of particular interest is the sensitivity of the
polarization signal to $v_\perp$, the component of the velocity 
perpendicular to the line of sight. 

Linear polarization $\propto (v_\perp/c)^2$ naturally appears as a
result of scattering of unpolarized isotropic {\bf continuum}
radiation with a Rayleigh phase function.  That is, for scattering
of isotropic CMB radiation in the Rayleigh-Jeans limit the net
polarization is given by ${\it P}\sim 0.1 (v_\perp/c)^2\tau$
\citep{Sun80}, where $\tau$ is the optical depth of the scattering
medium. In the same paper Sunyaev and Zeldovich presented an
approximate formula for the case of double scattering $P\sim 0.02 -
0.05 (v/c) \tau^2$, where the coefficient depends on the gas
distribution inside the cluster (it is $\sim 0.01$ for the King
distribution, \cite{Saz99}).  Both these effects are very
small. If there is an asymmetry in the initial angular distribution of
radiation with the dimensionless amplitude $a$, then additional
terms of order $a\times (v/c)$ can appear. But even these terms are
expected to be small since the value of $v/c$ itself is of order
$10^{-2}$ or less.

A much larger impact of the gas bulk motions is expected if {\bf line}
rather than {\bf continuum} emission is considered. If gas is moving
and along the direction of motion the Doppler shift of the photon
energy exceeds the line width (i.e. line leaves the resonance) then no
scattering occurs. This is illustrated in Fig.\ref{fig:geometry}. We
consider several gas lumps marked as A, B, C and D having different
bulk velocities. Region C correspond to the center of the cluster. Photons emitted in regions A, C and D
are scattered towards the observer in the region B. Shown in the
bottom panel of Fig.\ref{fig:geometry} are the line
  profiles in the reference system of the lump B. The rest energy of the
  line is 6.7 keV.  The energy dependent scattering cross section is
  schematically shown as a negative Gaussian and is marked with
  ``B''. Emission spectra coming from lumps A, C and D are marked as
  ``A'', ``C'', and ``D'' respectively. It will be observed
  (\S\ref{sec:resscat}) that a quadrupole asymmetry in the scattered
  radiation is needed to produce net polarization. This asymmetry
  arises both due to variations of the line flux coming to lump B from
  different directions and due to  
  Doppler shift of the photon energies caused by the differential bulk motions
  of the lumps. Indeed total (integrated over energies) line flux
  coming from region C is the largest because of the high gas
  emissivity in the cluster center. At the same time, gas lump A is
  moving away from lump B. Therefore, in the frame of lump B,
  the line profile is shifted towards lower energies. The line coming from
  lump D is, on the contrary, shifted towards higher
  energies. A comparison with the energy dependent scattering cross
  section clearly shows that asymmetry in the scattered emission
  coming from different directions (and therefore the polarization) is
  affected by the gas velocities.  What matters is the relative
  velocity of the gas lumps projected onto the line connecting the
  lumps. For example, if  lump B and C are only slowly moving towards (or away)
  from each other then line photons coming from region C will be
  effectively scattered in lump B and these photons will dominate
  in the scattered flux seen by the observer. If on the contrary lumps
  B and C are approaching each other (or receding) with the velocity
  larger than the width of the line profile then the lump B will not
  scatter photons coming from C and photons coming from other
  direction will dominate in the scattered flux. Therefore the
  intensity and polarization of the scattered flux will be different
  for these two cases. The mutual position of lumps B and C sketched in
  Fig.\ref{fig:geometry} shows that the polarization signal is
  in particular sensitive to the velocity component of lumps B and C
  along the X-axis, i.e. {\it perpendicular} to the line of sight.

  In general relative velocities of any two gas lumps affect the
  polarization. However, the central region of the cluster is very
  bright and therefore the motion of any gas lump relative to the
  center is the most important, since a large fraction of
  scattered photons were originally born near the cluster
  center. If the gas density rapidly drops off with radius then the
  largest contribution to the scattered flux (and hence to polarization)
  is provided by the regions along the line-of-sight which are most
  close to cluster center -- not far from the picture plane going though
  the center. Thus, the scattering by $\sim$90\deg is the most important,
i.e. photons, emitted in the cluster center, scatter in gas lump in the
picture plane and then come to observer. In this
  case the velocities perpendicular to the line-of-sight affect the
  scattering and polarization most strongly. Of course, in real systems
  different scattering angles make contributions to the scattered flux
  and therefore gas motions in different directions play a role. Below
  (Section 5) we make a full accounting of 3D velocity field using
  numerical simulations of galaxy cluster and take into account relative
  motions of the gas lumps in all directions.

At the same time, the presence of gas motions along the line of
sight leads to strong fluctuations of the centroid energy and the
shape of the resonant lines \citep[see][]{Ino03}\footnote{The change of
  the plasma density due to turbulent pulsations has orders of
  magnitude smaller amplitudes} and can be measured with high
resolution spectrometers. 

One can also consider the impact of small-scale random (turbulent) gas
motions on the line profiles and the polarization signal.  As has been
shown by \citet{Gil87}, micro-turbulence makes the lines broader,
decreasing the optical depth and therefore reducing the effects of
resonant scattering and polarization.  They pointed out that in galaxy
clusters the effect is especially strong for heavy elements, which
have thermal velocities much smaller than the sound velocity of the
gas. For example, for the 6.7 keV iron line in the spectrum of the Perseus
cluster, the inclusion of turbulent motions would reduce the optical
depth from $\sim 2.8$ to $\sim 0.3$ for a Mach number of 1
\citep{Chur04}.  Turbulent motions were also considered in the dense
cores of X-ray halos of giant elliptical galaxies. \citet{Wer09} placed tight 
constraints on turbulent velocities from the effect of resonant
scattering on the ratio of optically thin and thick lines. They
have shown that characteristic turbulent velocities in the center of
the giant elliptical galaxy NGC4636 are less than 100 km/s.

These points are illustrated in Fig.\ref{fig:intro} where the spectra
and the polarization signal coming from two gas clumps in a simulated
galaxy cluster are shown. Here we are using a model cluster as
described in section \ref{sec:3dprob}. All calculations for this plot were done assuming
a single-scattering by 90$^\circ$. The left panel shows the radiation
emerging from a region close to the cluster center. Due to it's
central location, the incident radiation field is almost isotropic and the
resulting polarization signal is week. Adding bulk motions in the
picture plane 
($v_x$, $v_y$ - blue and red lines, respectively)
immediately produces polarization in the perpendicular direction, but
does not affect the line shape. On the contrary, adding a line-of-sight
velocity component ($v_z$ - green line) shifts the line energy, but does not
affect the polarization signal. Finally, adding micro-turbulence makes
the line broader. The right panel of Fig.\ref{fig:intro} shows the
same effects for a small region located in the
cluster outskirts. For this region, the asymmetry in the incident
radiation is already present and non-zero polarization is expected even
without gas motions. Adding velocity in the ``horizontal'' direction
changes the sign of the polarization signal.

Thus combining the data from high resolution X-ray spectrometers and
X-ray polarimeters, one can constrain both line-of-sight and
perpendicular components of the velocity field - an exercise
hardly possible to do by any other means.

The structure of the paper is as follows. In Section 2 we discuss main
criteria for selection of X-ray emission lines, most promising for
polarization studies. In Section 3 we describe a Monte Carlo code used
to simulate resonant scattering. Two characteristic flow patterns (a
canonical cooling flow and an expanding spherical shock) in the
approximation of a spherical symmetry are considered in Section 4. In
Section 5 we present the results of resonant scattering simulations in
full 3D geometry, using several clusters taken from cosmological
simulations. The calculations described in Sections 4 and 5 take into
account only line photons. In Section 6 we summarize our results,
discuss additional effects which reduce the polarization degree in
galaxy clusters -- the most most important one is the contamination of
the polarized line flux with unpolarized continuum and line emission
from the thermal plasma.  Finally in Section 7 we outline requirements
for future X-ray polarimeters.

\section{Resonant scattering and  promising lines}
\label{sec:resscat}

For the most prominent lines (e.g. 6.7 keV line of He-like iron) the
scattering is characterized by the Rayleigh phase function
similar to the scattering phase function by free electrons.

If $E$ is the photon energy, $E_0$ is the energy of the transition
(line), ${\textit {\textbf v}}$ is the gas bulk velocity,
${\textit {\textbf m}}$ is the photon propagation direction, $\sigma$ is the
Gaussian width of the line set by thermal broadening and
micro-turbulent gas motions. The cross section for resonant scattering
in the lines in the rest frame of a given gas lump is a strong
function of energy 
\be
\disp s \propto \disp e^{-\frac{\left ( E \left [ 1-({\bf v
            m})/c\right ]-E_0\right )^2}{2\sigma^2}}. 
\label{eq:intro3}
\ee

If $\mu=\cos \theta$, where $\theta$ and $\phi$ are the polar angles in the frame set
by the photon direction of propagation after scattering, $I(E,\mu,\phi)$ is
the initial unpolarized radiation, $\sigma_0$ is the scattering
cross section by one ion and $n_i$
is the number density of corresponding ions, then according to
\cite{Chan50},
 one can write a simple expression for the integrated energy flux
scattered in a small volume of gas $dV$ and for the Stokes parameter
$Q$:
\begin{eqnarray}
I_{scat}= \sigma_0 n_i dV\int e^{-\frac{\left ( E \left [ 1-({\bf v
            m})/c\right ]-E_0\right )^2}{2\sigma^2}} \times \nonumber \\
\frac{3}{16\pi}(1+\mu^2) ~I(E,\mu,\phi) \d\mu\d\phi\d E \nonumber \\
Q=\sigma_0 n_i dV \int  e^{-\frac{\left ( E \left [ 1-({\bf v
            m})/c\right ]-E_0\right )^2}{2\sigma^2}} \times \nonumber  \\
\frac{3}{16\pi}(\mu^2-1)\cos 2\phi ~I(E,\mu,\phi) \d\mu\d\phi\d E.
\label{eq:intro}
\end{eqnarray}

The presence of the term $(\mu^2-1)\cos 2\phi$ in the expression for
$Q$ shows that net polarization arises when an angular function 
\be
F(\mu,\phi)\equiv \displaystyle \int
e^{-\frac{\left ( E \left [ 1-({\mathbf v 
           \mathbf
           m})/c\right]-E_0\right)^2}{2\sigma^2}}I(E,\mu,\phi)\d E
\ee
has non-zero quadrupole moment. As discussed in
\S\ref{sec:intro} this quadrupole moment can arise due to the
asymmetry in $\int I(E,\mu,\phi) dE$ and/or by angular dependence in
the cross section $s(E,\mu,\phi)$ caused by the factor $({\mathbf v
 \mathbf m})/c$.  Note that in the bottom panel of Fig.\ \ref{fig:geometry} all profiles
are plotted in the rest frame of a given gas lump, while the above
expressions are written in the laboratory frame.
 
The optical depth of a line is defined as 
\be
\tau=\int\limits
_0^{\infty}n_\i \sigma_0 \d r,
\ee
 where $n_i$ is the ion concentration and 
$\sigma_0$ is the cross section at the center of  the line
\be
\sigma_0=\frac{\sqrt{\pi}hr_\e cf}{\Delta E_\D},
\label{eq:sig0}
\ee
where $r_\e$ is the classical electron radius and $f$ is the 
oscillator strength of a given atomic transition. 
In plasma with a temperature characteristic of galaxy clusters, the line width is determined by
the velocities of thermal and turbulent motions, rather than by the radiative width. The Doppler width of the
line is defined as

\be
\Delta E_\D=E_0\left[\frac{2kT_\e}{Am_\p c^2}+\frac{V_\turb^2}{c^2}\right]^{1/2},
\label{eq:den}
\ee where $A$ is the atomic
mass of the corresponding element, $m_\p$ is the proton mass and
$V_\turb$ is the characteristic turbulent velocity. $V_\turb$ is
parametrized as $V_\turb=c_\s M$, where $M$ is the Mach number and the
sound speed in the plasma is $c_\s=\sqrt{\gamma k T/\mu m_\p}$, where
$\gamma=5/3$ is the adiabatic index for an ideal monoatomic gas and
$\mu=0.62$ is the mean atomic weight. We can rewrite
the previous expression as 
\be \Delta
E_\D=E_0\left[\frac{2kT_\e}{Am_\p c^2}(1+1.4AM^2)\right]^{1/2}.  
\ee
It is clear from the above expressions that to have large
optical depth in the line, the oscillator strength has to be large. If
the line width is dominated by thermal broadening then the lines
of the heaviest elements will be narrower than for the lighter elements,
and the optical depth for lines of heavy elements will be larger as
well.

Resonant scattering can be represented as a combination of two processes: 
isotropic scattering with a weight $w_1$ and dipole (Rayleigh) 
scattering with weight $w_2=1-w_1$ \citep{Ham47, Chan50}. Weights 
$w_1$ and $w_2$ depend on the total angular momentum $j$ of the 
ground level and on the difference between the total angular momentums
of excited and ground levels $\Delta j$ (=$\pm 1$ or $0$). 
The expressions for the weights were calculated by \citet{Ham47}.  
Isotropic scattering introduces no polarization, while Rayleigh
scattering changes the polarization state of the radiation field. Therefore,
to have noticeable polarization, we have to select lines with
a large optical depth and with large Rayleigh scattering weight $w_2$.

In the spectrum of a typical rich cluster of galaxies, often the strongest line 
is the He-like K$_\alpha$ line of iron with rest energy 6.7 keV. 
Ions of He-like iron are present in plasma with  temperature in the
range from $\sim$1.5 to $\sim$10 keV. 
The transition $1s^2(^1S_0)-1s2p(^1P_1)$ of He-like iron has an absorption oscillator strength $\sim 0.7$ 
and we can expect this line to be optically thick, leading to a
significant role of resonant scattering. Also this transition corresponds 
to the change of the total angular momentum from 0 to 1 and according 
to \citet{Ham47} the scattering has a pure dipole phase function. So, we can expect 
a high degree of polarization in this line. In the vicinity of this 
resonant line there are intercombination ($1s^2(^1S_0)-1s2p(^3P_{1,2})$), 
forbidden ($1s^2(^1S_0)-1s2s(^3S_1)$) lines of He-like iron and many satellite
lines. Forbidden and intercombination lines have an absorption oscillator
strength at least a factor 10 smaller than the resonant line
\citep{Por01} and a small optical depth accordingly. Satellite lines
correspond to the transitions from excited states and in the low
density environment have negligible optical depth.

In cooler clusters with temperatures below 4 keV, the Li-like
iron is present in sufficient amounts and the Fe XXIV line
$1s^22s(^2S_{1/2})-1s^23p(^2P_{3/2})$ at 1.168 keV is especially
strong. The dipole phase function has a weight of 0.5 for this line.
 
At even cooler temperatures $\sim 1~{\rm keV}$ the most promising line
for polarization studies is the Be-like line
($1s^22s(^1S_0)-1s^22s3p(^1P_1)$) of iron with rest energy 1.129
keV, which has an oscillator strength 0.43 and a unit weight of dipole
scattering. Other lines are discussed in detail by \citet{Saz02}.

\section{Monte-Carlo simulations}
\label{sec:mc}

We performed two types of numerical simulations for two different
initial conditions:
\begin{itemize}
\item spherically symmetric clusters.
\item full three-dimensional galaxy cluster models, taken
  from cosmological large-scale   structure simulations.
\end{itemize}

\subsection{Spherically symmetric clusters}
In the spherically symmetric simulations, we describe a cluster as a
set of spherical shells with given densities, temperatures and
radial velocities.  To calculate plasma line emissivities we use the
Astrophysical Plasma Emission Code (APEC, \cite{Smi01}). Line energies and oscillator strengths are
taken from
ATOMDB\footnote{http://cxc.harvard.edu/atomdb/WebGUIDE/index.html} and
the NIST Atomic Spectra
Database\footnote{http://physics.nist.gov/PhysRefData/ASD/index.html}. The
ionization balance (collisional equilibrium) is that of \cite{Maz98} - the same as used in APEC.

Multiple resonant scattering is calculated using a Monte-Carlo
approach \citep[see e.g.][]{Poz83,Saz02,Chur04}. We 
start by drawing a random position of a photon within the cluster,
random direction of the photon propagation ${\mathbf m}$ and the
polarization direction ${\mathbf e}$, perpendicular to ${\mathbf
  m}$. The photon is initially assigned a unit weight $w$.  Finding an
optical depth $\tau$ in the direction of photon propagation up to
the cluster edge and an escape probability $p_\esc=e^{-\tau}$, we draw
the optical depth of the next scattering as
$\tau_\next=-\ln(1-\xi(1-p_\esc))$, where $\xi$ is a random number
distributed uniformly on the interval $[0,1]$. Using the value of
$\tau_\next$ the position of the next scattering is identified.  The
code allows one to treat an individual act of resonant scattering in
full detail. The phase function is represented as a combination of
 dipole and isotropic scattering phase matrixes with weights $w_1$
and $w_2$. In the scattering process the direction ${\bf m'}$ of the
emerging photon is drawn in accordance with the relevant scattering
phase matrix. For an isotropic phase function, the new direction ${\bf
  m'}$ is chosen randomly. For a dipole phase function,  the probability
of the emerging photon to have new direction ${\bf m'}$ is $P({\bf
  m'}, {\bf e'})\propto \disp ({\bf e'},{\bf e})^2$, where the electric
vector ${\bf e'}=\disp\frac{{\bf e}-{\bf m'}\cos\alpha}
{\sqrt{1-\cos^2\alpha}}$ and $\alpha$ is an angle between the electric
vector ${\bf e}$ before scattering and the new direction of propagation
${\bf m'}$, i.e. $\cos\alpha=({\bf e}, {\bf m'})$. For the energy of
the photon,  we assume a complete energy redistribution. After
every scattering, the photon weight $w$ is reduced by the factor
$(1-e^{-\tau})$.  The process repeats until the weight drops below the
minimal value $w_{min}$. Thus, specifying the minimum photon weight,
we can control the number of multiple scatterings. A typical value of
the minimum weight used in the simulations is $w_{min}=10^{-9}$.

After every scattering for a photon propagating in the direction
${\mathbf m}$, a reference plane perpendicular to ${\mathbf m}$ is set
up for Stokes parameter calculations. One of the reference axes is
set by projecting a vector connecting the cluster center and the
position of the last scattering on the reference plane.  The Stokes
parameter $Q$ is defined in such a way that it is non-zero when the
polarization is perpendicular to the radius. From the symmetry of the
problem, it is obvious that the expectation value of the $U$ parameter
in our reference system is $\equiv 0$. Then the projected distance $R$
from the center of the cluster is calculated and the Stokes parameters
$I$ and $Q$ are accumulated (as a function of $R$) with the weight
$\propto \epsilon\times w \times e^{-\tau}$, where $\epsilon$ is the
volume emissivity at the position where the initial photon was
born. Finally the degree of polarization is calculated as $P(R)=Q(R)/I(R)$.
  
\begin{table}
 \centering
  \caption{Oscillator strength and optical depth of the most 
promising X-ray lines in Virgo/M87 and Perseus clusters.} 
  \begin{tabular}{@{}rccccc@{}}
  \hline
 Ion & $E,~{\rm keV}$ & $f$ & $w_2$ &$\tau$, Virgo/M87& $\tau$, Perseus\\
\hline 
 Fe XXII & 1.053 & 0.675 & 0.5 & 0.65 & 0.02 \\
 Fe XXIII & 1.129 & 0.43 & 1 & 1.03 & 0.16 \\
 Fe XXIV & 1.168 & 0.245 & 0.5 & 1.12 & 0.73\\
 Si XIV & 2.006 & 0.27 & 0.5 & 0.6 & 0.24 \\
 S XV & 2.461 & 0.78 &  1 & 0.68 & 0.03 \\
 Fe XXV & 6.7 & 0.78 & 1 & 1.44 & 2.77 \\
 Fe XXV & 7.881 & 0.15 & 1 & 0.24 & 0.45\\
\hline
\label{tab:lines}
\end{tabular}
\end{table}

\subsection{Full 3D clusters}
In the full 3D simulations,  cluster parameters (gas density and
temperature) are computed on a 3D Cartesian grid and the velocity field is
represented as a 3D vector field.  In contrast to a symmetric code, the
3D code precisely takes into account the changes of photon energy due
to scattering, assuming Gaussian distributions  of thermal and
turbulent ion velocities.

For every simulation we choose the viewing direction and setup a
reference system in the plane, perpendicular to the viewing
direction. The Stokes parameters $I$, $Q$ and $U$ are calculated with
respect to this reference system and are accumulated in a form of 2D
images. Finally the degree of polarization is calculated as
$P=\sqrt{Q^2+U^2}/I$.

\section{Spherically symmetric problems}

We have performed spherically symmetric calculations to simulate 
resonant scattering polarization effects in two galaxy clusters: 
Perseus and Virgo/M87, as examples of  cooling flow clusters. 

\subsection{Perseus cluster}
\begin{figure}
\plotone{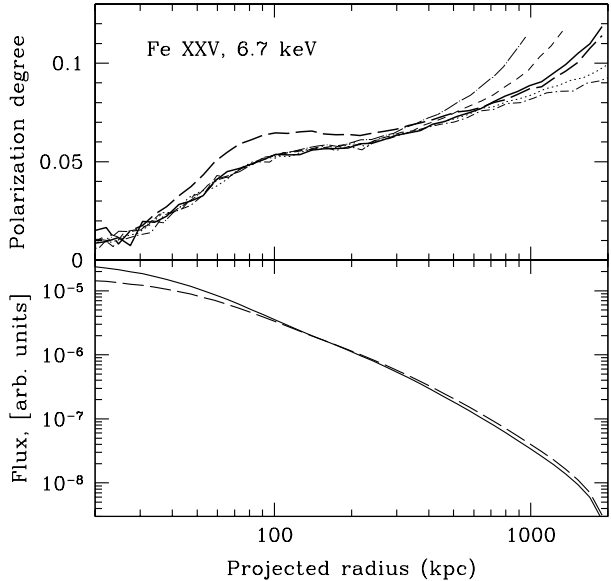}
\caption{Top panel: polarization degree as a function of projected 
distance from
  the center of the Perseus cluster in the most prominent resonant 
line of Fe XXV with  energy 6.7 keV, assuming flat iron abundance (the thick solid curve). 
Other thin curves show results of calculations done for different
sizes of the cluster (maximal radius $r_\max$) assumed in the simulations: the upper dot-dashed curve for $r_\max=1000$ kpc, 
the dashed curve is for $r_\max=1400$ kpc, the dotted line is for $r_\max=3000$ kpc and the lower dot-dashed line  
for the cluster size $r_\max=4000$ kpc. Clearly 
the  polarization degree in the cluster outskirts strongly depends on
the assumed extent of the cluster. For reference, the 
virial radius of the Perseus cluster is $r_{200}\approx 2100$ kpc.
Depending on $r_\max$ the degree of polarization at a distance of 1
Mpc can vary from 8 to 12\%. At smaller distances  the degree of
polarization is less sensitive to the cluster properties at large
radii.
Thick long dashed curve shows polarization degree in the case of
peaked abundance. 
\newline
Bottom panel: simulated surface brightness of Perseus in the 6.7
    keV line, calculated with and without resonant scattering - dashed
    and solid lines, respectively. Resonant scattering diminishes the flux
in the cluster center and increases the flux at larger radii.
\label{fig:polPer}
}
\end{figure}

\begin{figure}
\plotone{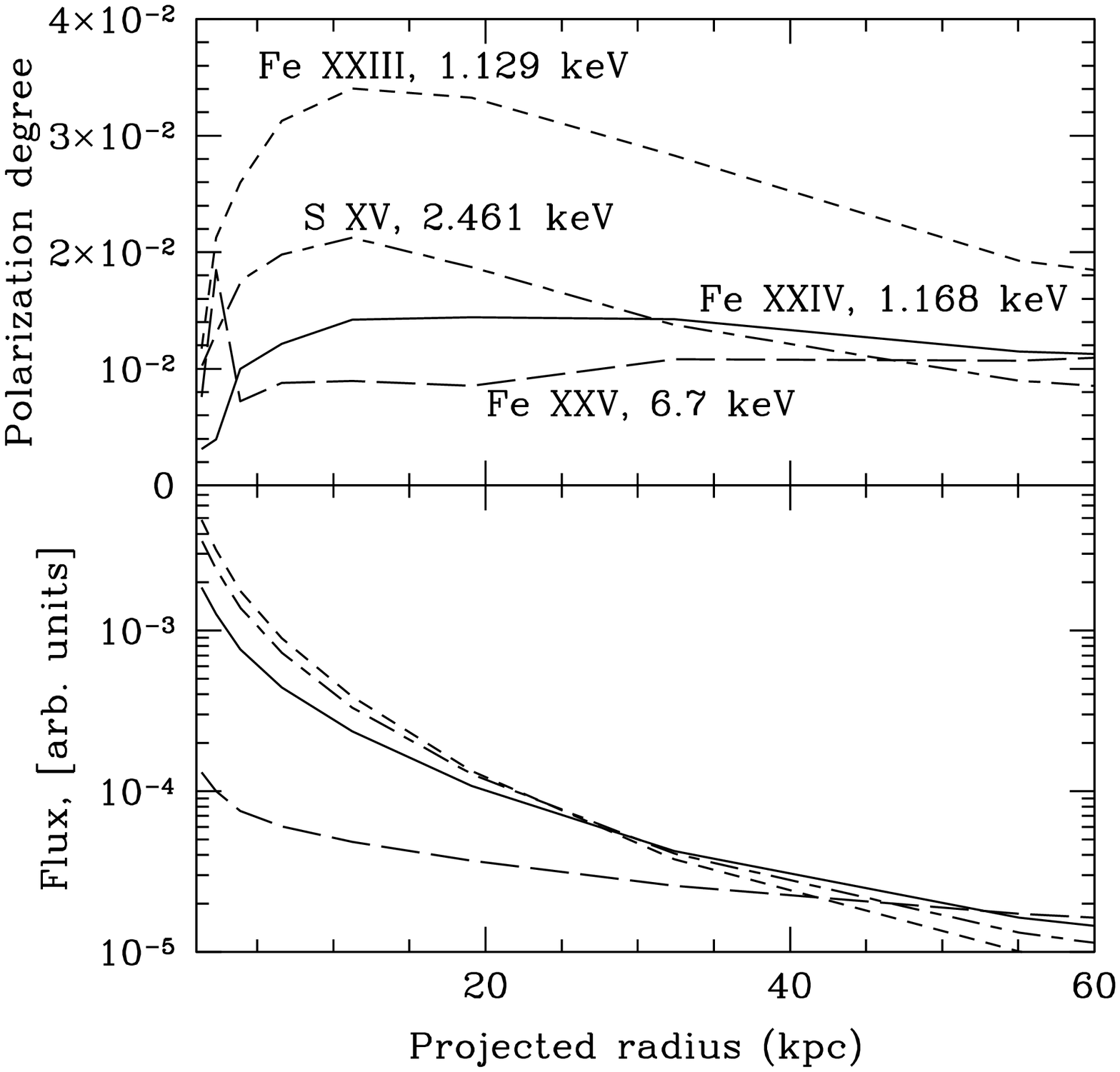}
\caption{Top panel: polarization degree as a function of projected distance from 
the center of the Virgo/M87 cluster in the most prominent resonant 
lines, presented in Table \ref{tab:lines}.\newline
Bottom panel: simulated surface brightness on the Virgo/M87 cluster in
the lines presented on the top panel. Scattering is included. 
\label{fig:polM87}
}
\end{figure}

The electron density distribution for the Perseus
cluster was adopted from \citet{Chur03}. They describe the density
profile as a sum of two $\beta-$models and they assume a Hubble constant
of $50~\rm{km/s/Mpc}$. Correcting the density distribution to the value of
the Hubble constant $H_0=72~{\rm km/s/Mpc}$ we find
\begin{eqnarray}
n_e=\frac{4.68\times10^{-2}}{\left [1+\left (\frac{r}{56}\right )^2 \right]^{\frac{3}{2}\times 1.2}}+
\frac{4.86\times10^{-3}}{\left [1+\left (\frac{r}{194}\right )^2\right ]^{\frac{3}{2}\times 0.58}}~~~{\rm cm}^{-3}
\label{ne}
\end{eqnarray}
The temperature distribution is described as
\begin{eqnarray}
T_e=7\frac{\left [1+\left (\frac{r}{100}\right )^3 \right ]}{\left [2.3+\left (\frac{r}{100}\right )^3 \right ]}~~~{\rm keV},
\label{te}
\end{eqnarray}
where $r$ is measured in kpc.

The iron abundance is assumed to be constant over the whole cluster
and equal to 0.5 solar using the  Anders and Grevesse abundance table
\citep{And89}. This value is equivalent to 0.79 solar if the newer
solar photospheric abundance table of \cite{Lod03,Asp06} is used.

\citet{Saz02} produced a list of the strongest X-ray lines in the
Perseus cluster with optical depth $\tau > 0.5$.  Only three lines 
have dipole scattering weight larger than the weight of isotropic
scattering: the He-like K$_\alpha$ line with energy 6.7 keV, the
He-like K$_\beta$ line ($1s^2-1s3p(^1P_1)$) at 7.88 keV and the
L-shell line of Li-like iron with energy $1.168~{\rm keV}$. The
optical depths of these lines are $\sim 3$, $\sim 0.45$ and $\sim 0.73$
respectively (see Table \ref{tab:lines}). Therefore, for the Perseus
cluster we performed calculations for the permitted line at 6.7 keV as the
optical depth in this line is the largest, weight $w_2=1$, and hence
the polarization of scattered radiation in this line is expected to be
the most significant.

\subsection{M87/Virgo cluster}

Another example of a cooling flow cluster is the M87/Virgo cluster.
In this cluster the temperature in the center is lower than in
Perseus. The temperature and number electron density profiles in the
inner region were taken from \citet{Chur08}. The electron density is
described by $\beta$-law distribution with
\begin{eqnarray}
n_e=\frac{0.22}{\left [1+\left (\frac{r}{0.93}\right )^2 \right]^{\frac{3}{2}\times 0.33}}~~~{\rm cm}^{-3},
\label{nem87}
\end{eqnarray}
where $r$ is in kpc. Temperature variations can be approximated as \be
T=T_0\left[1+\left(\frac{r}{r_\c}\right)^2\right]^{0.18}, \ee where
the central temperature is parameterized by $T_0=1.55~{\rm keV}$ and $r_\c=10.23~{\rm
  kpc}$. In clusters with such low central temperatures the
lines of Li-like and Be-like iron become very strong.  In Table
\ref{tab:lines} we list potentially interesting lines in M87, which
have weight $w_2\ge 0.5$. The most
interesting lines are those at 1.129 keV and at 6.7 keV as the
optical depth is large and the scattering phase function is pure
dipole. Also notable are the lines of Li-like iron at 1.168 keV and 
line of S XV at 2.461 keV, which have 
weights $w_2=0.5$ and $w_2=1$ respectively. The line at 1.168 keV has
two components. The first component has
zero weight of dipole scattering, while the second at 1.168 keV has
equal weights of dipole and isotropic scattering. Therefore, we
consider only the second component in our simulations.

The Fe and S abundances are assumed to be constant at $r < 10$ kpc,
being Z(Fe)=1.1 and Z(S)=1 (relative to the solar values of \cite{Lod03}), then gradually falling with radius to reach the values
Z(Fe)=0.56 and Z(S)=0.63 at $r=40$ kpc, and remain constant from there
on. These approximations are in a good agreement with 
observations \citep[e.g.][]{Wer06}.

One can notice, that optical depths, presented in
Table \ref{tab:lines}, are systematically lower than in
\cite{Saz02}. This is caused by differences in number density and
temperature profiles. We are using newer profiles and, for example, in
case of Virgo cluster number density is systematically lower and
temperature in center is higher, leading to the changes in ion
fractions. All this factors together reduce the optical depths of lines.

\subsection{Degree of polarization without bulk motions}

In Fig.\ref{fig:polPer} the results for the Perseus cluster are
shown. As already discussed the calculations were done for the He-like
K$_\alpha$ line at 6.7 keV. The polarization is zero at the center of
the cluster (as expected from the symmetry of the problem) and
increases rapidly with increasing distance from the center. 
At large distances from the center the degree of polarization depends
strongly on the maximum radius $r_\max$ used in the simulations. This
is an expected result. Indeed for steep density profiles at the
periphery of a cluster the largest contribution to the scattered
signal is due to small region along the line-of-sight (near the
closest approach to the cluster center). Since most of the photons are
coming from the central bright part of the cluster, the role of
90\deg scatterings increases. Recall that for the Rayleigh phase
function, 90\deg scattering produces 100\% polarized
radiation. Therefore near the edges of the simulated volume the degree
of polarization is expected to increase, even although the polarized
scattered radiation will always be diluted by locally generated line
photons. Obviously the final value of polarization degree at a given
projected radius depends also
on the cluster properties at larger radii. The virial radius of
the Perseus cluster is $r_{200}\approx 2.1$ Mpc. Our simple
approximations for the gas density and temperature certainly break
there (perhaps at a fraction of the virial
radius). Nevertheless, the gas beyond virial radius has a very low
density and it is likely not hot enough to produce strong emission in
the 6.7 keV line. From this point of view an effective cut-off radius
$r_\max$ in the gas distribution may suffice as an illustration of
the polarization signal sensitivity to the structure of the cluster
outskirts. In Perseus simulations as $r_\max$ varies from 1 Mpc
to 4 Mpc (Fig.\ref{fig:polPer}, the top panel), i.e. the boundary of
the simulated volume extends to larger distances, the degree of
polarization at the projected distance of 500 kpc decreases from
$\sim$8 to 7\%. At the very edge of the cluster one can expect the
maximum degree of polarization in the Perseus cluster of order 10\% .
Also
    shown in  Fig.\ref{fig:polPer} are the surface brightness profiles in
    the 6.7 keV line, calculated with and without resonant
    scatterings (the bottom panel). 

 As already mentioned above, a flat abundance profile ($Z=0.79$)
was assumed in these calculations. We also tried a peaked abundance
profile based on the XMM-Newton observations of the Perseus cluster
\citep{Chur03}: the iron abundance is $\sim 1.1$ in the
center ($r<60$ kpc), gradually decreases to $\sim 0.7$ at $r\sim200$
kpc and is flat at larger radii. The polarization degree,
corresponding to the peaked abundance profile is shown in
Fig.\ref{fig:polPer} with the thick long dashed curve. In general the agreement
with the calculations for a flat abundance profile is good. The
maximum polarization does not change, while at distances $r<200$ kpc
a small bump in polarization degree appears. At $r=100$ kpc the
polarization degree increases from $5\%$ to $6\%$, staying constant
till 200 kpc. We further discuss the effect of the peaked abundance
profile in Section 5.

\begin{figure}
\plotone{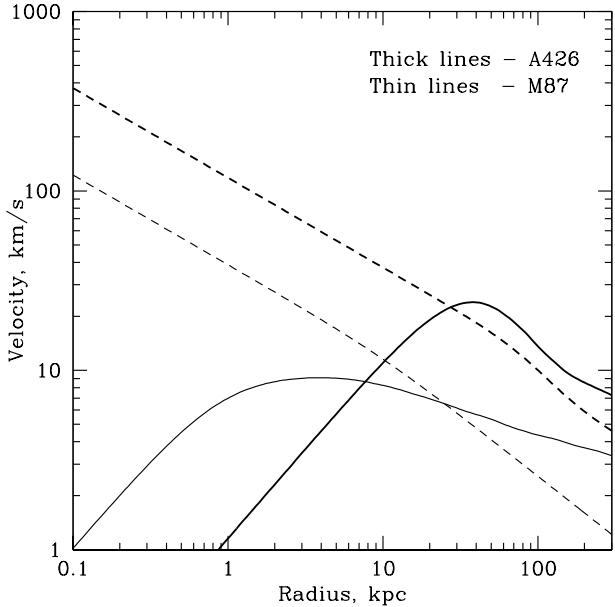}
\caption{Inflow velocity $v_\r$ in homogeneous cooling flow models for
  Perseus (thick lines) and Virgo (thin lines). The solid lines show
  the radial velocities estimated from eq.(\ref{eq:vcf}) and the
  dashed lines show velocities calculated from
  eq.(\ref{eq:sarazin}). The cooling rates for Perseus and Virgo
  are assumed to be constant and are 183 M$_{\odot}/{\rm yr}$
  and  10 M$_{\odot}/{\rm yr}$ correspondingly.
\label{fig:cf}
}
\end{figure}

Fig.\ref{fig:polM87} (top panel) shows the simulated radial profiles of the degree
of polarization for Virgo  in several prominent lines. The
highest polarization is achieved in the Fe XXIII 1.129 keV line,
reaching a maximum of 3.5$\%$ at the distance $\sim$15 kpc from the
center and falling off from this maximum with increasing distance from the
center. A similar behaviour pertains for the lines at 1.168 keV and
2.461 keV. In spite of higher optical depth of line at 1.168 keV, the
polarization is lower due to the smaller weight of dipole
scattering. It is also important to note that close to the 1.168 keV
line there is a second component at 1.163 keV, which produce
unpolarized emission and contaminate polarized signal from 1.168 keV
line unless the energy resolution is better that 5 keV. Polarization in the line of He-like (at 6.7 keV) iron  
increases slowly with projected radius
and is of order 1\%. Surface brightness profiles are also shown on the
Fig.\ref{fig:polM87} (bottom panel) for all lines shown on the upper
panel. We see that lines of Fe XXIII and S XV have not only the largest
polarization degree, but also the largest intensity in Virgo cluster.

Comparing profiles of the polarization degree in Perseus and M87 one
can notice a different behavior of the curves: in Perseus
cluster the polarization is an increasing function of radius, while
in M87/Virgo cluster the degree of polarization decreases with
radius.  This is caused by different radial behavior of the
density profiles, since the polarization degree strongly depends on
the $\beta$-parameter. It is clear, that the polarization degree is
smaller for smaller $\beta$, since for small $\beta$ the cluster
emission becomes less
centrally peaked, leading to a more isotropic radiation
field. According to the analytical solution given by \cite{Saz02}, at
large distances from the cluster center $\disp P= Q/I\propto
\frac{r^{-3\beta -1}}{r^{-6\beta +1}}=r^{3\beta -2}$. It is clear that if
$\beta>2/3$ (case of the Perseus cluster) the polarization
degree will increase with distance, while if $\beta<2/3$ (case of
the M87/Virgo cluster) emission is less peaked and the polarization
degree diminishes far from the cluster core.

\subsection{Canonical cooling flow model}

The polarization degree was calculated for several patterns of gas 
motions. First we consider a canonical cooling flow model (a slow flow
of cooling gas toward the center of the cluster). While we understand
the canonical cooling flow model has now been transformed into a
picture of feedback from central supermassive black holes, the model
provides a baseline for exploring symmetric gas flows. The
velocity of the flow can be estimated, assuming that the flow time is
approximately equal to the cooling time: 
\be 
r/v\approx t_\cool,
\label{eq:vcf}
\ee where $\displaystyle t_\cool=\frac{5/2nkT}{n^2\Lambda(T)}$, $n$ is
the gas density, and $\Lambda(T)$ is the gas cooling function. The
radial velocities estimated from eq. (\ref{eq:vcf}) using the observed
temperature and density profiles in Perseus and M87 are shown
in Fig.\ref{fig:cf} with solid lines. In both clusters the
velocities are very small and do not exceed 30 ${\rm km~s^{-1}}$.

Larger velocities are anticipated in a {\bf homogeneous} cooling flow
model which assumes that 
the mass flow rate is constant at all radii: $\displaystyle \dot{M}_\cool(r)=4\pi
r^2\rho v_\r={\rm const}$. \cite{Sar96} provides the following
estimate of the inflow velocity for such a model:
\begin{eqnarray}
&v_\r= \nonumber 9\left(\frac{r}{100kpc}\right)^{-1/2}\times \\ &\left[\frac{\dot
    M_\cool(r)}{300
    M_{\odot}yr^{-1}}\right]^{1/2}\left[\frac{T(r)}{10^8
    K}\right]^{-1/2}~~~~{\rm km~s^{-1}} .  
\label{eq:sarazin}
\end{eqnarray} 
According to \citet{Fab94} the cooling rates in M87 and Perseus
 are 10  M$_{\odot}{\rm /yr}$ and 183 M$_{\odot}{\rm /yr}$
respectively.  The radial velocities in Perseus and M87, estimated
from eq.(\ref{eq:sarazin}) using the observed temperature
profiles\footnote{We note here that the observed density and temperature
  profiles are not consistent with the homogeneous cooling flow
  model.}, are shown in Fig.\ref{fig:cf} with the dashed lines.  Even
if velocities are calculated using eq.(\ref{eq:sarazin}) they are not
large: at a distance of 1 kpc from the center they are about 30 km/s
for M87 and about 100 km/s for the Perseus cluster.

\begin{figure}
\plotone{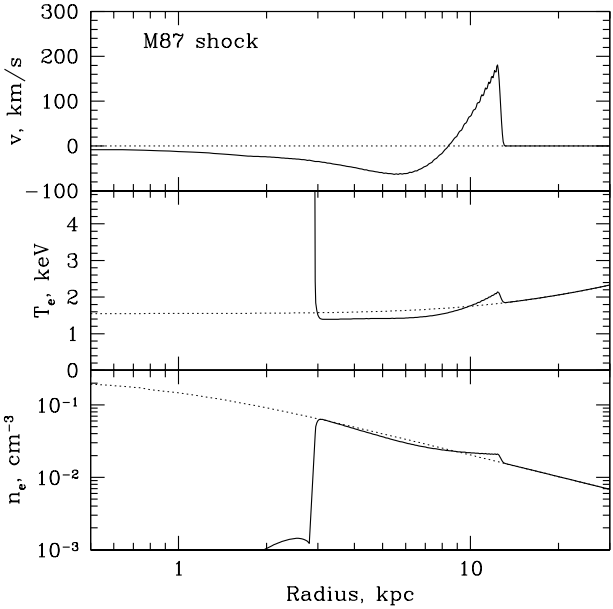}
\caption{Density, temperature and gas velocity profiles in 1D
  simulations of an expanding shock wave produced by an AGN outburst
  in the giant elliptical galaxy M87 with the total energy release of
  $E_0\sim 5\times10^{57}{~~\rm ergs}$ and the duration of the
  outburst $\Delta t\sim 2\times10^{6} {~~\rm yr}$ \citep{For09}.
  Dotted lines show the assumed initial density and temperature
  distributions. Solid lines in each panel show the density,
  temperature and velocity distributions about 12 Myr after the
  beginning of the outburst.
\label{fig:m87vp}
}
\end{figure}

Inclusion of velocities of cooling flows does not change the degree 
of polarization in either M87 or Perseus since the flow velocities 
are small. The only appreciable effect is in the very central 
parts of the clusters (Fig.\ref{fig:cf}) where the degree of polarization 
is very small anyway, with or without the flow.

\subsection{Spherical shock model}

\begin{figure}
\plotone{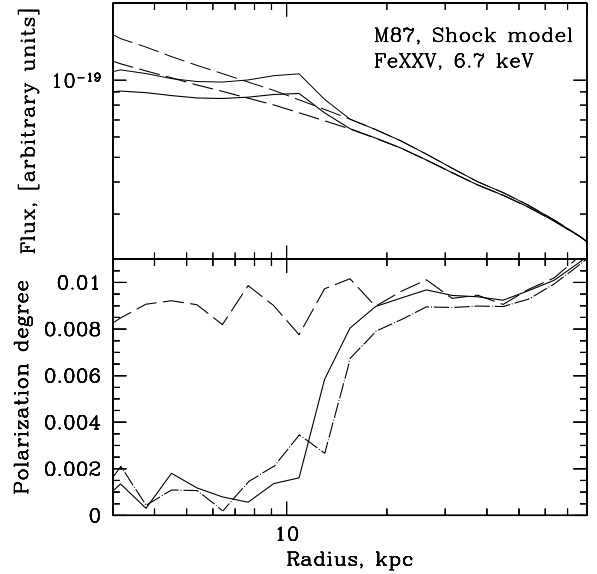}
\caption{Top panel: surface brightness profile in the resonant K$_{\alpha}$
  line of He-like iron at 6.7 keV for Virgo/M87. The dashed curves
  correspond to the initial temperature and electron number density
  distributions and the solid curves are for the case of a propagating
  shock wave. Here, the upper curves (dashed and solid) show profiles
  without scattering while the other two curves show surface
  brightness taking into account scattering.\newline
 Bottom panel: expected degree of polarization in the 6.7 kev line. The dashed curve corresponds to the initial gas density and temperature distributions without the shock. The solid curve corresponds to the density, temperature and velocity profile expected in the case of a Mach 1.2 shock at r=13 kpc propagating through the ICM. The dashed-dotted line correspond to the same case, but with the gas velocity set to zero. The differences between the last two
  profiles is not large. Therefore, changes in polarization degree
  are most sensitive to changes in number density and temperature
  profiles, rather than to the gas velocity.
\label{fig:m87sbr6.7}
}
\end{figure}


As another possible pattern of gas motions, we consider an expanding
spherical shock, using M87 as an example \citep{For05,For07}.  This
can be considered as a prototypical case of a cool core cluster,
albeit less luminous than the Perseus cluster. All cool core clusters contain a
supermassive black hole (an AGN) in a giant elliptical galaxy at the
cluster center which is believed to be the source of energy for the
cooling gas. AGN activity is also a natural candidate for generation
of gas motions: it can either cause turbulent motions by stirring the
gas or produce larger scale gas motions in a form of an expanding
shock wave. Following \citet{For07,For09}, we used 1D simulations of
an expanding shock in M87, which is produced by the AGN outburst with
the total energy release $E_0\sim 5\times10^{57}{~~\rm ergs}$ and the
duration of the outburst $\Delta t\sim 2\times10^{6} {~~\rm
  yr}$. These parameters were derived based on the comparison of the
shock simulations with the data of a 0.5 Ms long Chandra observations
of M87 by \citet{For07,For09}. Observed position of the shock is at
$\sim 2.8'$ (13 kpc) from the center of M87 and the Mach number of the
shock is $\sim 1.2$. Shown in Fig.\ref{fig:m87vp} are results of 1D
simulations of shock propagation through the M87 atmosphere about 12
Myr after the beginning of the outburst: solid lines in each panel
show the density, temperature and velocity distributions. Dotted lines
show the assumed initial density and temperature distributions. The
energy in the simulations was released in a small volume near the
center which by the end of the simulations has expanded into a $\sim$3
kpc sphere filled with a very high entropy gas. This is of course the
result of our adopted 1D geometry. In reality, the structure of the
inner 3 kpc is much more complicated \citep[see][]{For07}. We expect
however that our modelling of the outer part is sufficiently
accurate and can be used as an illustrative example of the impact of
a weak shock on the polarization signal. The gas properties
(Fig.\ref{fig:m87vp}) show all the features of a weak spherical shock, in
particular a ``sine-wave'' structure of the velocity distribution. The
maximum positive velocity (expansion) is $\sim 180~{\rm km~s^{-1}}$,
while maximal negative velocity (contraction) is $\sim 60~{\rm
  km~s^{-1}}$.

For the Virgo cluster we then carried out three radiative transfer
simulations: {\bf A}) using initial undisturbed profiles (dotted lines in
Fig.\ref{fig:m87vp}), {\bf B}) using density and temperature distributions
as in the shock, but setting the gas velocity to zero and {\bf C}) using density,
temperature and velocity distributions as  in  the shock (solid lines in
Fig.\ref{fig:m87vp}). The reason for doing simulations {\bf B} is that
we want to see the impact of the nonzero velocity separately from
other effects. We preformed calculations for the line with energy 6.7 keV
(see Table \ref{tab:lines}) as the influence of  the shock
wave at such energy is more noticeable.

The results of radiative transfer calculations are
shown in Fig.\ref{fig:m87sbr6.7}.  Shown
in the top panel are the surface brightness profiles for cases A
(the dashed curves) and B (the solid curves). Profiles in  the simulation
C are the same as in  case B. Here, the upper lines correspond to
the profiles without scattering and the lower lines show surface
brightness, taking into account scattering.  Due to resonant
scattering the surface brightness becomes weaker in the center and
stronger outside. Because of shock wave propagation the number density
in the center becomes smaller leading to the lower surface brightness
in the cluster center. Also, due to  the shock wave, we see a characteristic
peak $\sim$10 kpc from the center.  Bottom panel in
Fig.\ref{fig:m87sbr6.7} shows
 the polarization degree for cases A, B, C. We see that after the shock
wave propagates, the gas is less dense, therefore  the  line has smaller
optical depth and  the polarization decreases. The velocities of gas are
larger than in the case of cooling flows and lead again to the
decrease of  the polarization degree.

\section{Three-dimensional problem}
\label{sec:3dprob}

We now consider full three-dimensional models of galaxy
clusters, taken from large-scale structure formation. The density,
temperature distributions and the velocity field are taken from a set of
high-resolution simulations of galaxy clusters \citep[see e.g.][]{Dol05,Dol08} which are based on Gadget-2 SPH simulations \citep{Spr01}
and include various combinations of physical effects. We use the
output of non radiative simulations. All data are adaptively
smoothed and placed in a cube with half size 1000 kpc and cell
size 3 kpc. The resonant scattering is calculated using a Monte-Carlo
approach, which is described above in section 3.

\begin{figure*}
{\centering \leavevmode
\epsfxsize=0.65\columnwidth \epsfbox[40 160 610 650]{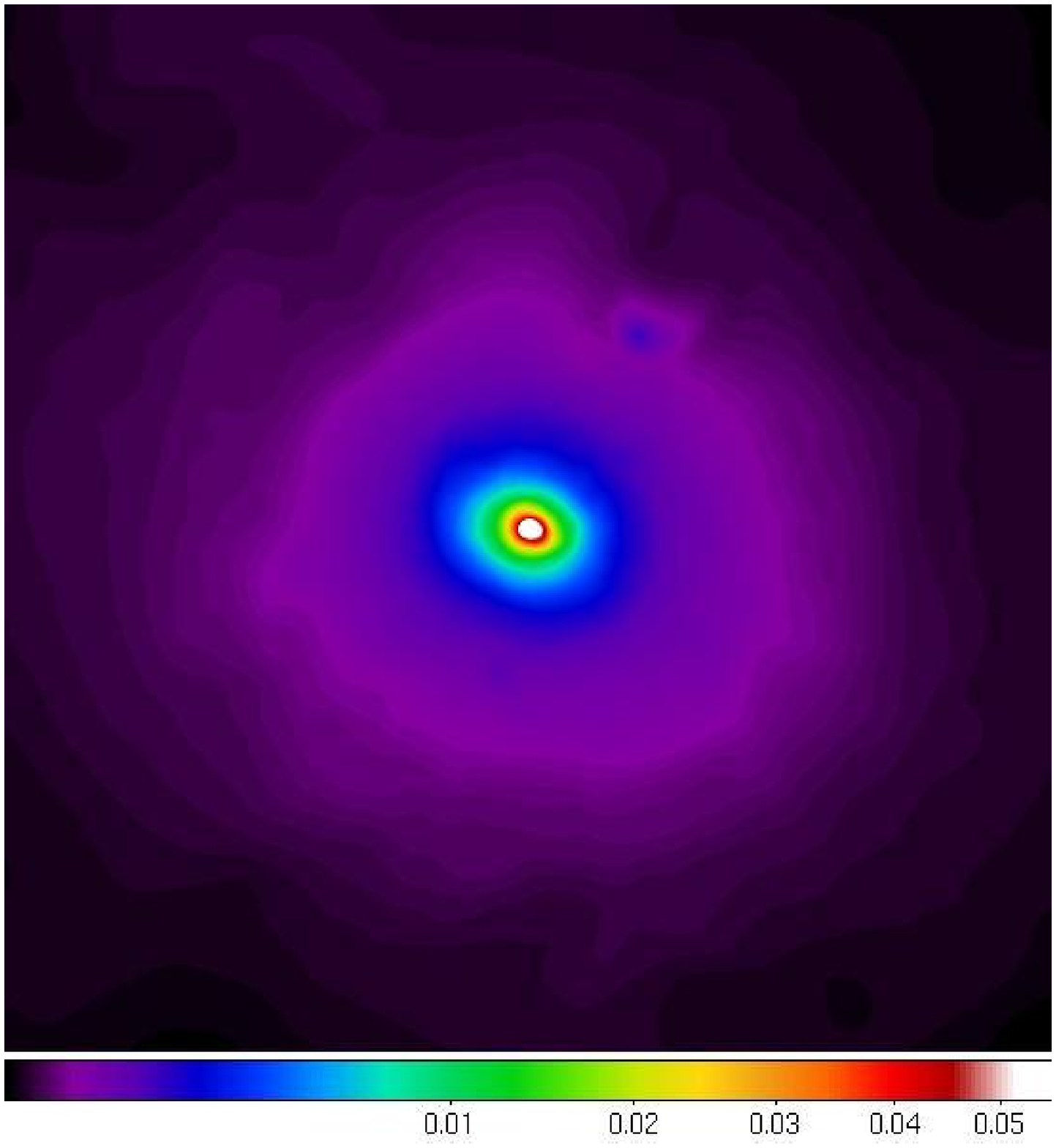}
\epsfxsize=0.65\columnwidth \epsfbox[20 160 590 650]{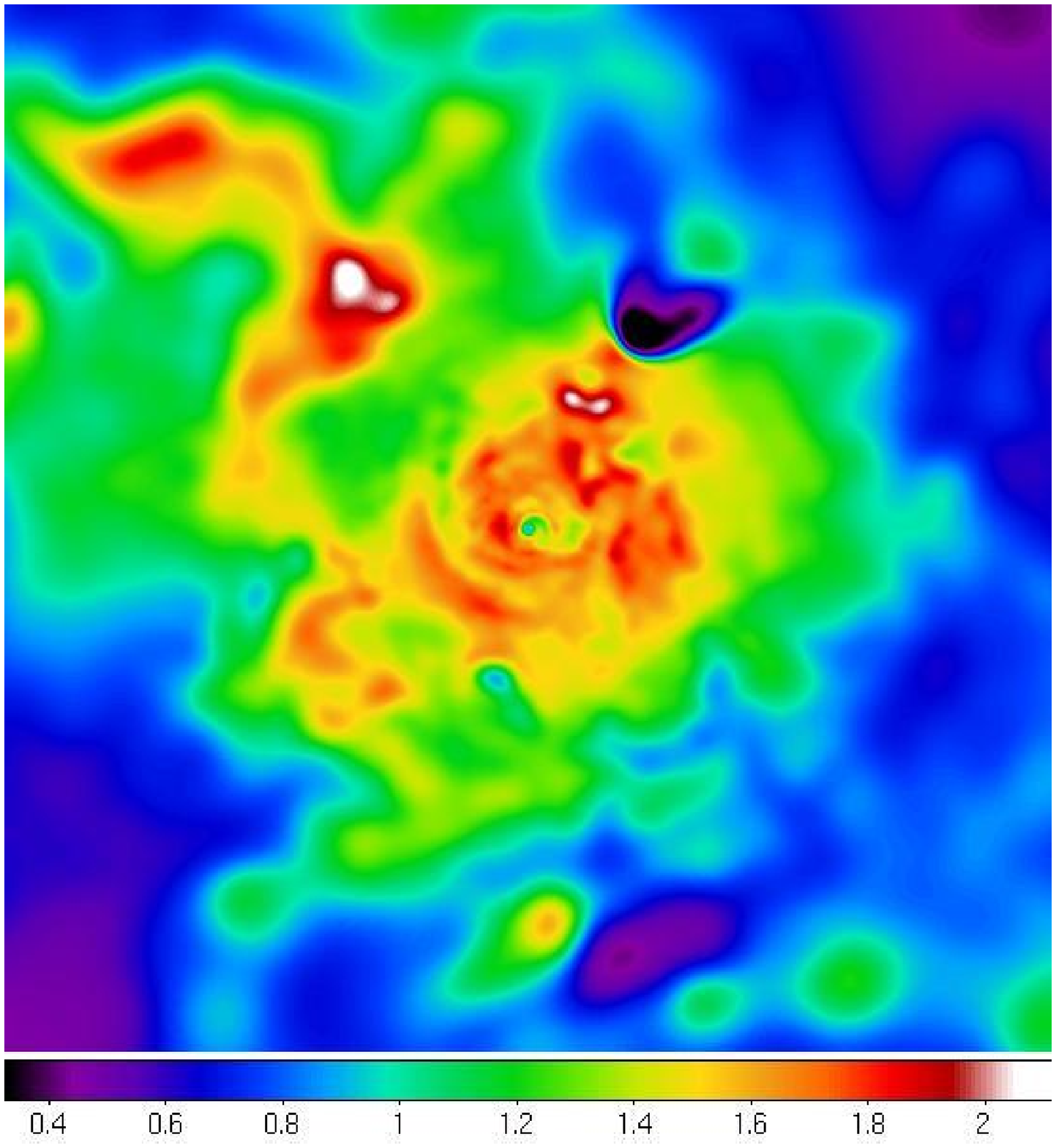}
\epsfxsize=0.65\columnwidth \epsfbox[0 160 570 650]{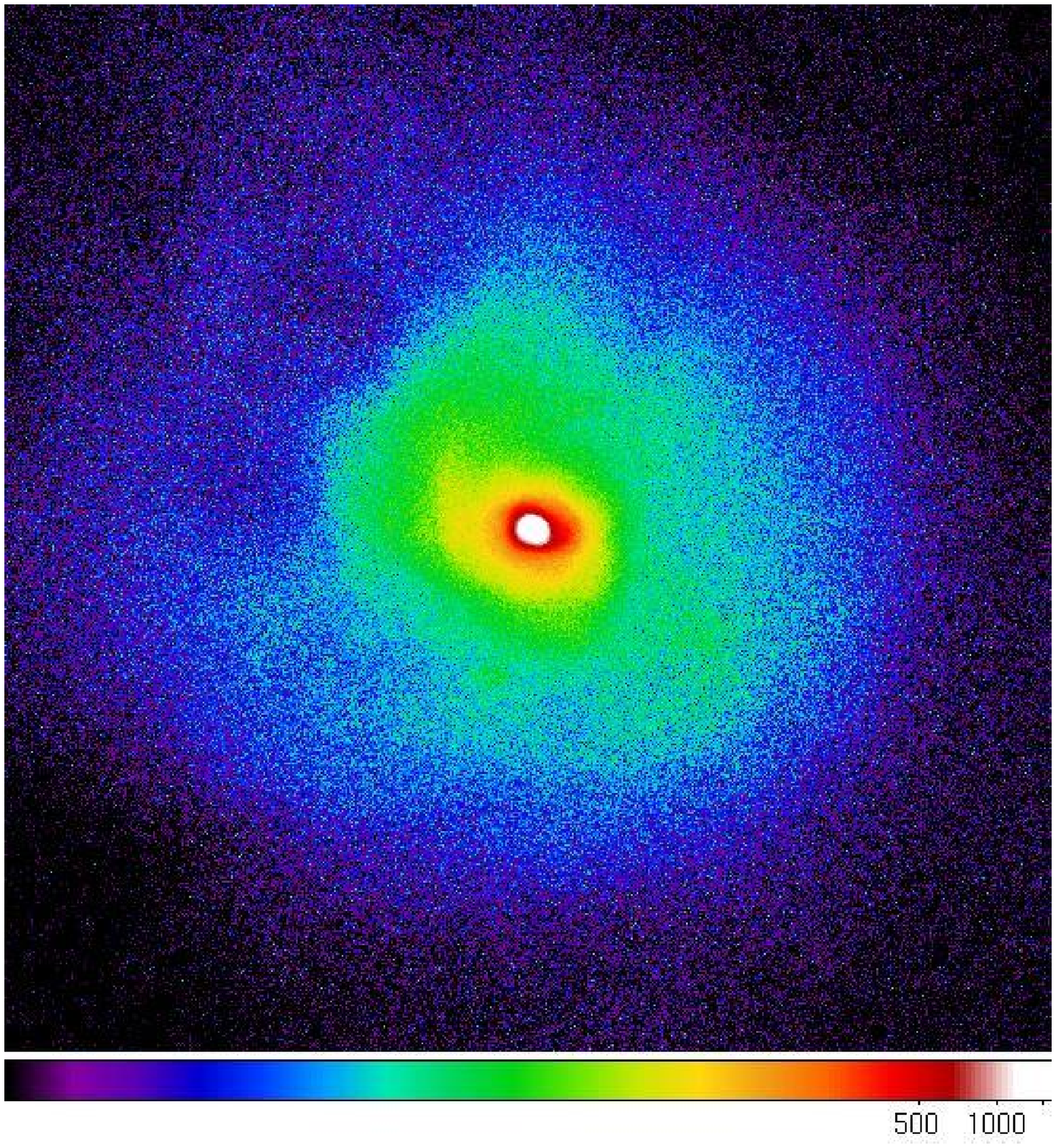}\hfil
\epsfxsize=0.65\columnwidth \epsfbox[40 300 610 780]{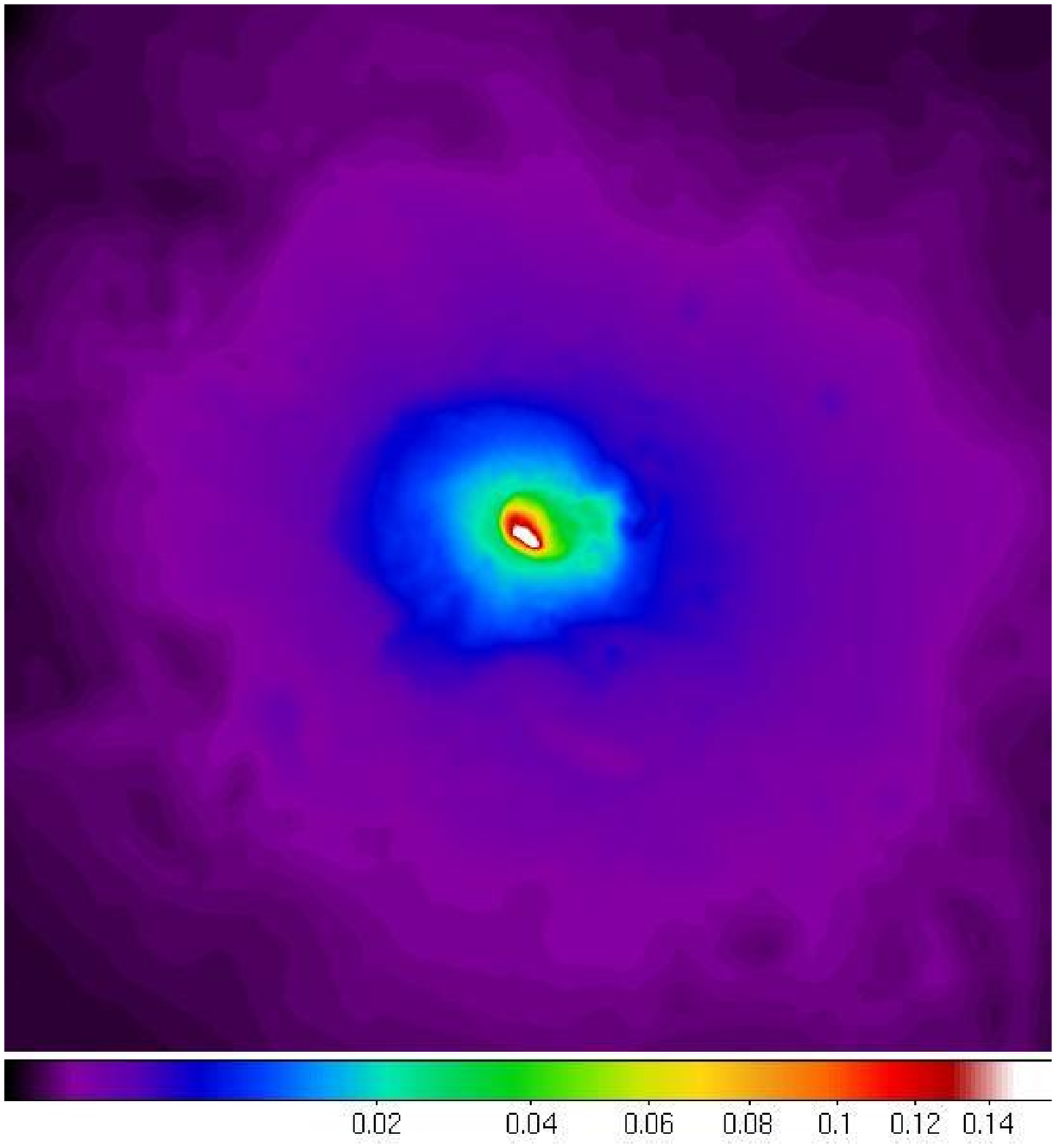}
\epsfxsize=0.65\columnwidth \epsfbox[20 300 590 780]{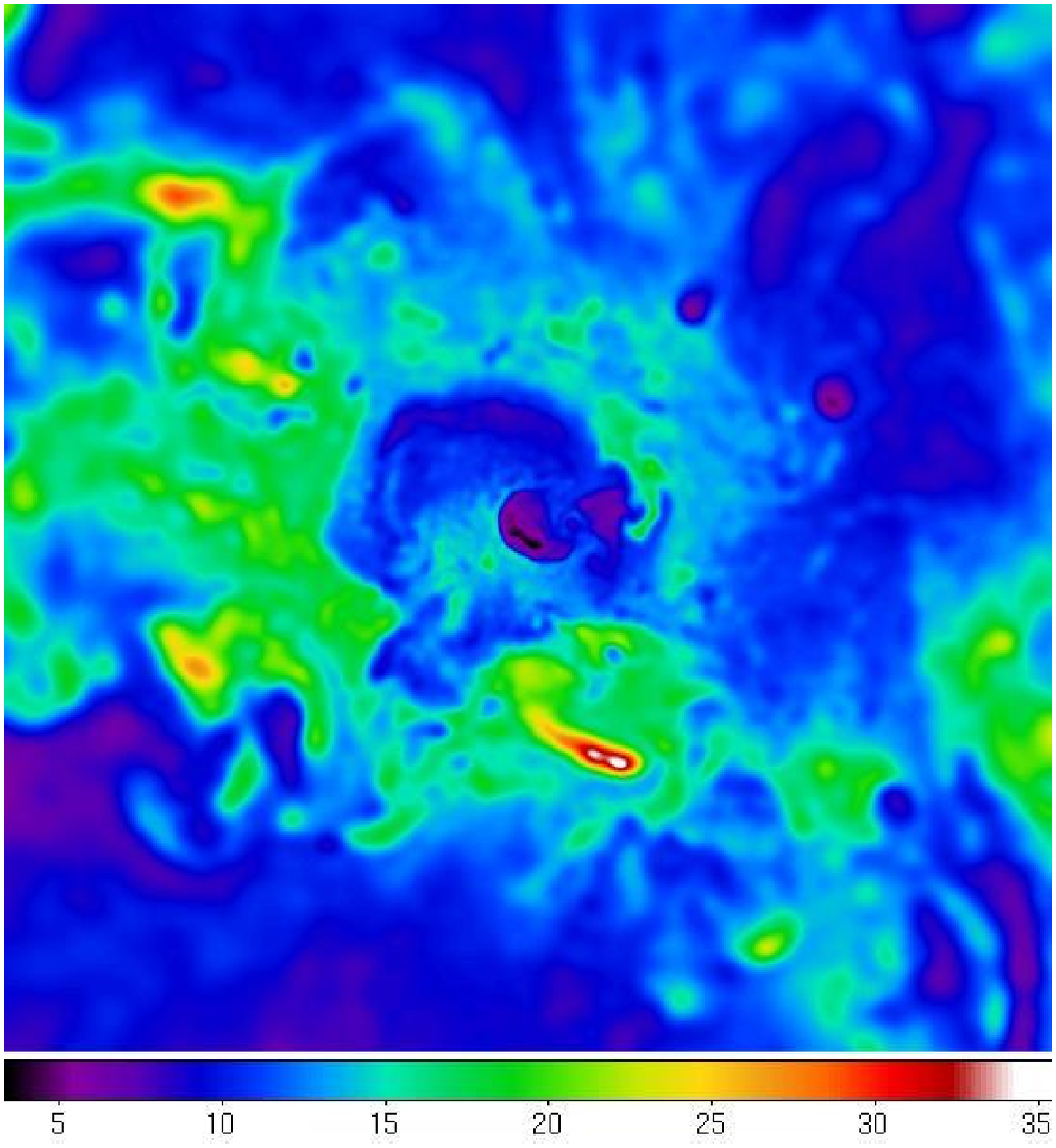}
\epsfxsize=0.65\columnwidth \epsfbox[0 300 570 780]{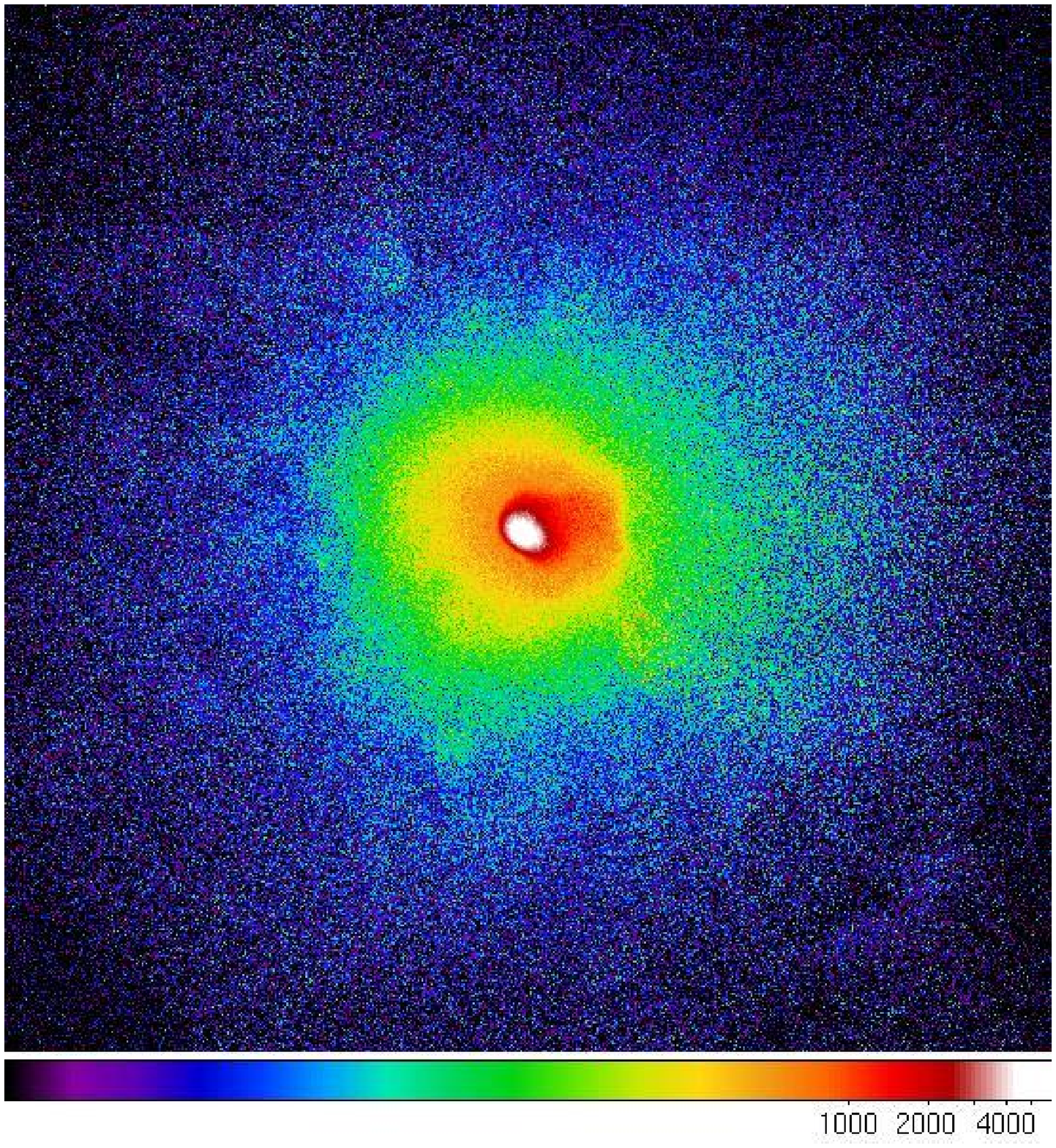}\hfil
\epsfxsize=0.65\columnwidth \epsfbox[40 10 610 920]{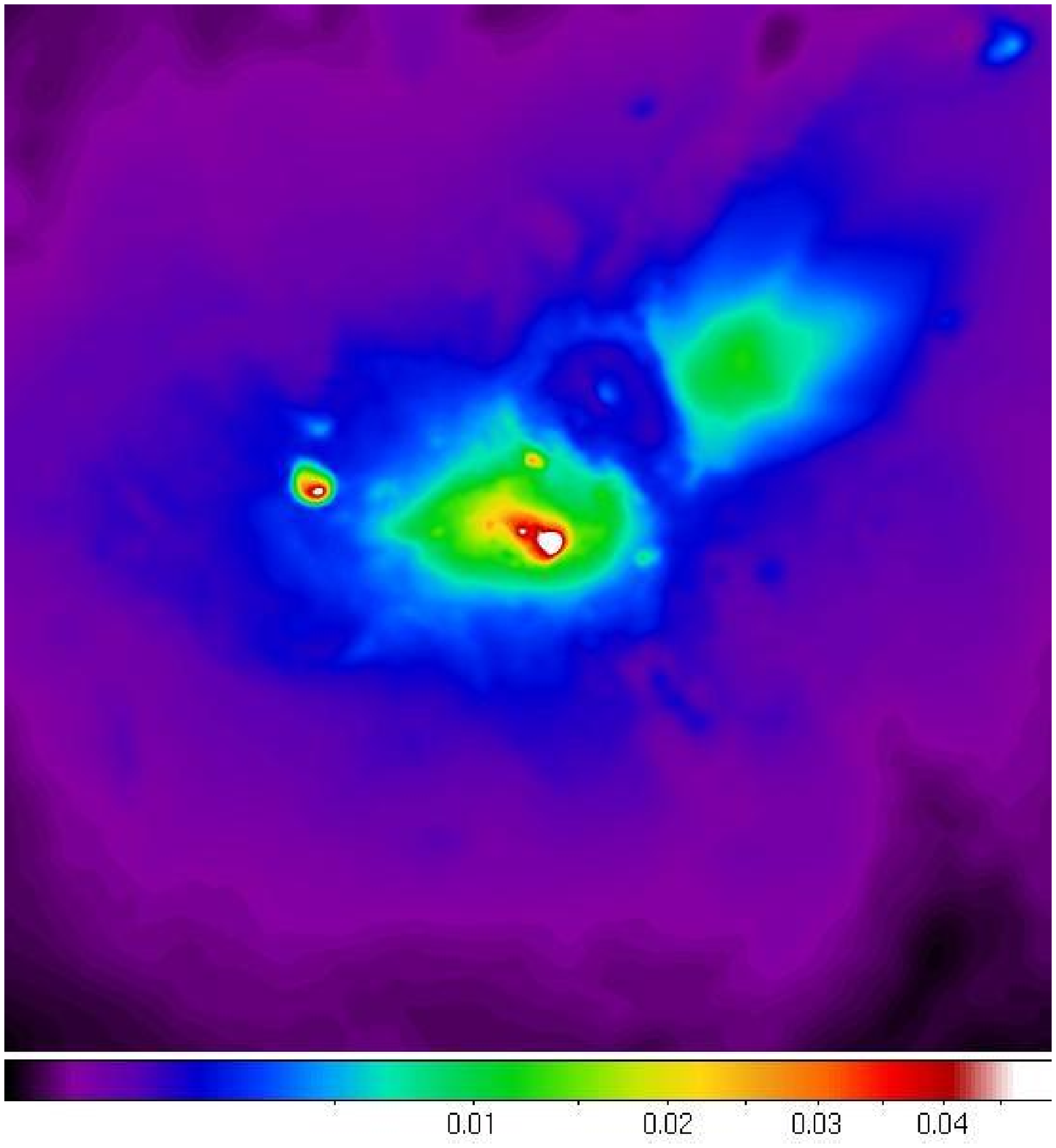}
\epsfxsize=0.65\columnwidth \epsfbox[20 10 590 920]{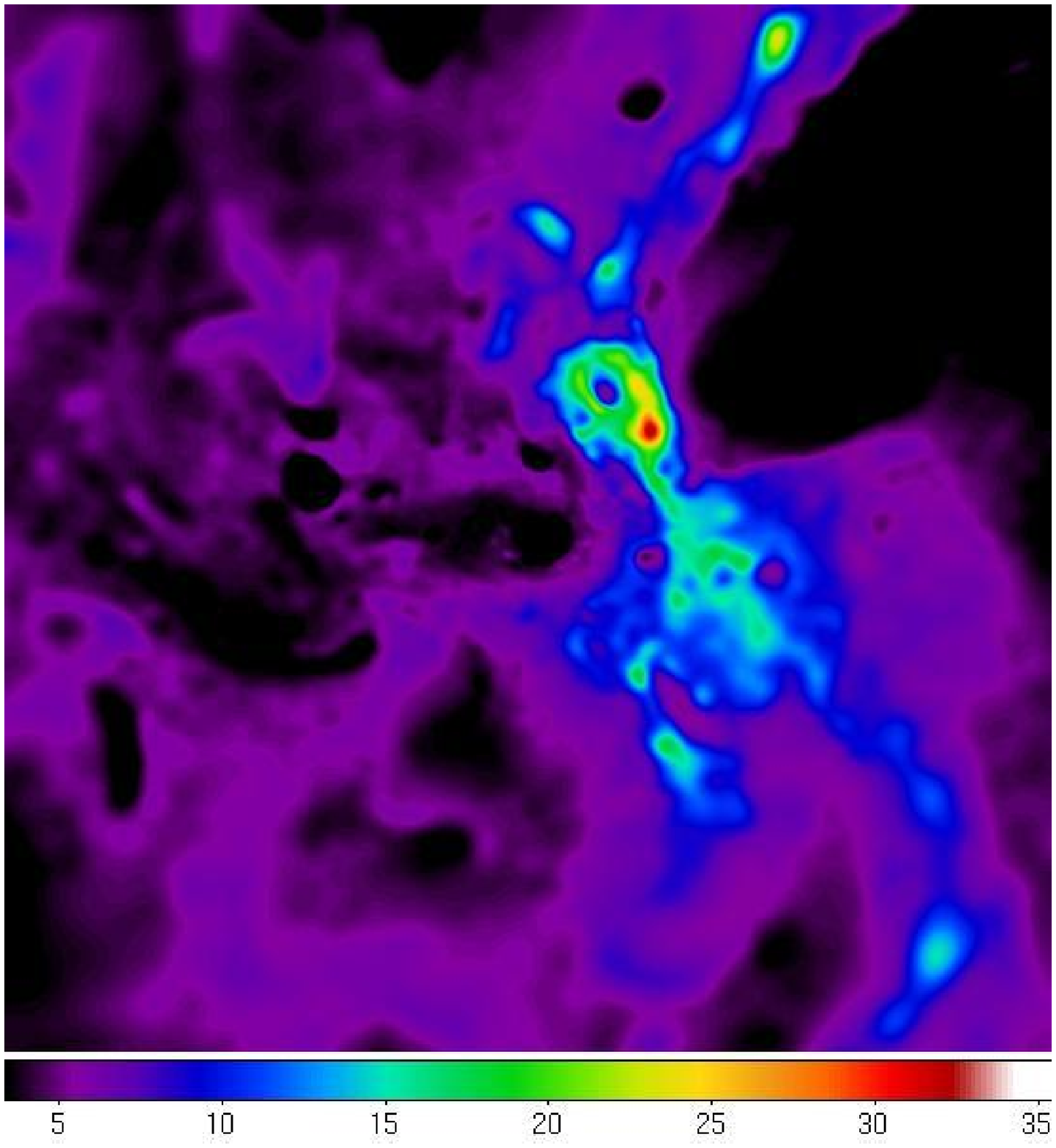}
\epsfxsize=0.65\columnwidth \epsfbox[0 10 570 920]{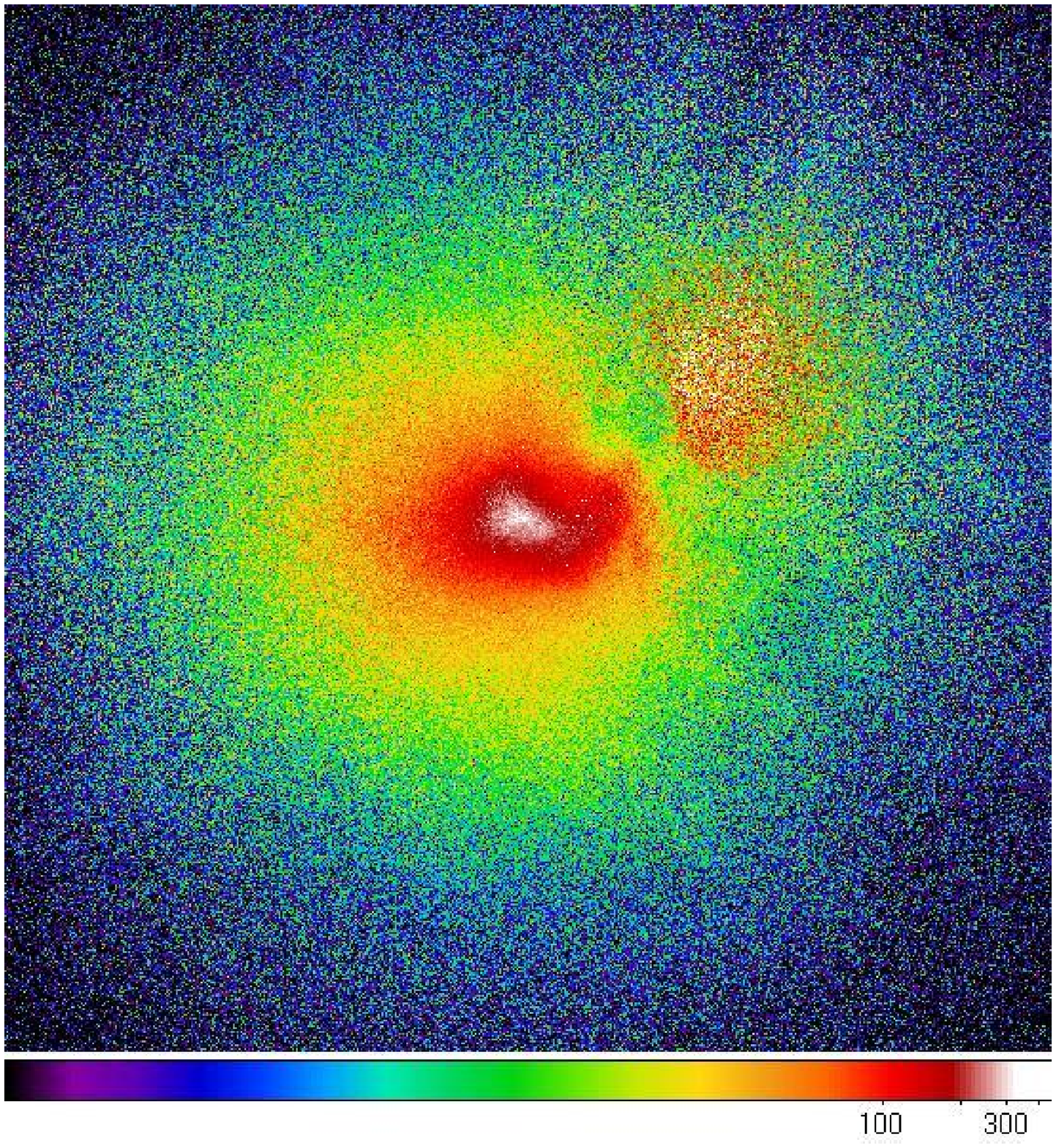}}
\caption{Simulated clusters g6212 (the top panels), g8 (the middle
  panels) and g72 (the bottom panels). Slices of the electron density in
  ${\rm cm}^{-3}$ are in the left column and the gas temperature in keV
  are in the middle
  column. The slices go through the center of the clusters and have an
  effective thickness of 3 kpc, corresponding to the size of one cell.
The right panels show the projections of surface brightness (${\rm
  photons/s/cm^2/arcmin^2}$) on the plane perpedicular to the line of
sight. For g8 and g72 clusters we consider the He-like iron line at
6.7 keV and for g6212 cluster the line of Fe XXI at 1.009 keV.
 The image size is 2$\times$2 Mpc, resolution is 3.6 kpc. 
\label{fig:inputcl}
}
\end{figure*}

We discuss here results of calculations for three simulated galaxy
clusters: g6212, g72 and g8. Table \ref{tab:clpar} shows the basic
parameters of  the chosen clusters. The slices of the density and
temperature distributions for all clusters and their projected images
in X-ray lines are shown in Fig.\ref{fig:inputcl}.

\begin{figure*}
{\centering \leavevmode
\epsfxsize=0.55\columnwidth \epsfbox[40 160 610 650]{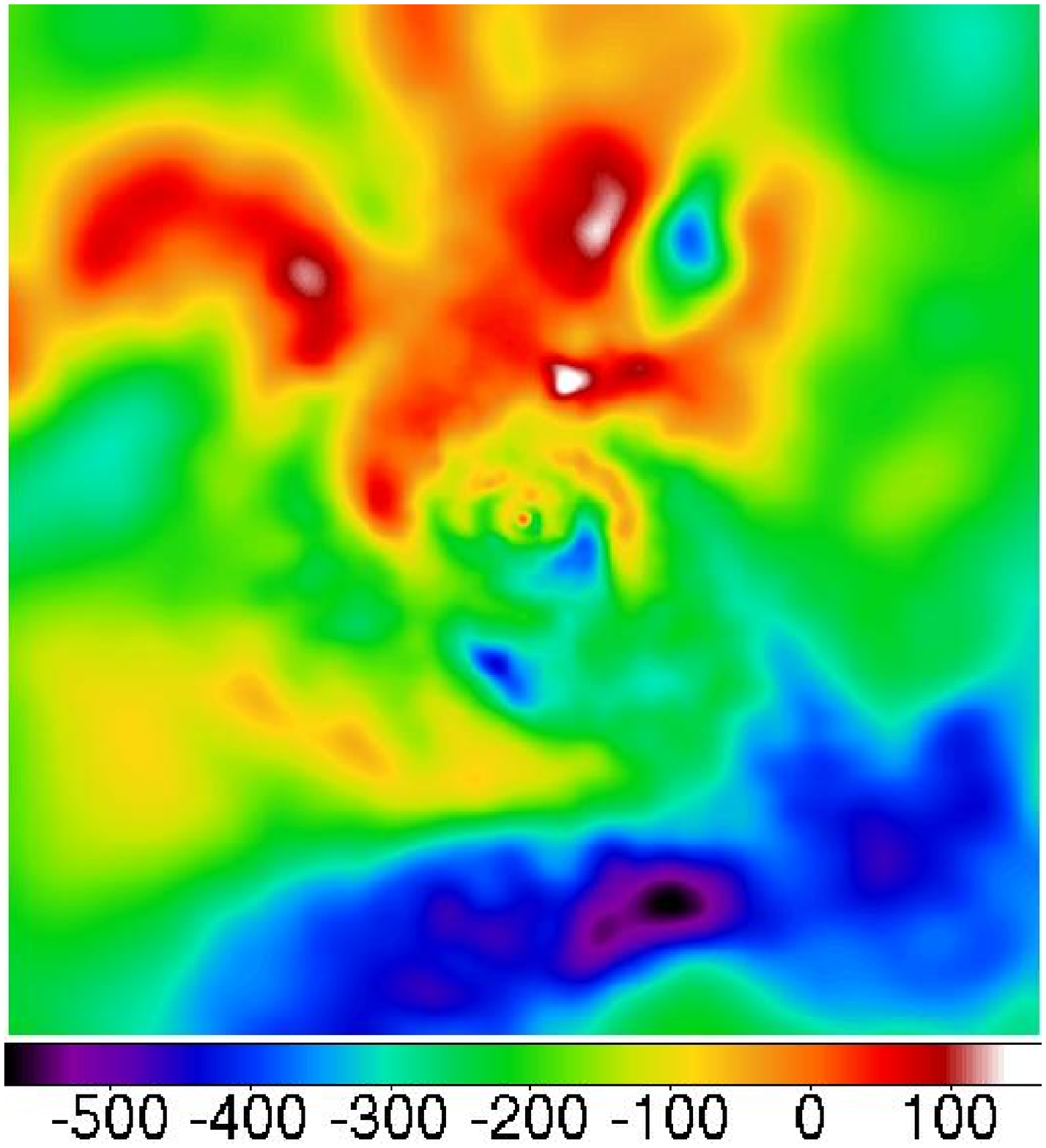}
\epsfxsize=0.55\columnwidth \epsfbox[20 160 590 650]{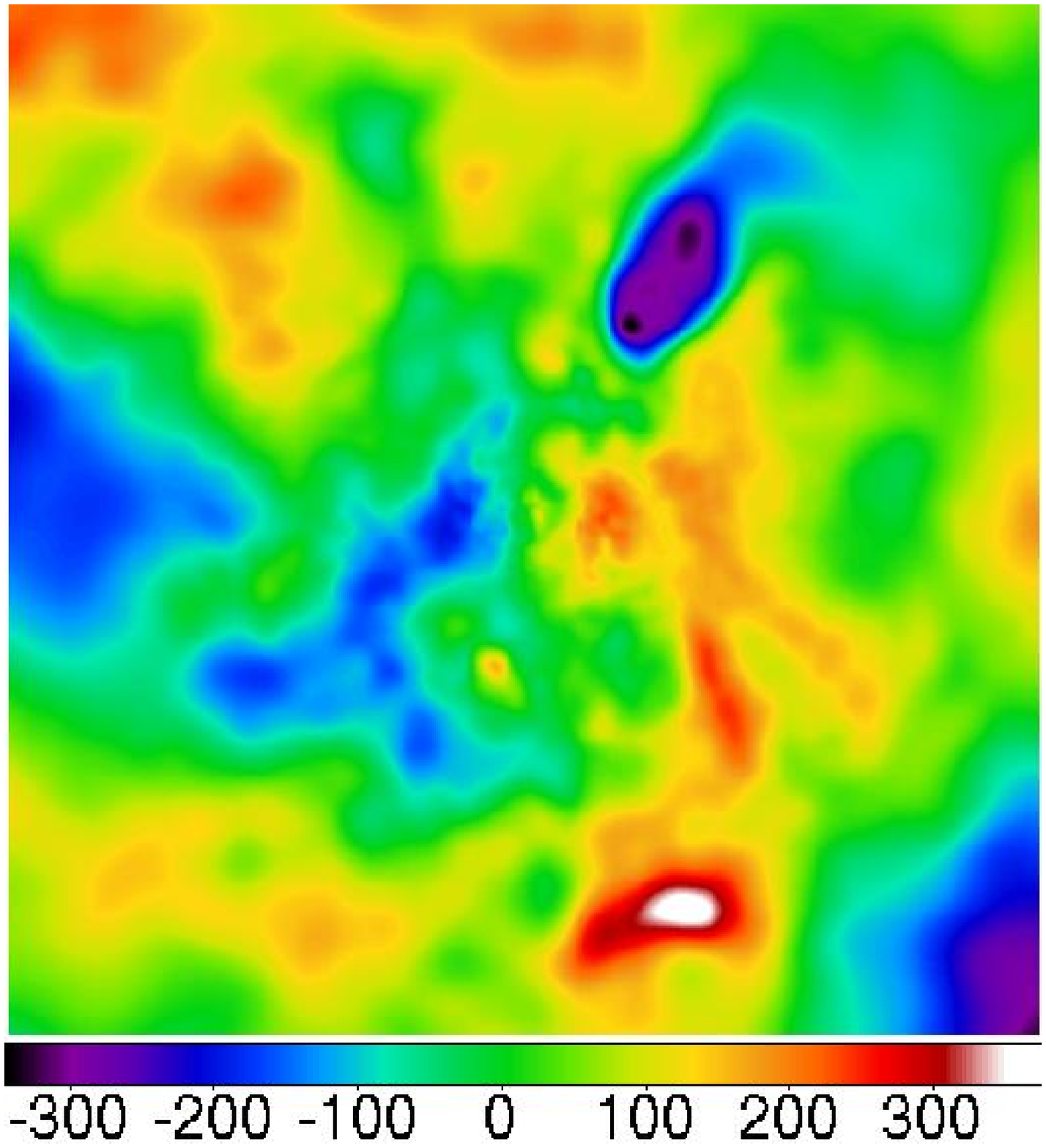}
\epsfxsize=0.55\columnwidth \epsfbox[0 160 570 650]{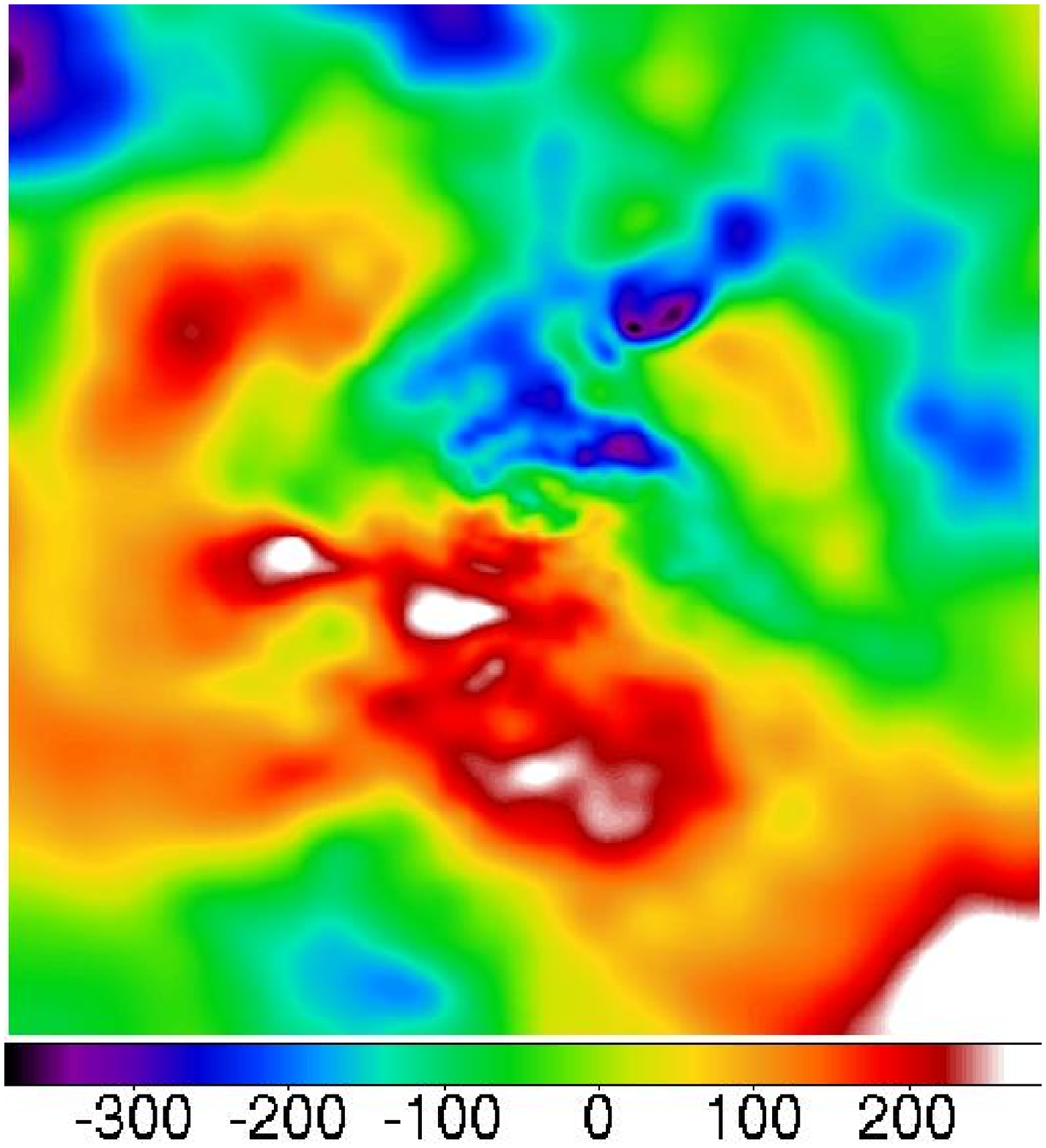}\hfil
\epsfxsize=0.55\columnwidth \epsfbox[40 300 610 780]{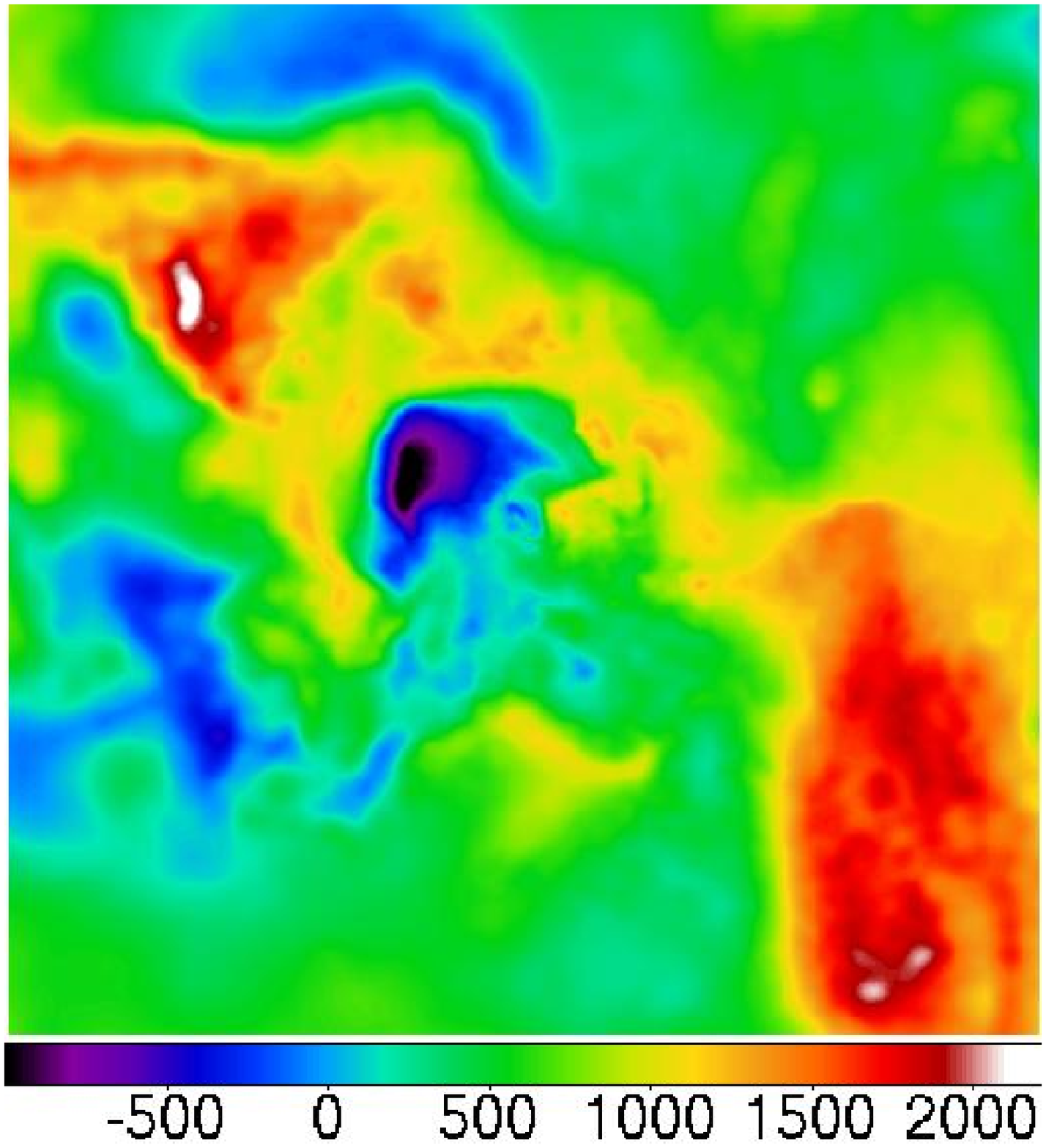}
\epsfxsize=0.55\columnwidth \epsfbox[20 300 590 780]{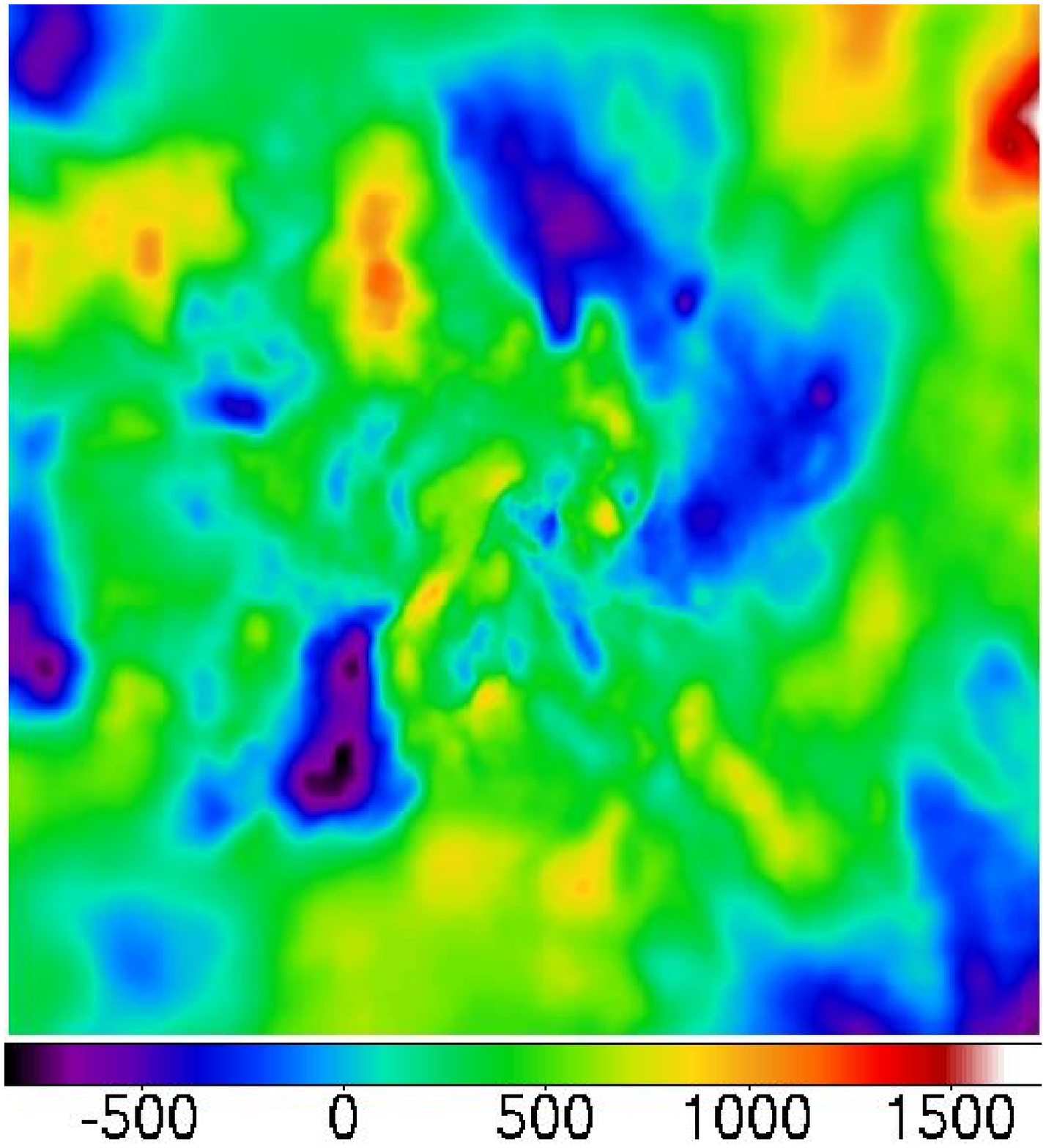}
\epsfxsize=0.55\columnwidth \epsfbox[0 300 570 780]{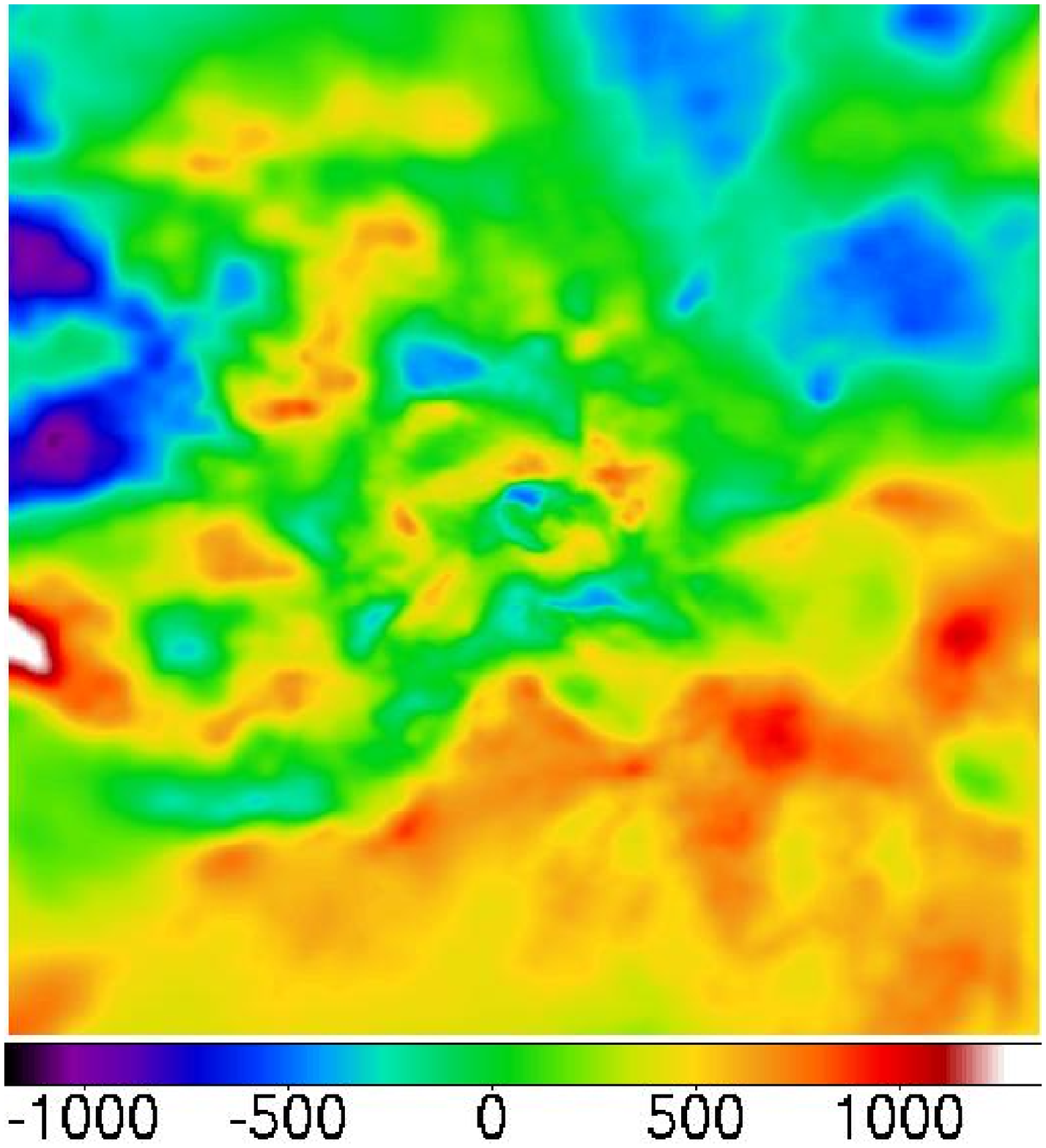}\hfil
\epsfxsize=0.55\columnwidth \epsfbox[40 10 610 920]{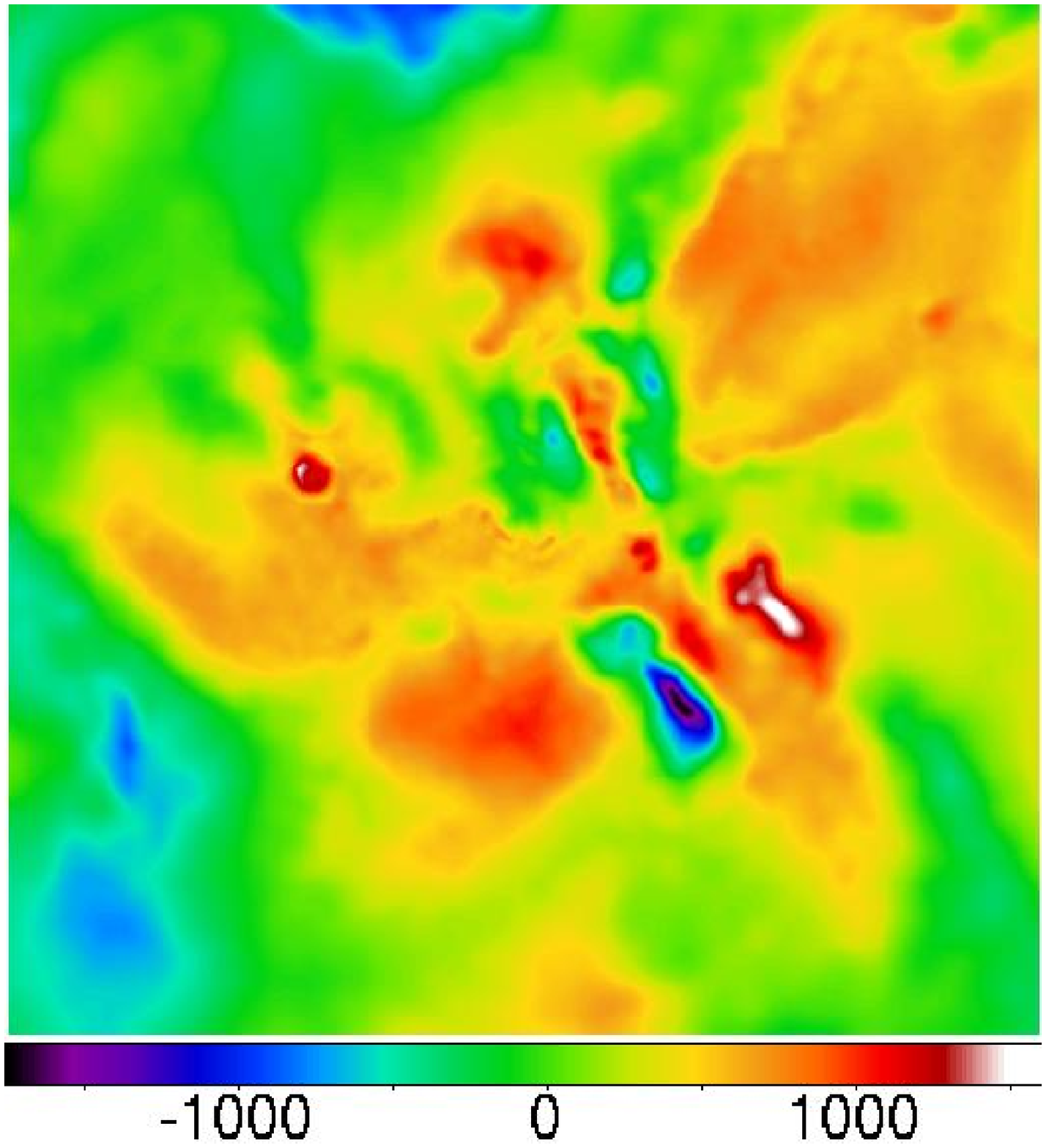}
\epsfxsize=0.55\columnwidth \epsfbox[20 10 590 920]{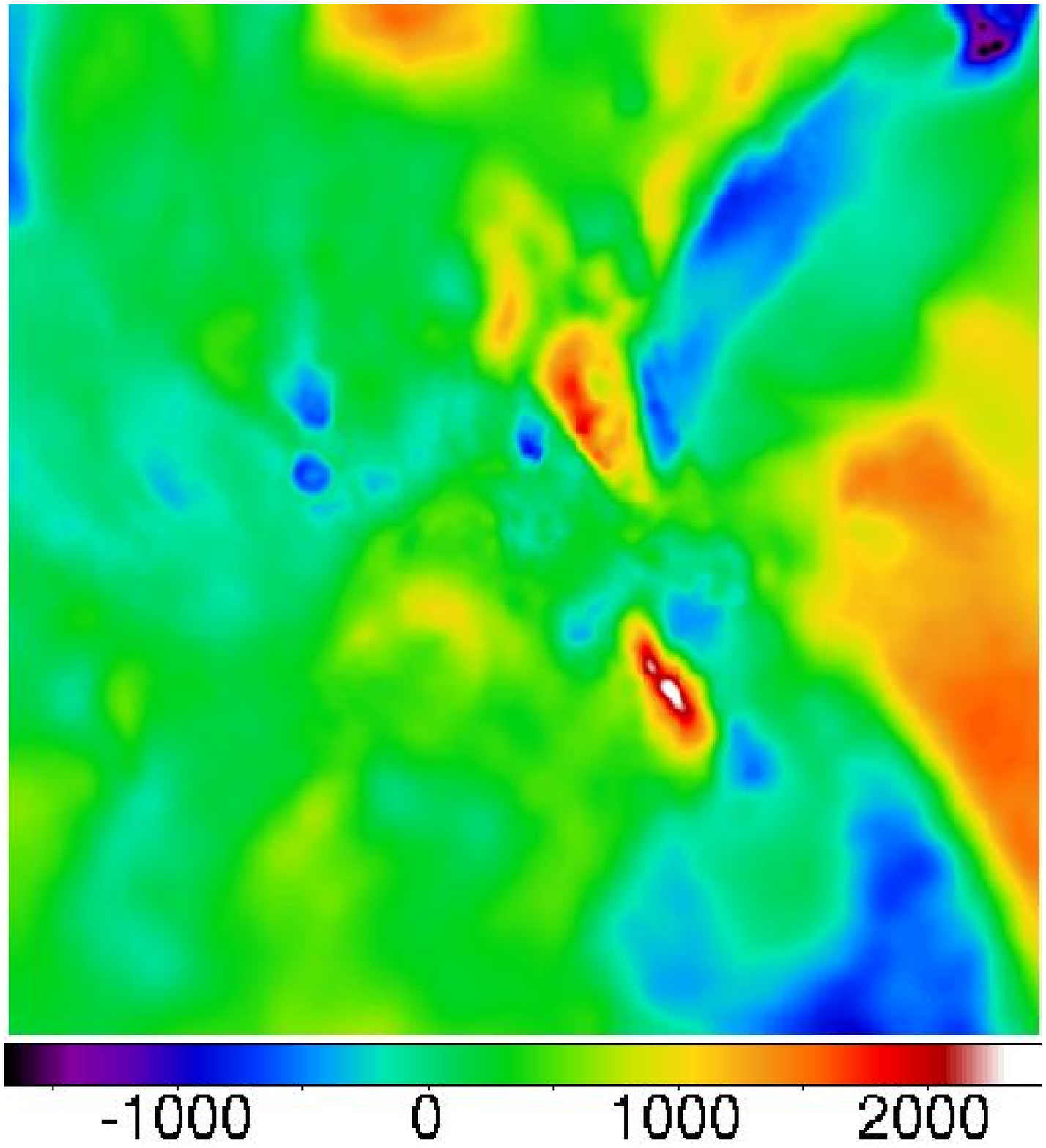}
\epsfxsize=0.55\columnwidth \epsfbox[0 10 570 920]{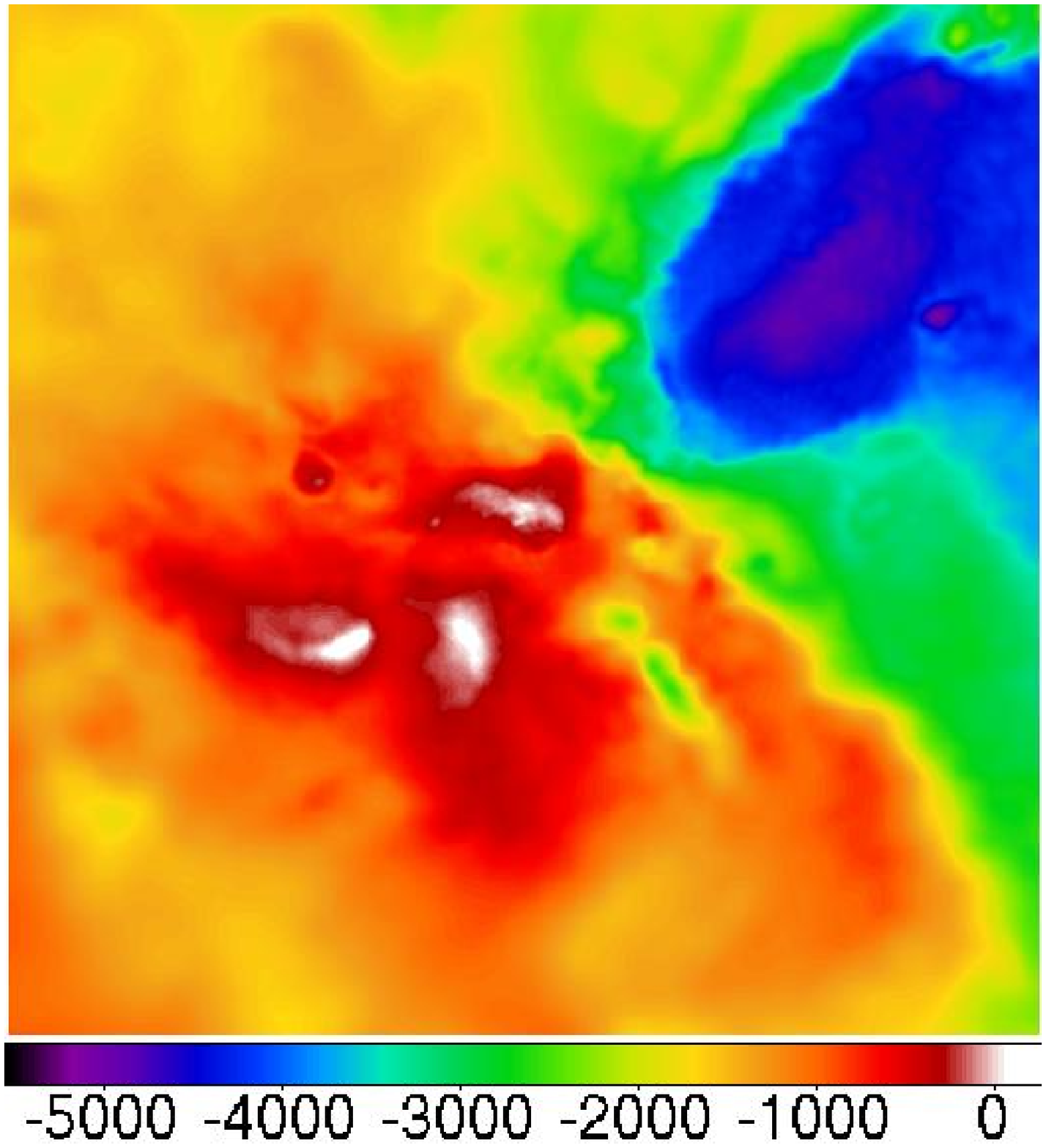}}
\caption{Slices of the velocity field (in km/s) in simulated clusters g6212 (the
  top panels), g8 (the middle panels) and g72 (the bottom panels). The
  right column shows x component of velocity, the middle column
  y component and the left column z component. The slices go through
  the center of the clusters and have an effective thickness of 3 kpc, 
  corresponding to the size of one cell. The image size is 2$\times$2 Mpc.  
\label{fig:velcl}
}
\end{figure*}

\begin{figure*}
{\centering \leavevmode
\epsfxsize=0.62\columnwidth \epsfbox[40 40 610 690]{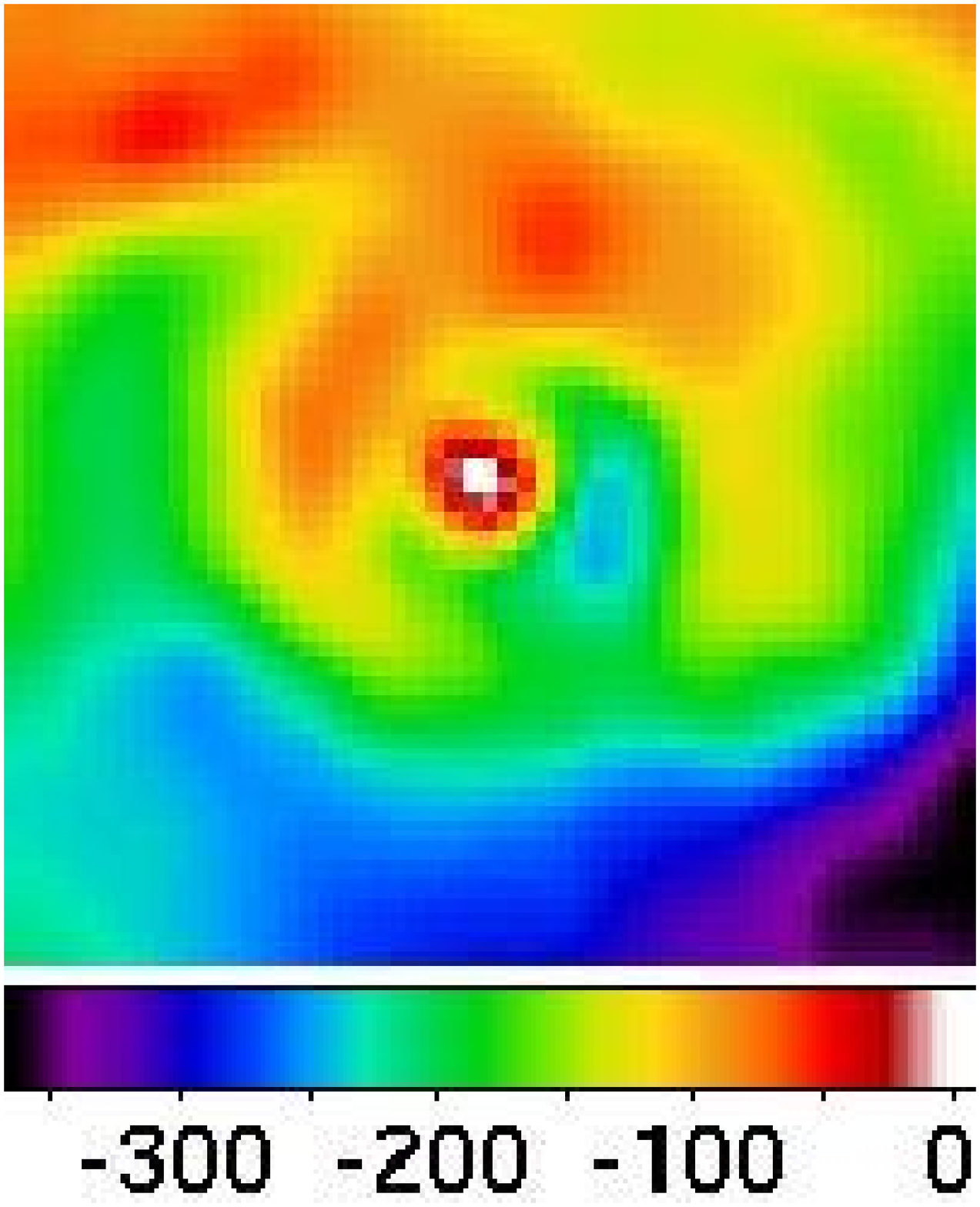}
\epsfxsize=0.62\columnwidth \epsfbox[20 40 590 690]{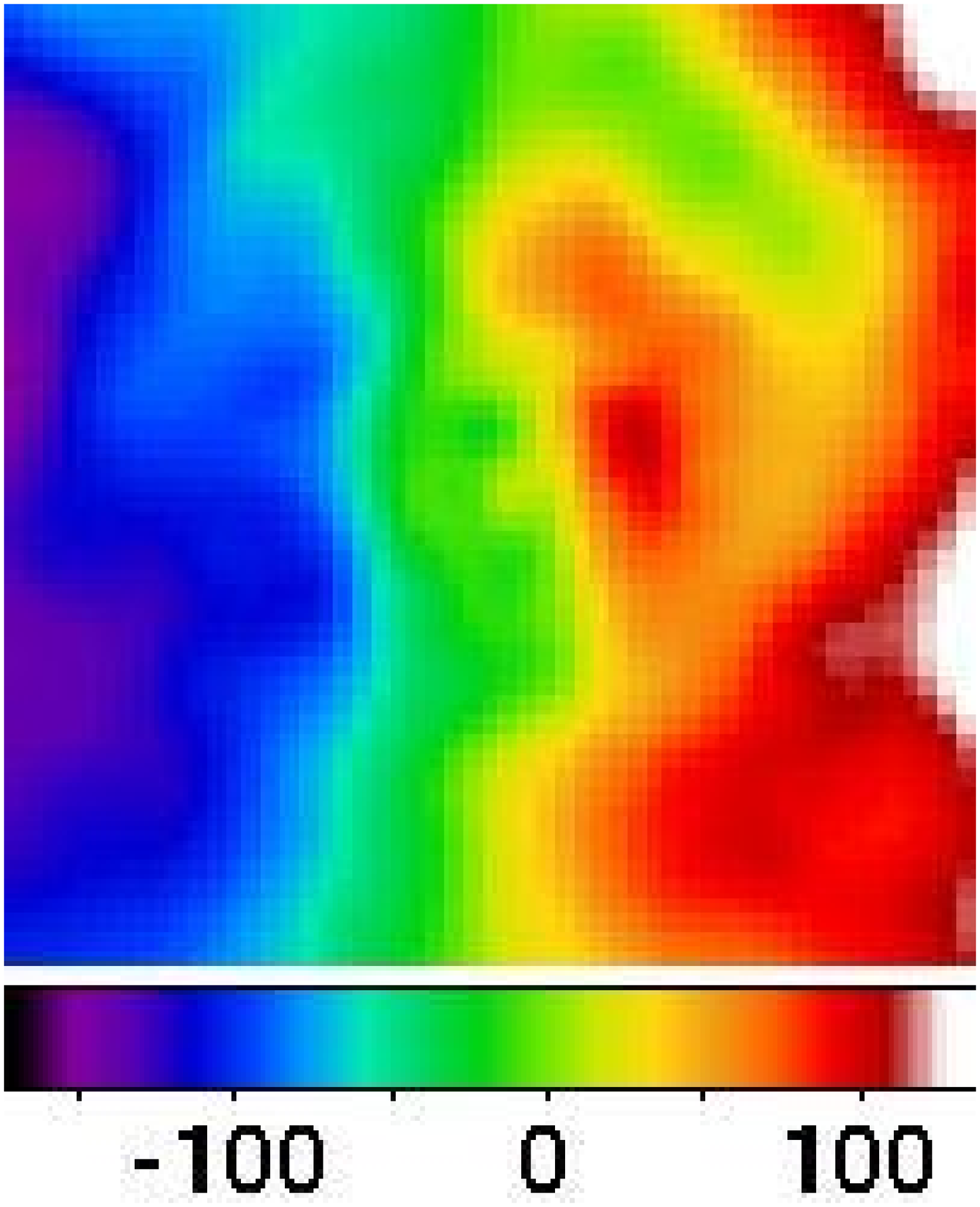}
\epsfxsize=0.62\columnwidth \epsfbox[0 40 570 690]{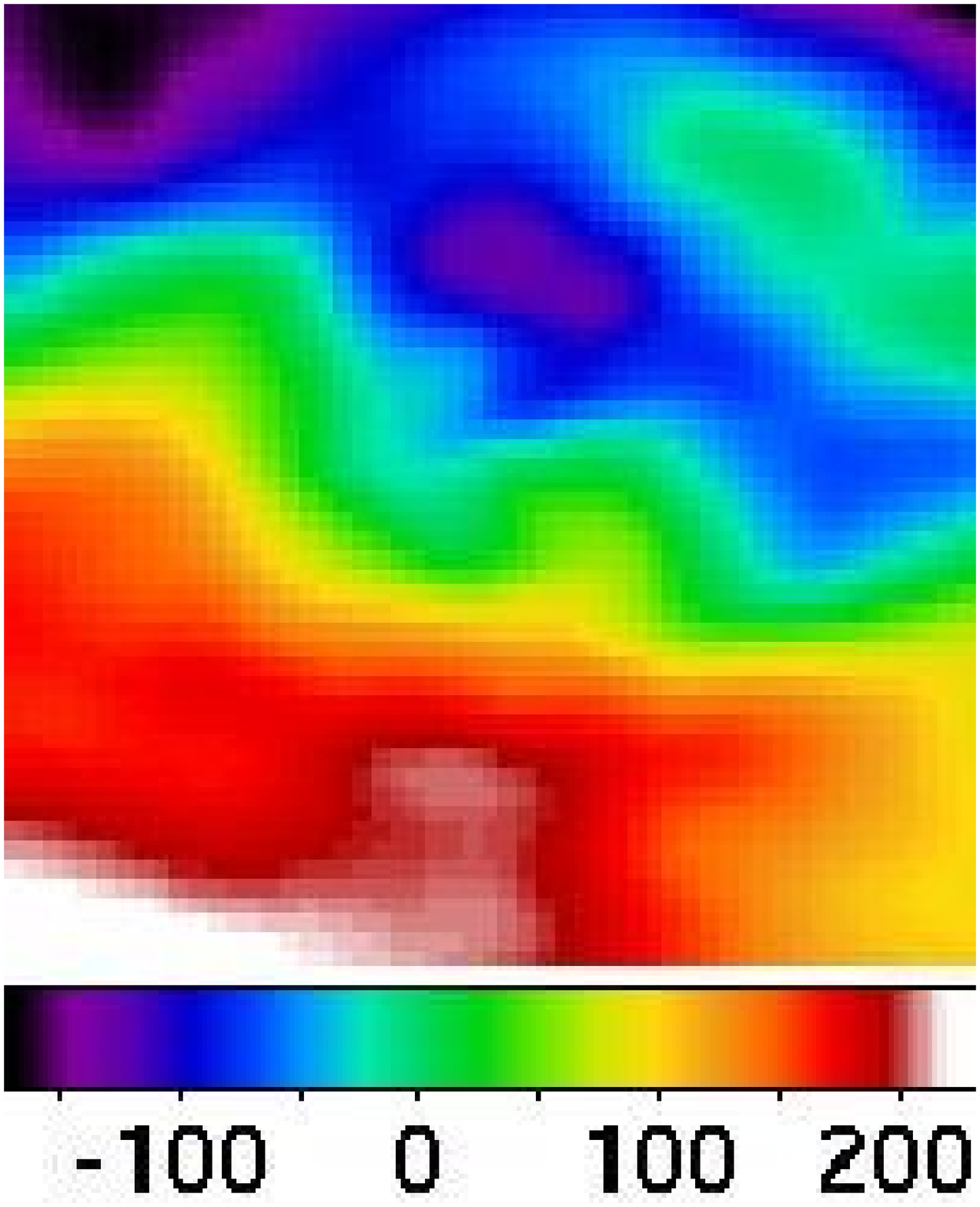}\hfil
\epsfxsize=0.62\columnwidth \epsfbox[40 100 610 800]{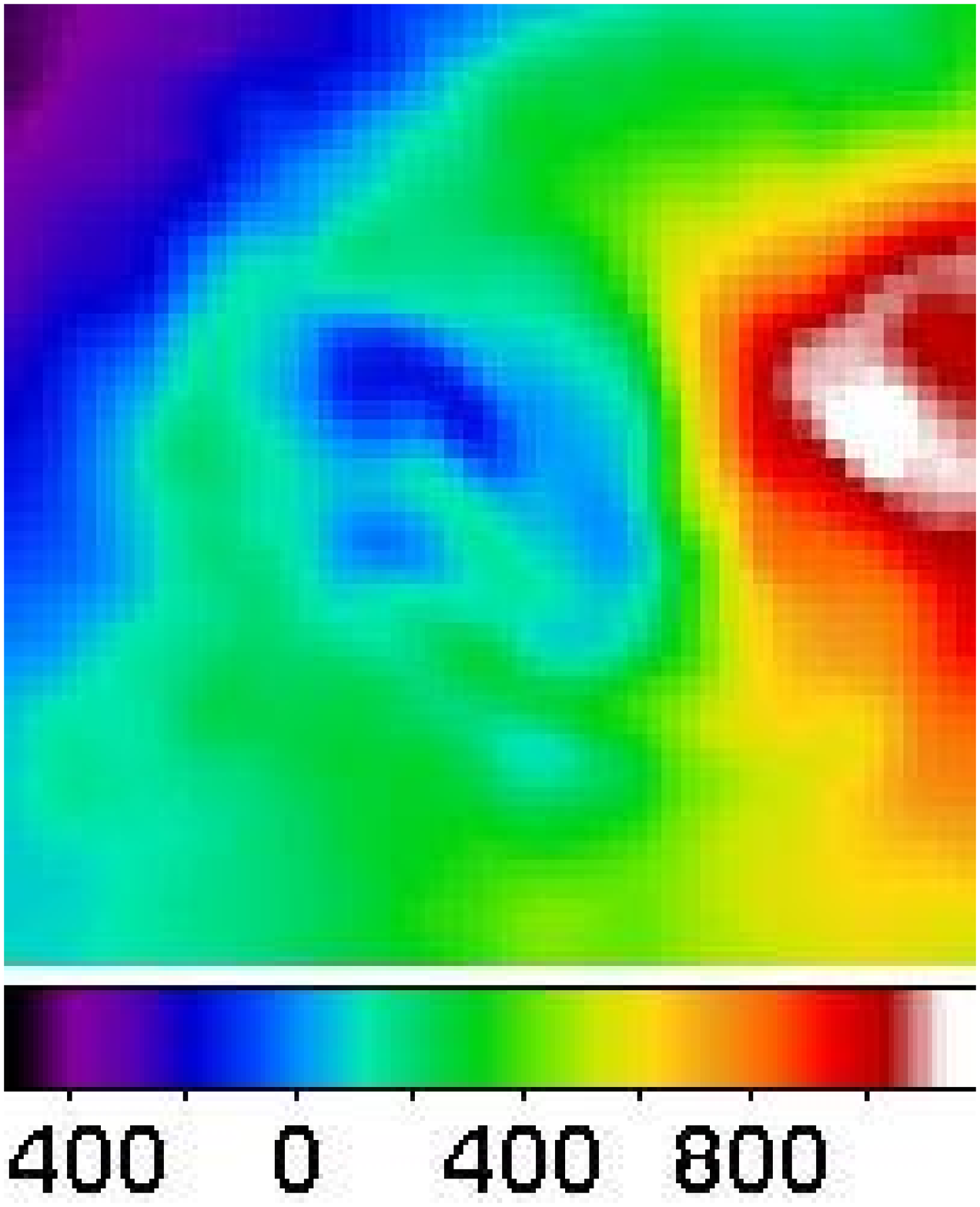}
\epsfxsize=0.62\columnwidth \epsfbox[20 100 590 800]{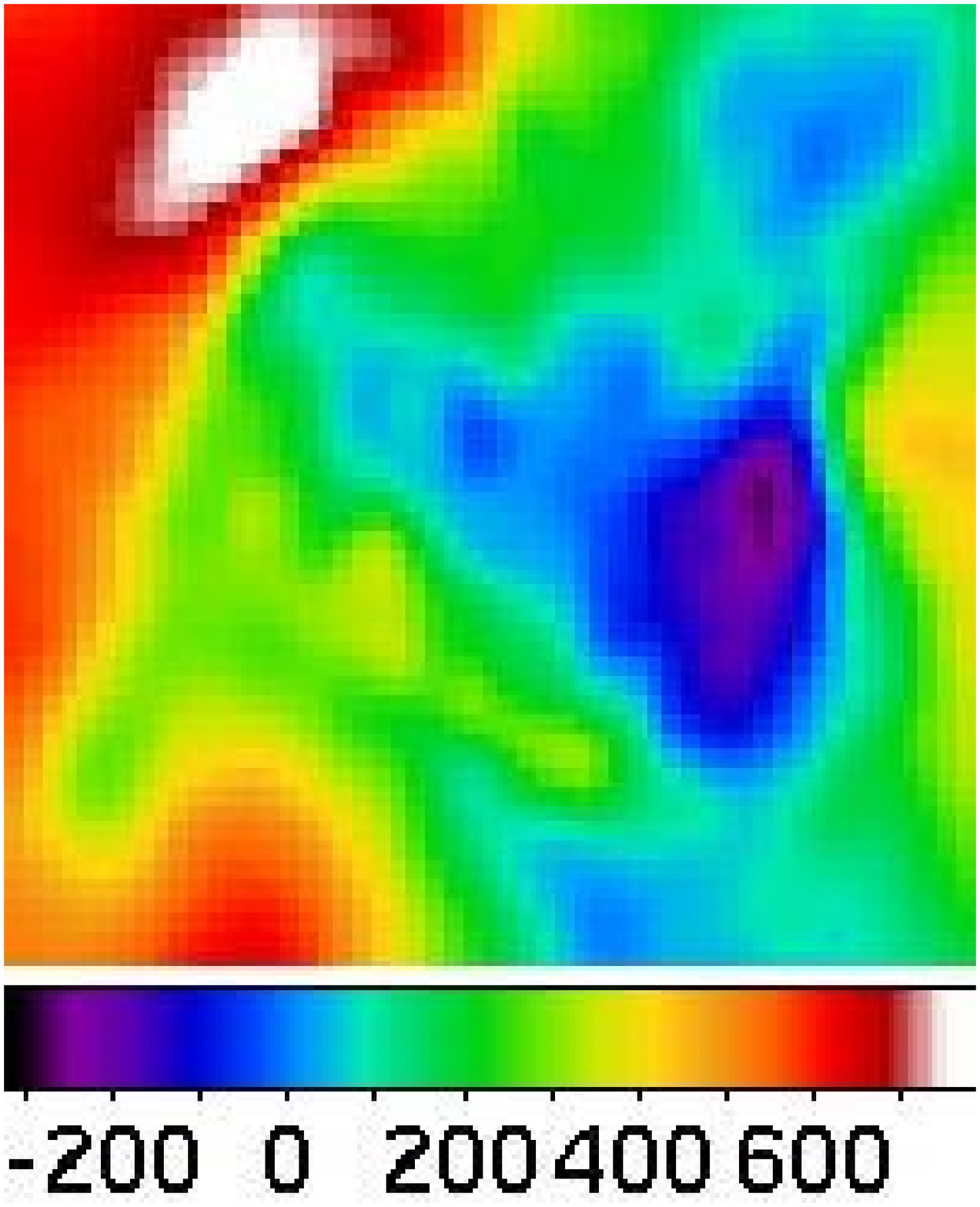}
\epsfxsize=0.62\columnwidth \epsfbox[0 100 570 800]{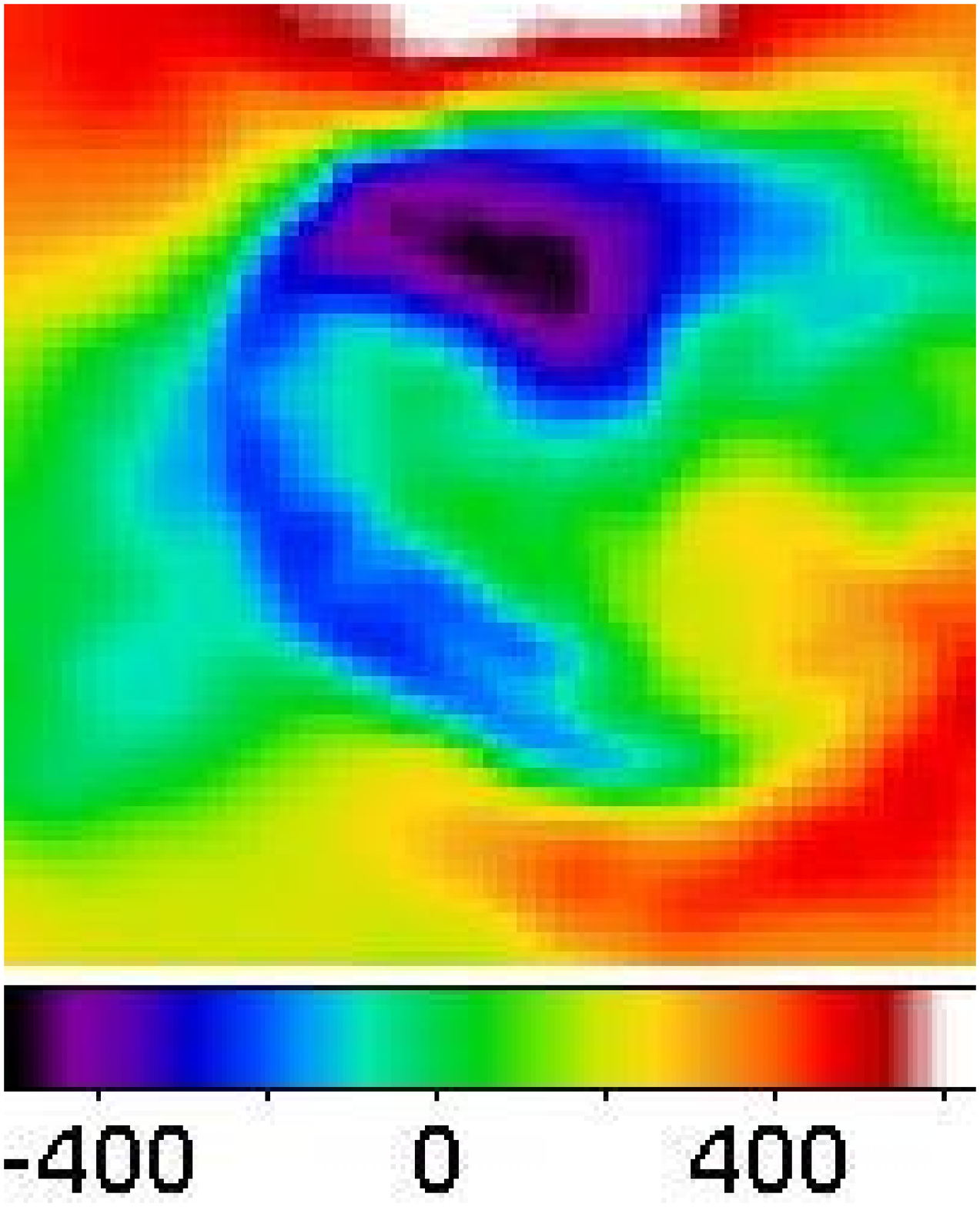}\hfil
\epsfxsize=0.62\columnwidth \epsfbox[40 20 610 860]{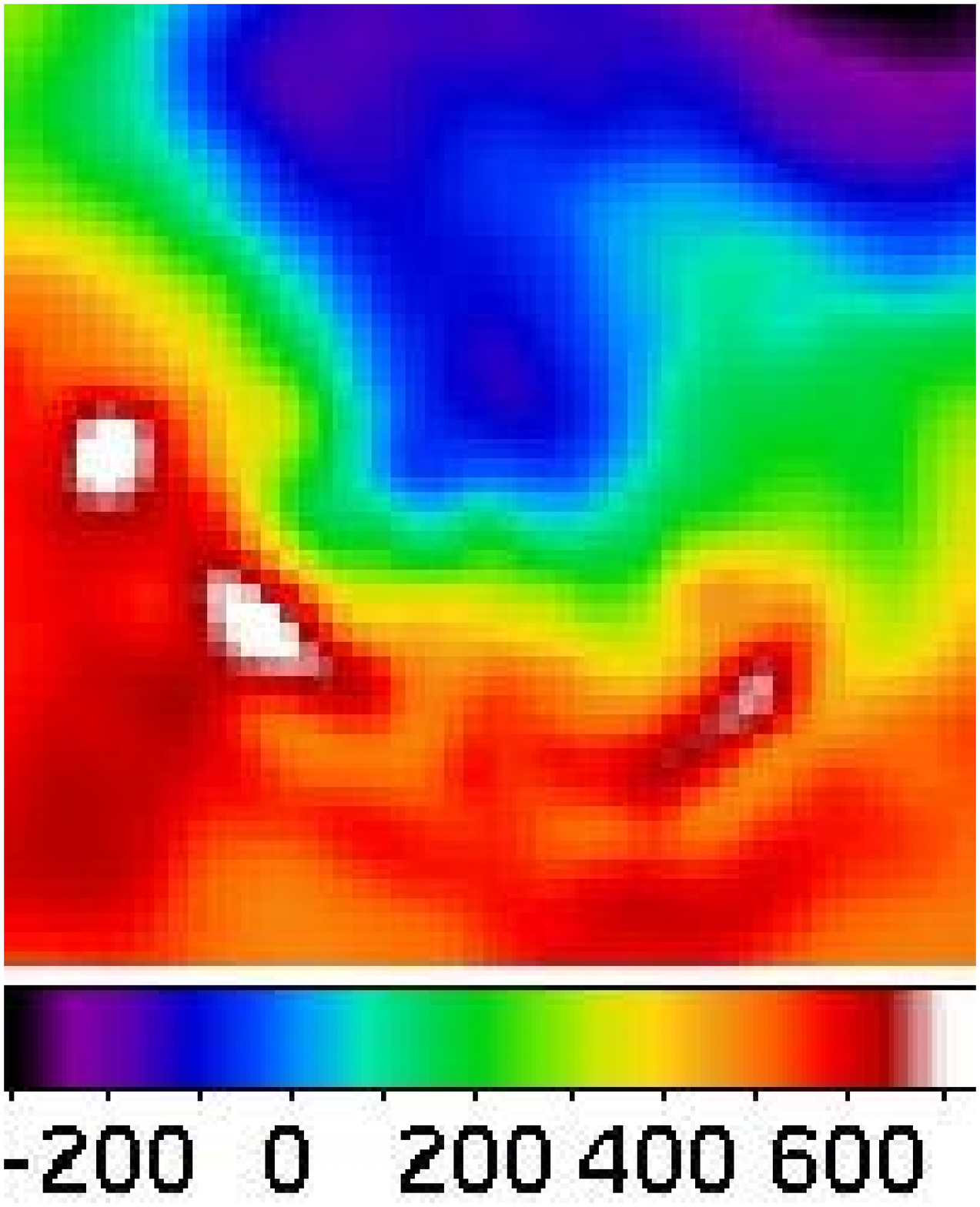}
\epsfxsize=0.62\columnwidth \epsfbox[20 20 590 860]{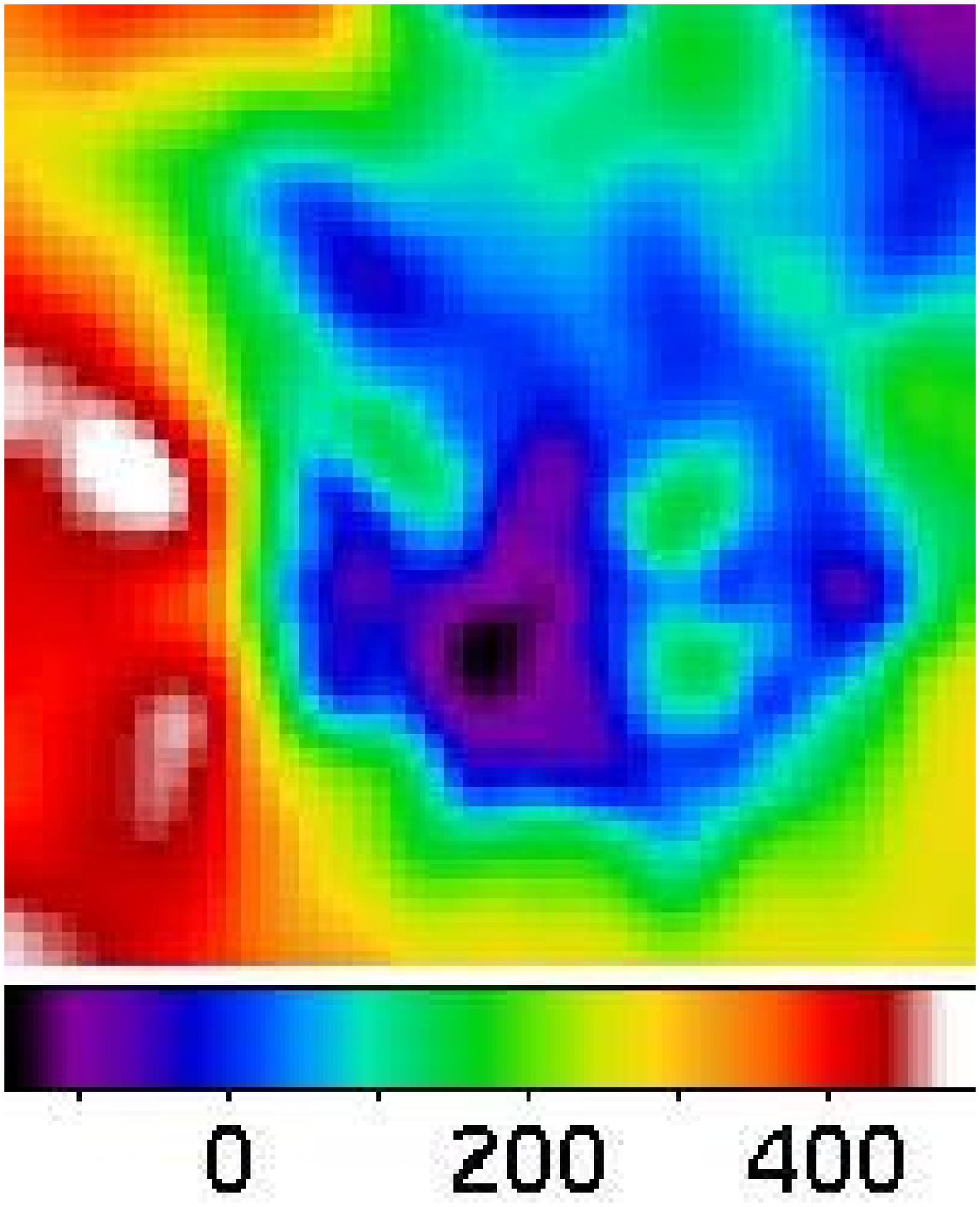}
\epsfxsize=0.62\columnwidth \epsfbox[0 20 570 860]{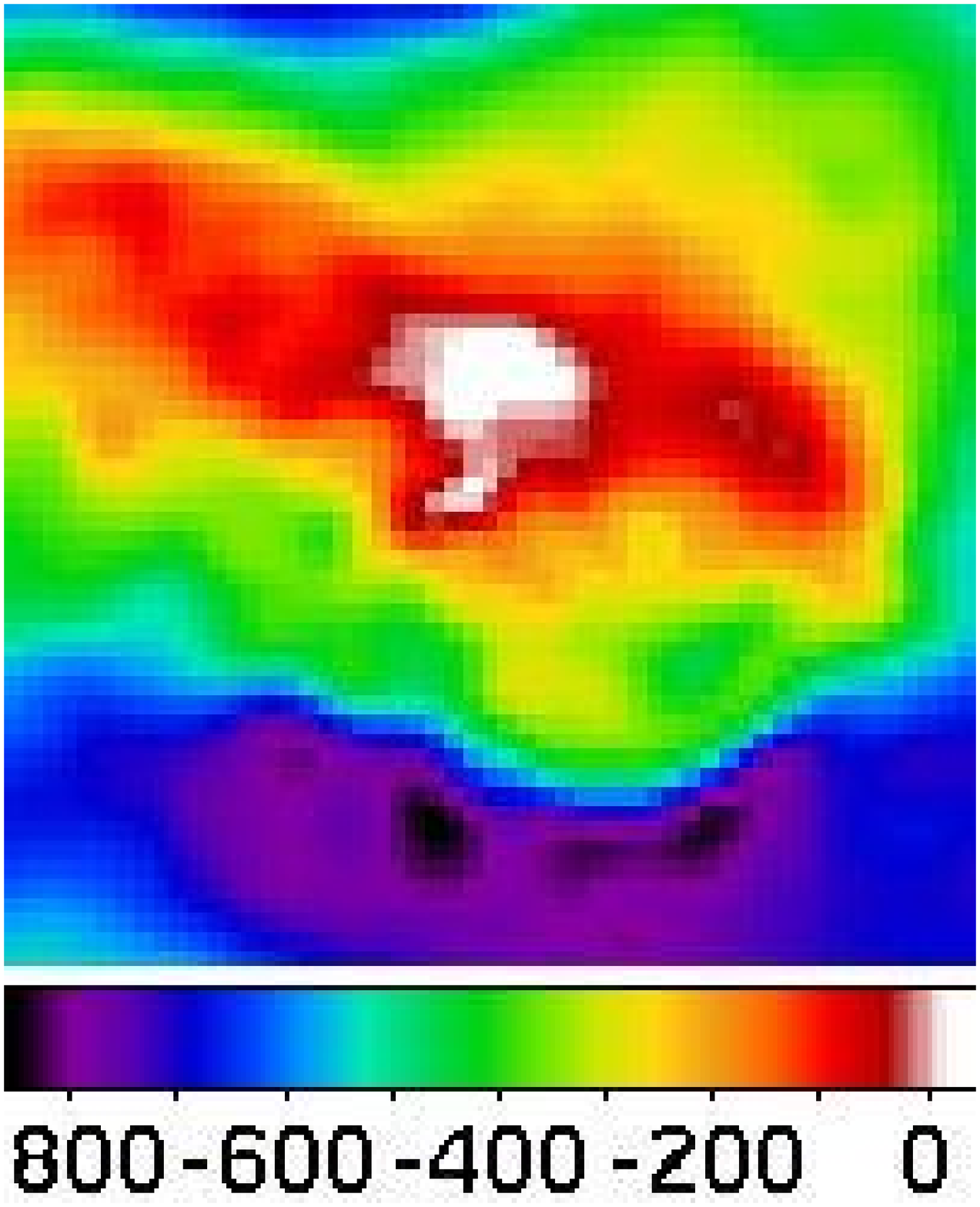}}
\caption{The same as in Fig.\ref{fig:velcl}, but for the central part of each
  cluster. The image size is 100$\times$100 kpc.
\label{fig:velcl100}
}
\end{figure*}

From Fig.\ref{fig:inputcl}, we see that  the g6212 cluster is relatively cool
with a maximum temperature $\sim$ 2 keV. In such clusters the lines of
B-like and C-like iron and lines of lighter elements become
strong. But only few of these lines have optical depth large enough for
a signficant resonant scattering effect. In Table \ref{tab:3dline} we show the list
of lines, which are the strongest in a typical cluster with  mean
temperature less then 2 keV and have a weight of dipole scattering
$\ge0.5$. The optical depth presented in Table \ref{tab:3dline} was 
calculated in the line center as
\be
\tau=\int\limits_{r_0}^{\infty}n_\i\sigma_0\d r
\ee
 along an arbitrary
chosen radial direction, where the distance from the center $r_0\sim
10$ kpc. We avoided integration in the very center of the cluster since the
number density in few central cells and their contribution to the optical depth are very high in some of the simulated clusters. We see that for the cluster g6212 the most
interesting line is the one at 1.009 keV, which corresponds to the
$2s^22p^2(^3P_0) - 2s^22p3d(^3D_1)$ transition in C-like iron and has a
large optical depth and a pure dipole scattering phase function.

In the g8 cluster, the mean temperature is $\sim 15$ keV and the
most promising lines for significant resonant scattering are: the
K$_\alpha$ line of He-like iron at 6.7 keV and the K$_\alpha$ line of
H-like iron at 6.96 keV. The H-like line has two components,
$1s(^2S_{1/2})-2p(^1P_{1/2})$ at 6.95 keV and
$1s(^2S_{1/2})-2p(^1P_{3/2})$ at 6.97 keV. The first component has an
oscillator strength 0.135; the total angular momentum is equal to 1/2
both for the ground and excited states, and therefore \citep{Ham47}
the scattering phase function is isotropic and scattered emission is
unpolarized. The second component has an oscillator strength 0.265 and
equal weights of dipole and isotropic scattering. The optical depth is
$\sim$3.56 for the line at 6.7 keV and $\sim$1.14 for the line at
6.97.  Therefore the highest polarization degree in the g8 cluster is
expected in the 6.7 keV line of Fe XXV.

The g72 cluster is an example of a merging cluster, as is clearly
seen from Fig.\ref{fig:inputcl}. The mean temperature is similar to the temperature
in Perseus, i.e. about 5-6 keV. Therefore  the polarization degree
was calculated in the 6.7 keV line of He-like iron.

\begin{table*}
 \centering

  \caption{Basic properties of simulated clusters g6212, g72
    and g8 at $z=0$.}
  \begin{tabular}{@{}rcccccc@{}}
  \hline
 Cluster & $M_\vir$, $10^{14} $M$_{\odot}$ & $R_\vir$, Mpc & $T_\min$, keV
 &$T_\max$, keV & $n_\min,~{\rm cm}^{-3}$ & $n_\max,~{\rm cm}^{-3}$\\
\hline
 g6212 & 1.61 & 1.43 & 0.2 & 2.2 & 1$\cdot 10^{-5}$ & 0.1 \\
 g72 & 19.63 & 3.29 & 0.3 & 32 & 7$\cdot 10^{-5}$ & 0.1 \\
 g8 & 32.70 & 3.90 & 3 & 35 & 7$\cdot 10^{-5}$ & 0.2  \\
\hline
\label{tab:clpar}
\end{tabular}

\end{table*}

 The velocity field was taken directly from simulations. The slices of
velocity components in all three clusters are shown in
Fig.\ref{fig:velcl}. The slices go through the centers of the clusters
and have a thickness of 3 kpc. The motion of a cluster as a whole was
compensated for by subtracting the velocity vector corresponding to
the central cell. This choice is convenient since it shows the gas
velocities relative to the densest (and therefore brightest) part of
the cluster, which is responsible for much of the line flux to be
scattered in outer regions. To characterize the spread of velocities
within the cluster we calculated mass-weighted value of the RMS of the
velocity (relative to the velocity of the central cell) over entire
volume of a $2\times 2\times2$ Mpc cube. The RMS values are $\sim 230$
km/s in g6212 cluster, $\sim 1000$ km/s in g8 cluster and $\sim 2400$
km/s in g72. If instead we calculate the RMS relative to the
mass-weighted mean velocity the value of RMS decreases to $\sim 800$
km/s for g8 cluster, $\sim 200$ km/s for g6212 cluster and to $\sim
1400$ km/s for g72 cluster. We note here that
for radiative transfer calculation the value of the subtracted
velocity vector is not important since the polarization signal is only
sensitive to the relative velocities of different gas lumps and
does not depend on the motion of the cluster as a whole.
 In Fig.\ref{fig:velcl100}
velocities of gas motions in the cluster centers are shown. The size of
the central region was chosen to be 100$\times$100 kpc. Notations are
the same as in Fig.\ref{fig:velcl}. The spread of velocities in the
centers is lower than at the edges. 

In order to see
the influence of different gas motions on polarization degree, we made
a number of radiative transfer simulations keeping the same density
and temperature distributions, but varying the characteristic
amplitude of the gas velocities. This was done by introducing a
multiplicative factor $f_v$ which is used to scale all velocities
obtained from the hydrodynamical simulations. In particular $f_v=0$
means no motions, $f_v=1$ means the velocity field as computed in the simulations,
$f_v=2$ means doubled velocities compared to hydro simulations,
etc. The second parameter characterizing the radiative transfer
simulation is the level of micro-turbulence, parametrized via the
effective Mach number $M$ (see equations \ref{eq:sig0} and
\ref{eq:den}). As discussed above, increasing $M$ causes the optical
depth to decrease, leading to the decrease of the polarization
signal. Thus the pair of parameters $(f_v, M)$ completely specifies
the scaling of the velocity field in a given radiative transfer
run, compared  to  the original hydrodynamical simulations. 
In our simulations the values of $f_\v$ and $M$ are treated as 
independent parameters, since our goal is to examine separately the 
impact of the large scale motions and the micro-turbulence on scattering 
and polarization. Strictly speaking, a more self-consistent approach 
would be to relate $M$ and the spread of velocities on larger scales, 
resolved by the simulation, by assuming e.g. Kolmogorov scaling in a 
turbulent cascade.

 The resolution of the SPH simulations used here is not fully
sufficient to resolve small scale motions of the ICM. Typical resolved
scales vary from few tens kpc in the center to few hundred kpc at
the outskirts of simulated clusters. Our approach of "hiding"
unresolved ICM motions as a micro-turbulence (i.e. as line
broadening) is valid if the mean free path of the photons is larger
than the resolved spatial scales. We tested this approximation for
radial distances $\sim$300 kpc from the center and found that this
criterion is safely satisfied. We further address this issue elsewhere
(Zhuravleva et al., in preparation).

\begin{table*}
 \centering
  \caption{Oscillator strengths and optical depths of the strongest
    X-ray lines with the weight of dipole scattering $\ge$0.5 in the
   simulated clusters g6212, g72 and g8.}
 \begin{tabular}{@{}rrccccccc@{}}
  \hline
 Cluster & Ion & $E$, keV & $f$ & $w_2$ &$\tau$, ($f=0,M=0$)&$\tau$,
 ($f=1,M=0$)& $\tau$, ($f=0,M=0.25$)& $\tau$, ($f=1,M=0.25$)  \\
\hline
 g6212 & Fe XXV & 6.7 & 0.78 & 1 & 2.45 & 2.02 & 1.01 & 0.96\\
 & Si XIII & 1.86 & 0.75 & 1 & 1.78 & 1.47 & 0.96 & 0.88 \\
& Fe XXII & 1.0534 & 0.675 & 0.5 & 2.86 & 2.27 & 1.17 & 1.1 \\
& Fe XXI & 1.009 & 1.4 & 1 & 2.8 & 2.1 & 1.16 & 1.04 \\
\hline
g72 & Fe XXV & 6.7 & 0.78 & 1 & 3.19  & 0.73  & 1.32  & 0.51 \\ 
\hline
g8 & Fe XXV & 6.7 & 0.78 & 1 & 3.56 & 1.83 & 1.47 & 1.11 \\
 & Fe XXVI & 6.97 & 0.265 & 0.5 & 1.14 & 0.52 & 0.47 & 0.34 \\
\hline
\label{tab:3dline}
\end{tabular}
\end{table*}

\begin{figure*}
{\centering \leavevmode
\epsfxsize=0.68\columnwidth \epsfbox[40 160 610 650]{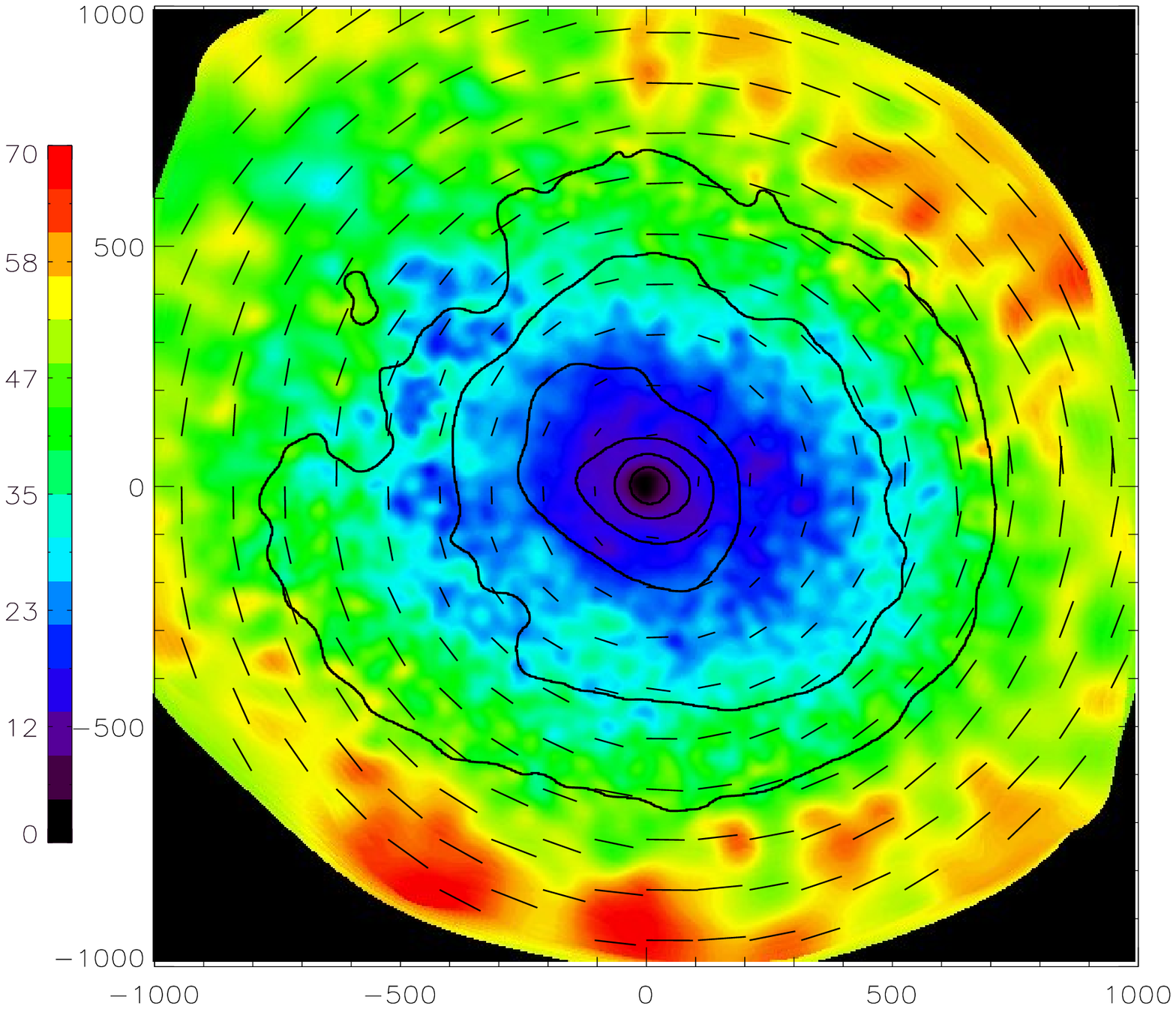}
\epsfxsize=0.68\columnwidth \epsfbox[40 160 610 650]{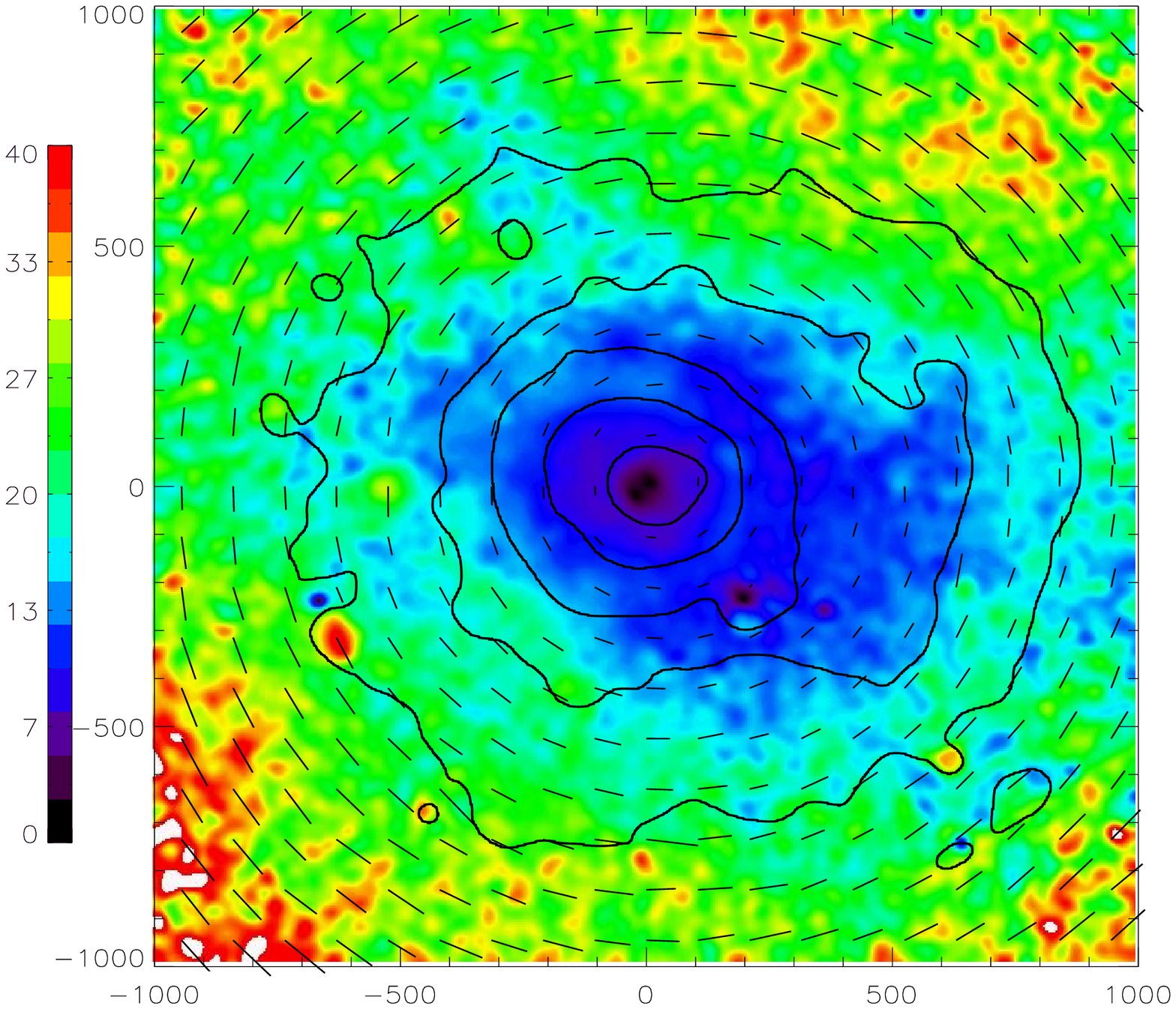}
\epsfxsize=0.68\columnwidth \epsfbox[40 160 610 650]{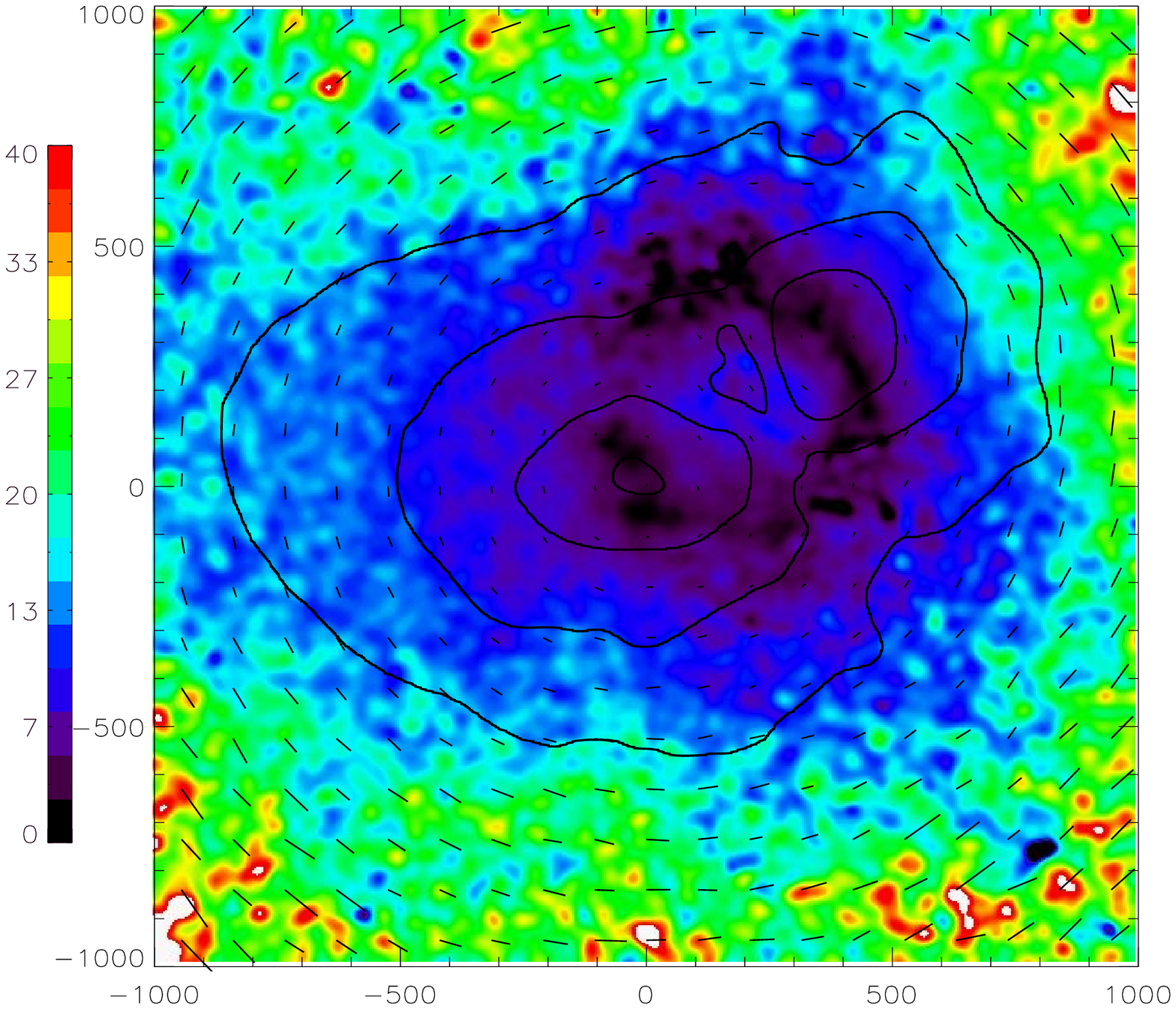}\hfil
\epsfxsize=0.68\columnwidth \epsfbox[40 160 610 650]{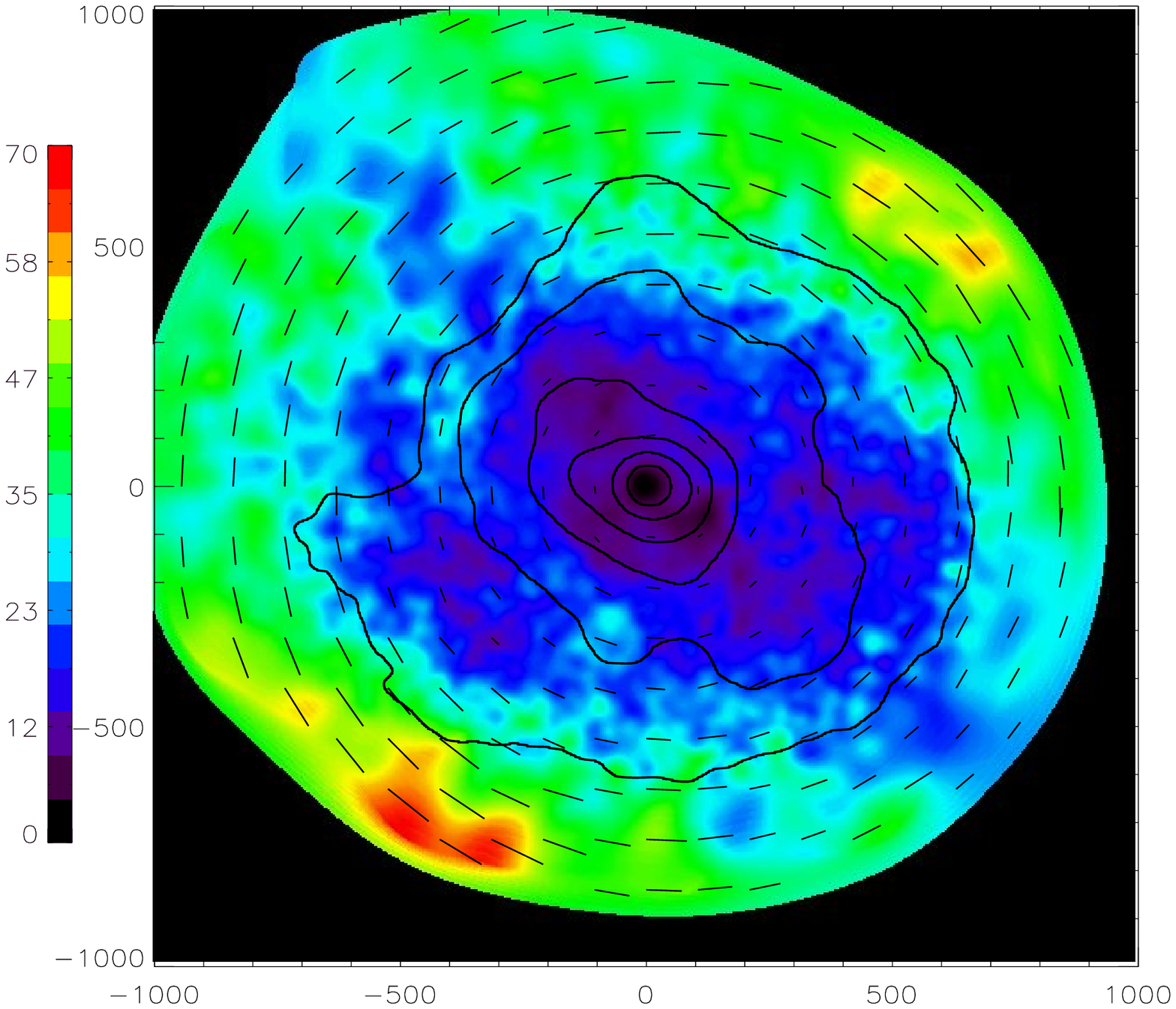}
\epsfxsize=0.68\columnwidth \epsfbox[40 160 610 650]{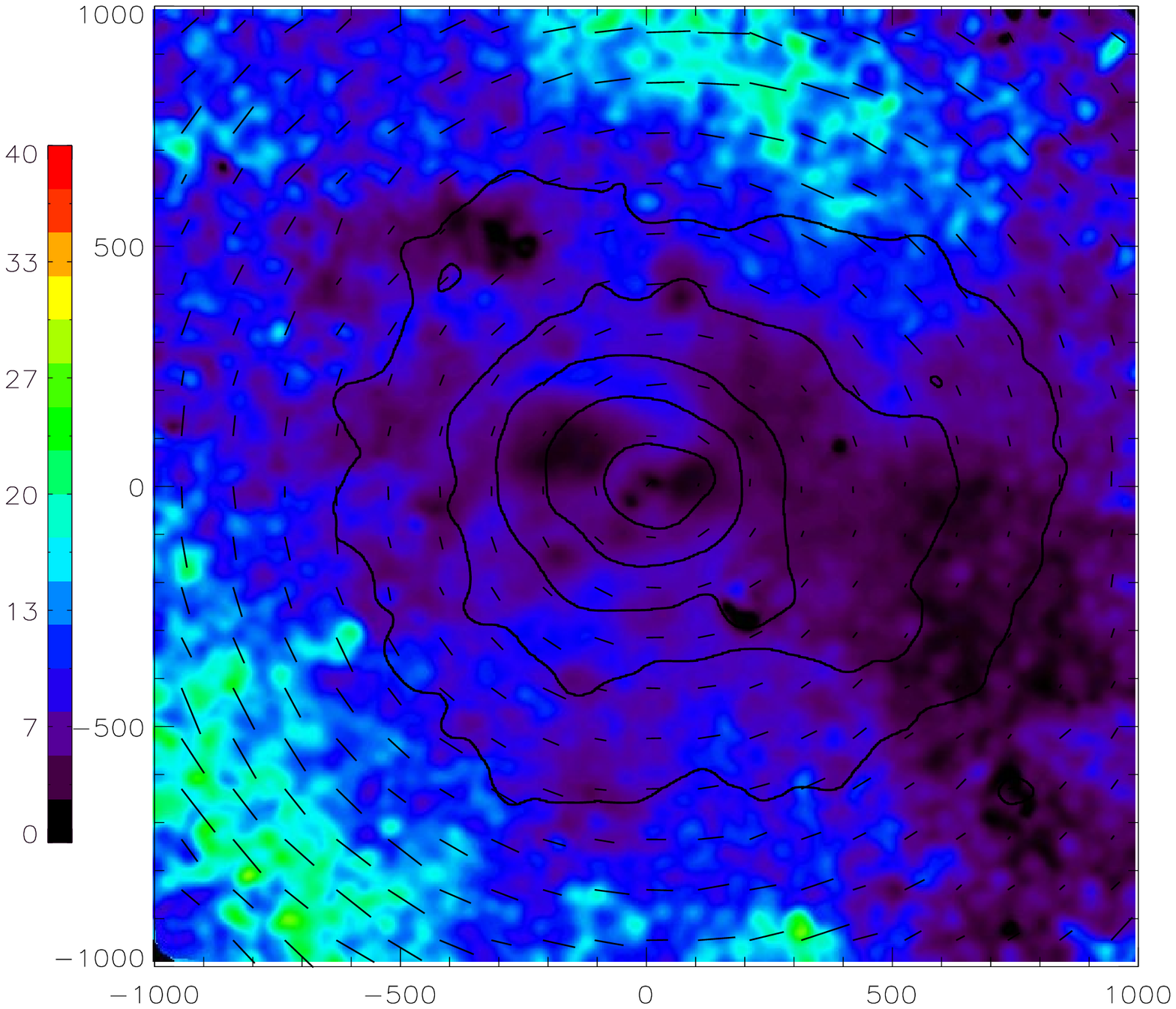}
\epsfxsize=0.68\columnwidth \epsfbox[40 160 610 650]{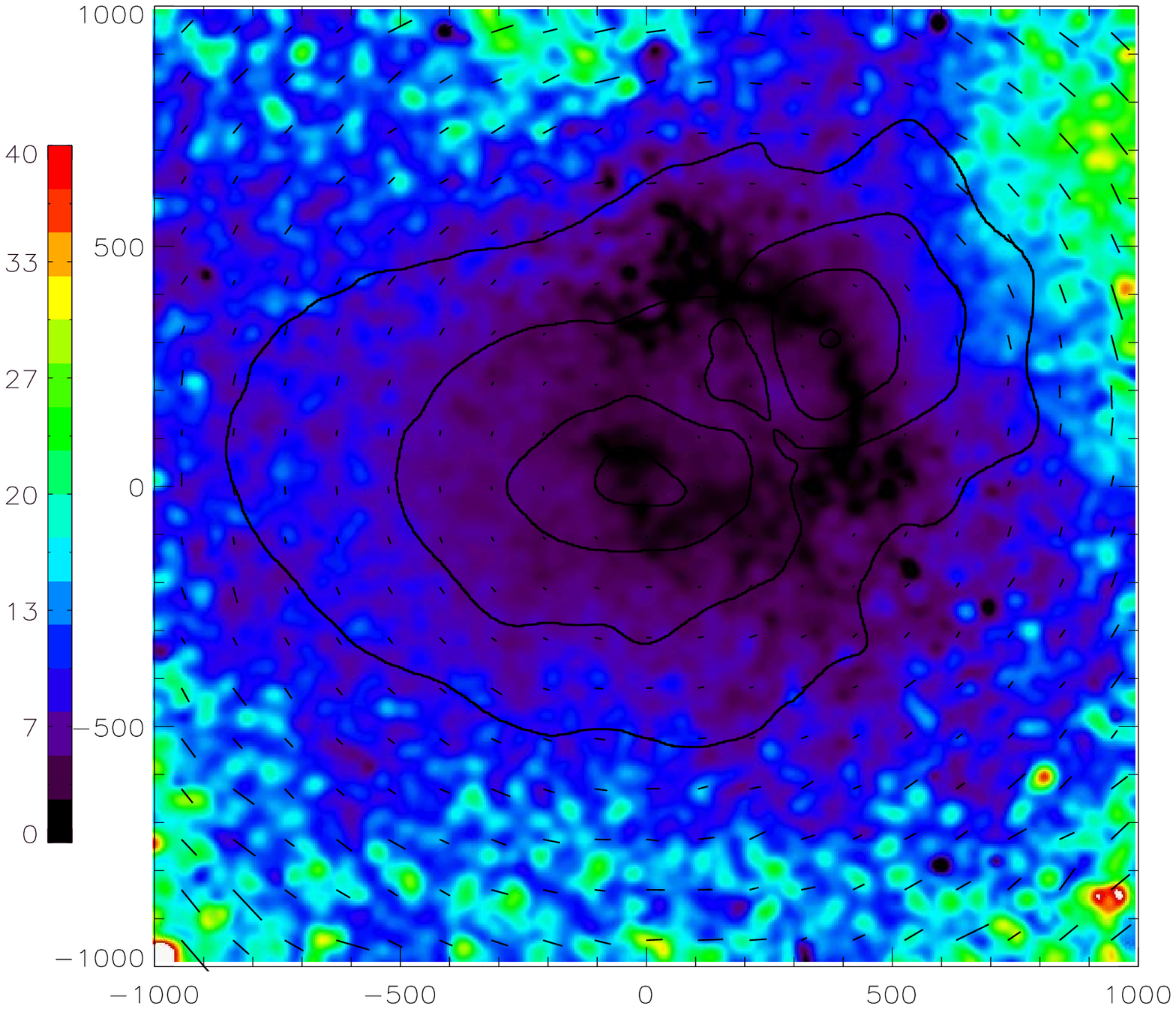}\hfil
\epsfxsize=0.68\columnwidth \epsfbox[40 160 610 650]{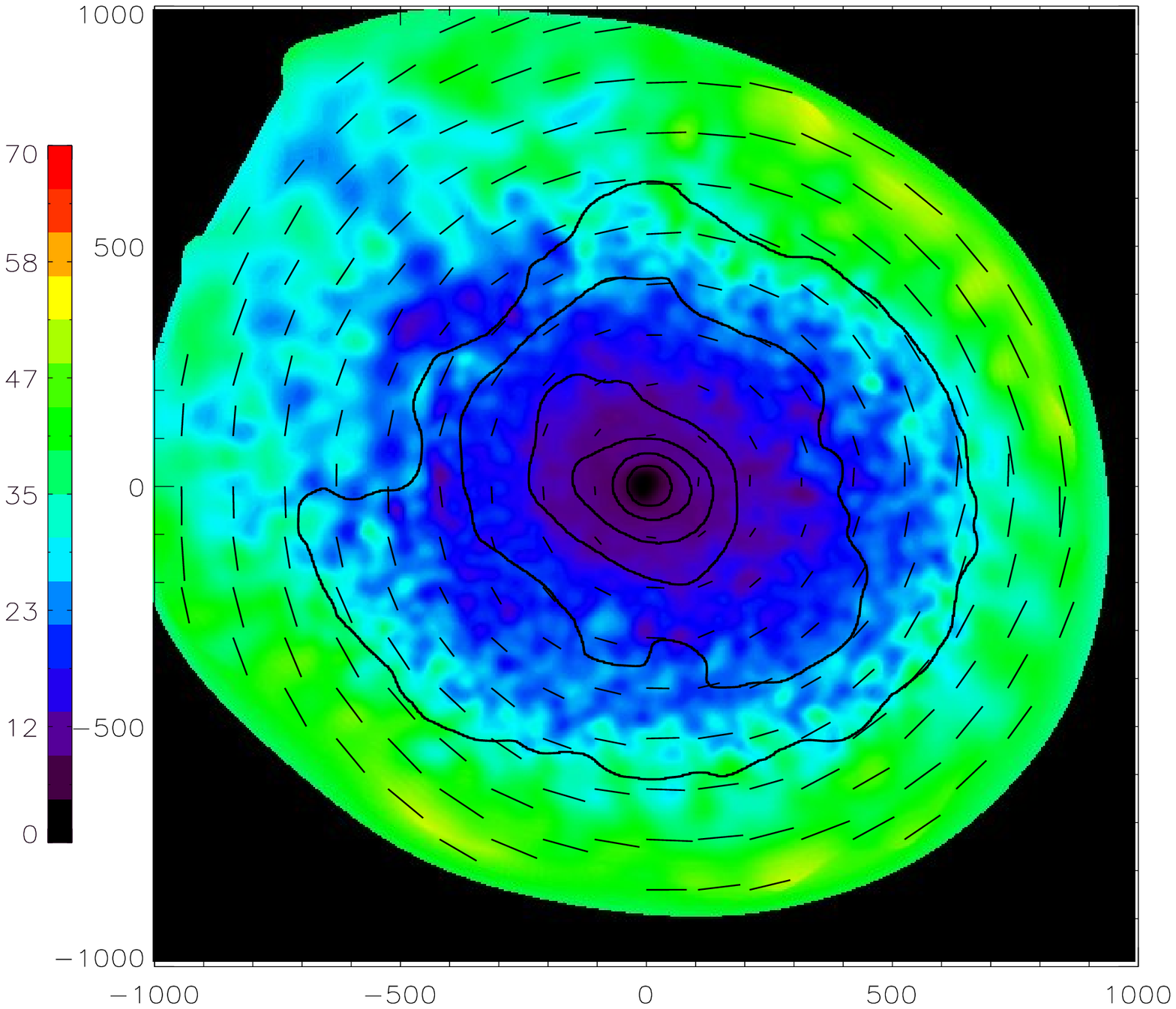}
\epsfxsize=0.68\columnwidth \epsfbox[40 160 610 650]{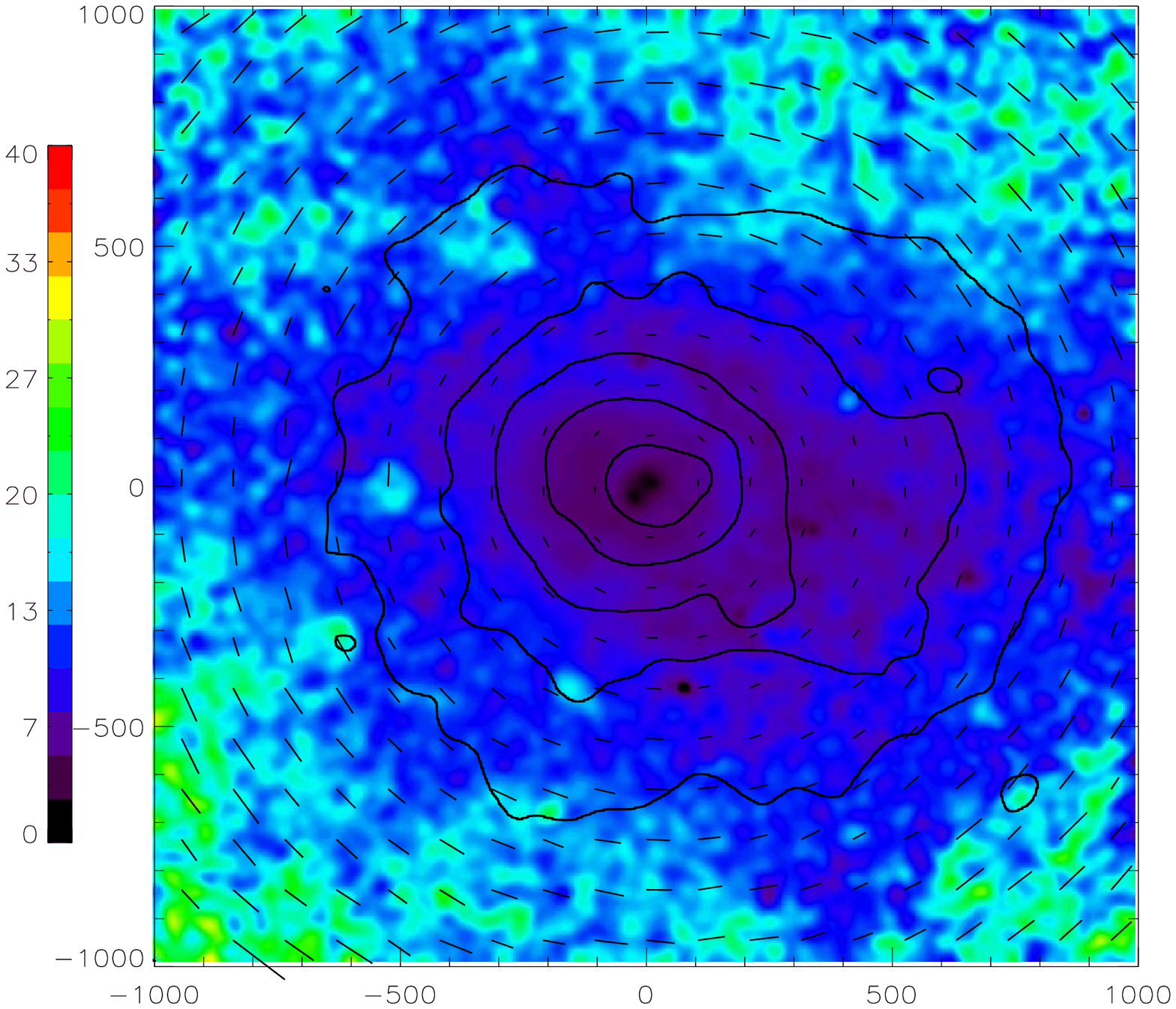}
\epsfxsize=0.68\columnwidth \epsfbox[40 160 610 650]{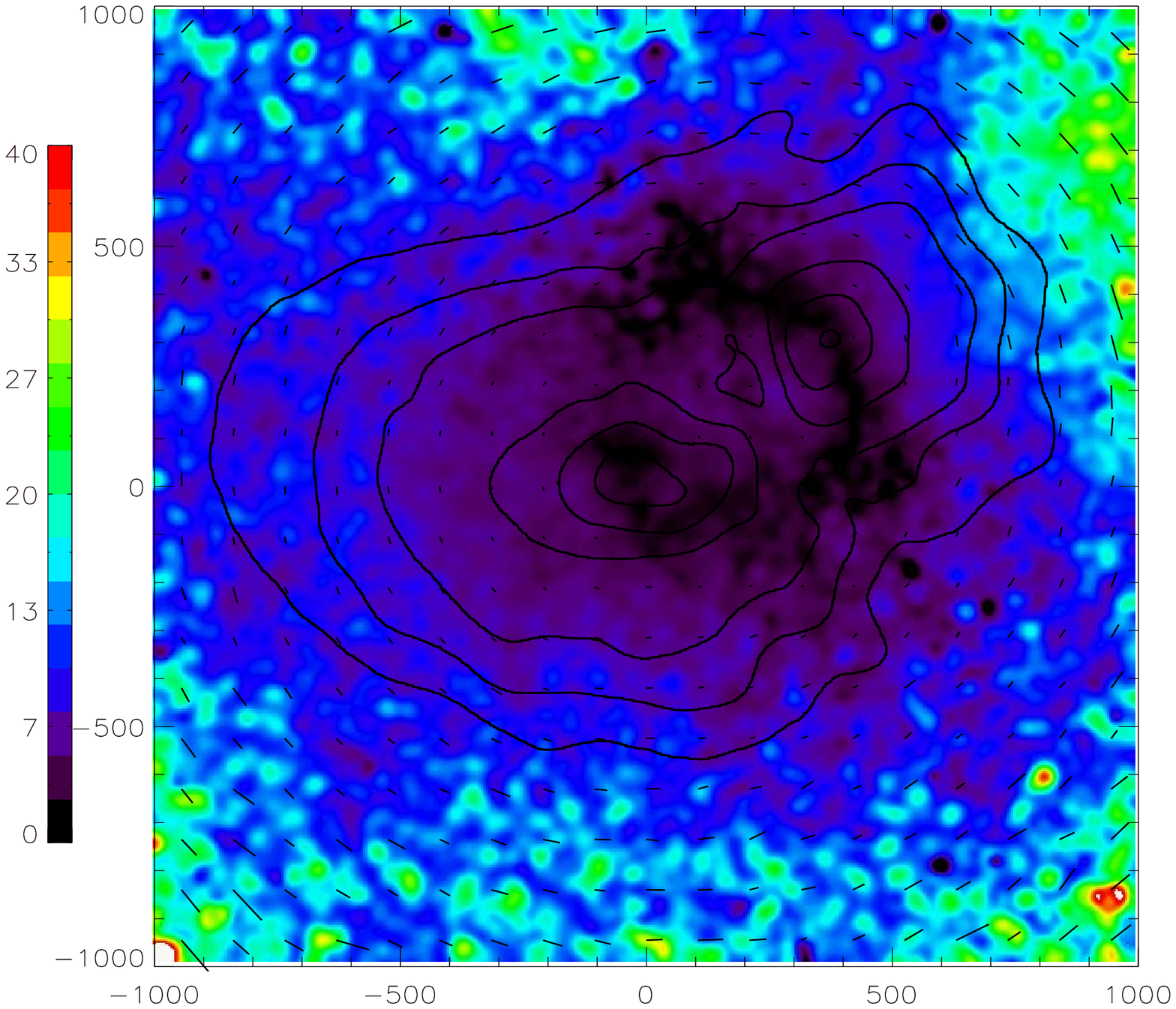}\hfil
\epsfxsize=0.68\columnwidth \epsfbox[40 160 610 650]{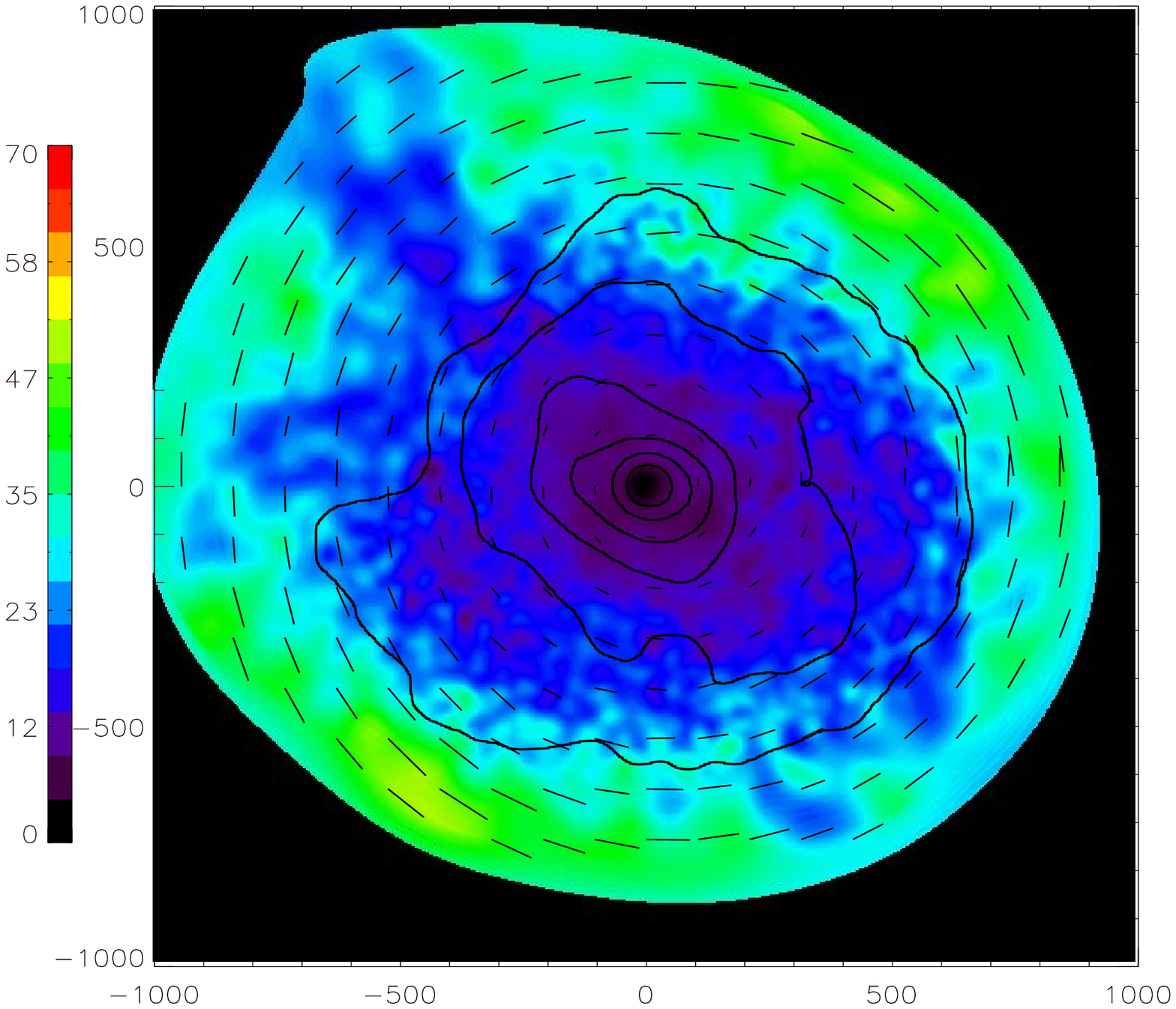}
\epsfxsize=0.68\columnwidth \epsfbox[40 160 610 650]{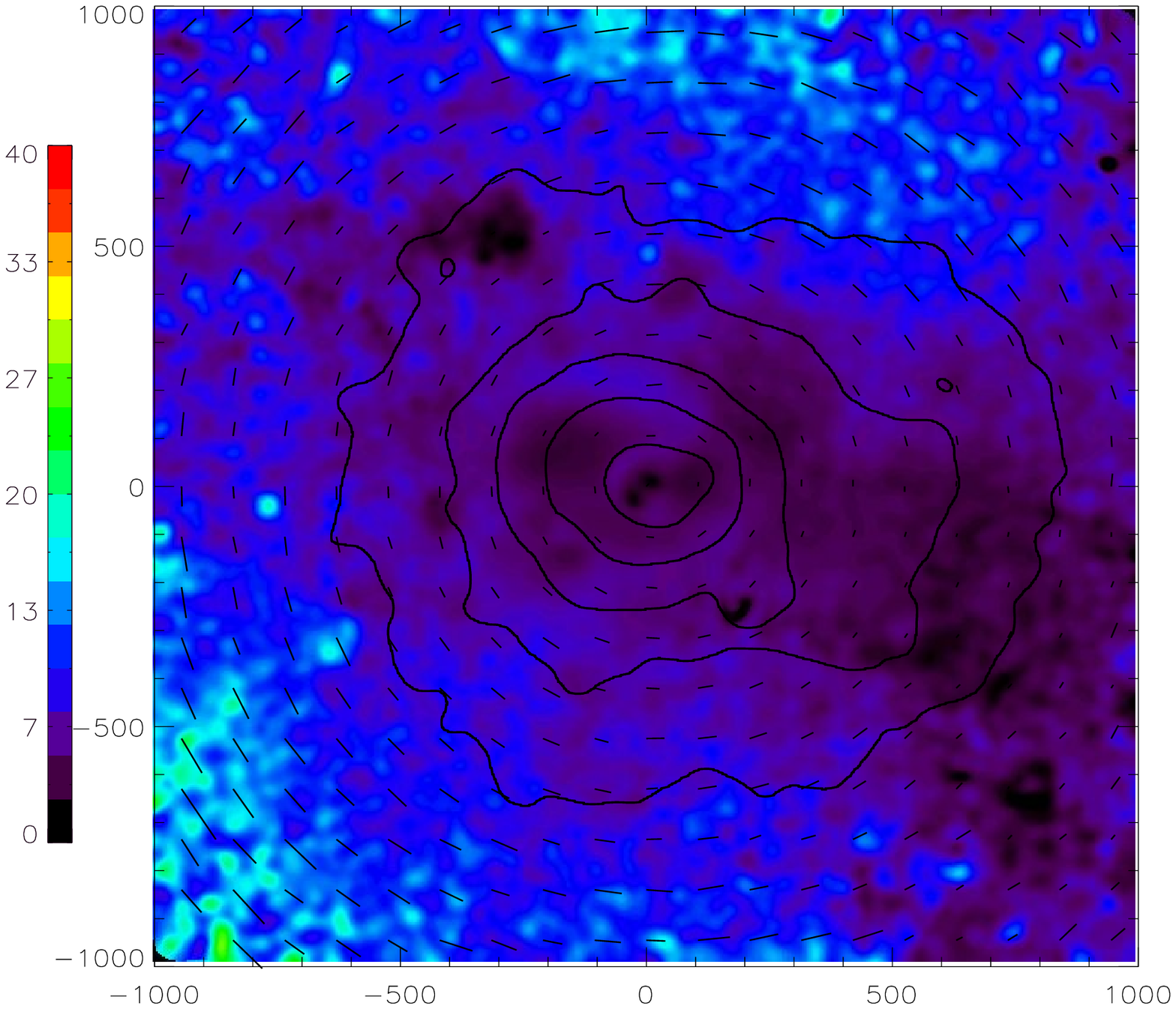}
\epsfxsize=0.68\columnwidth \epsfbox[40 160 610 650]{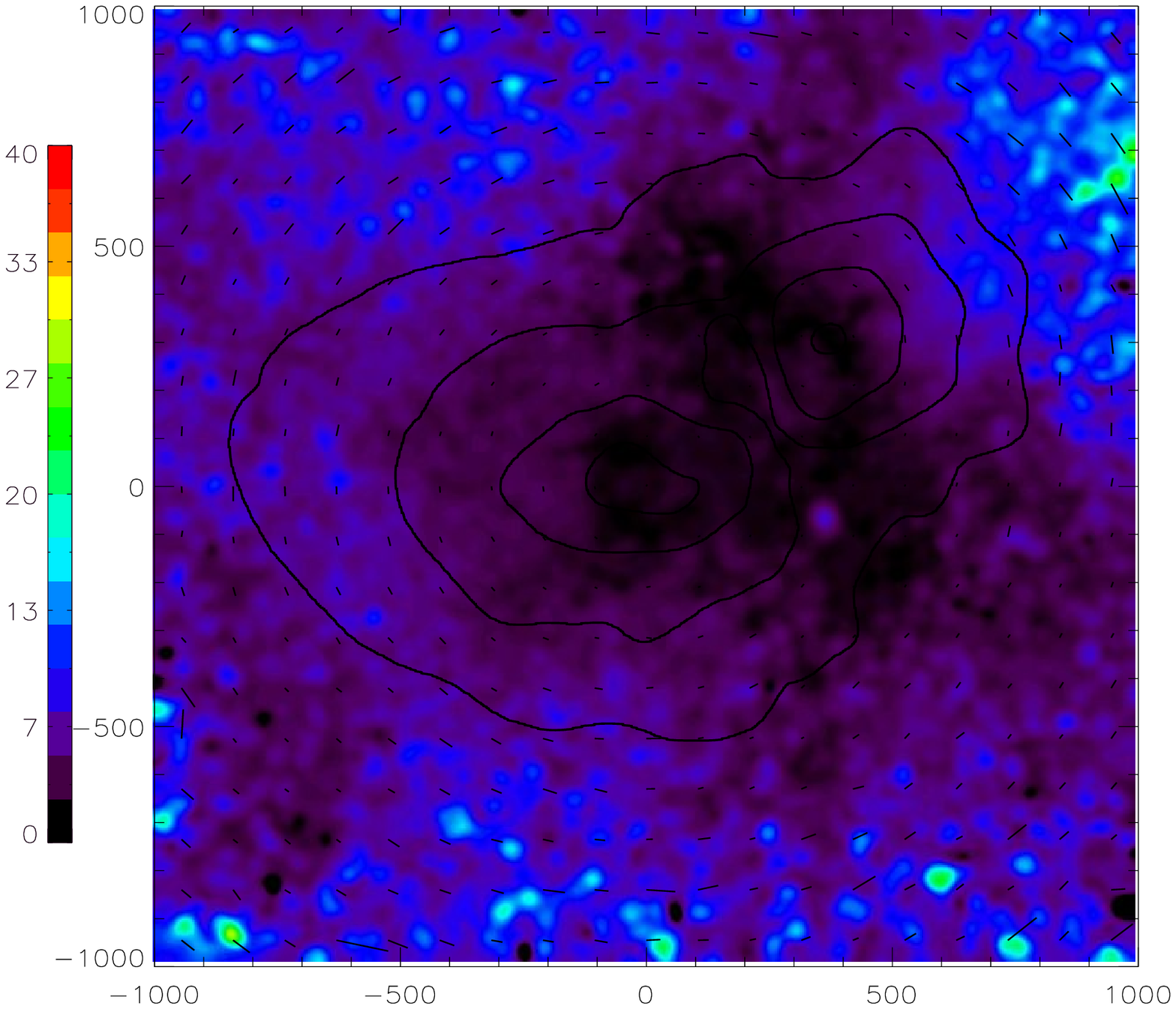}}
\caption{ Polarization degree in simulated clusters g6212 (the left
  panels), g8 (the middle panels) and g72 (the right panels). For
  the g8 and g72
  clusters,  the polarization degree was calculated in the K$_\alpha$ line
  of Fe XXV at 6.7
  keV and for  the g6212 cluster calculations were done in Fe XXI line at
  1.009 keV. The polarization degree was evaluated as
  $P=\sqrt{Q^2+U^2}/I$, the value of which is shown in colorbar in
  percent. $I$ is the total intensity, including scattered and direct
  emission. The colors in the images show  polarization
  degree, the short dashed lines show the orientation of the electric vector. The
  contours (factor of 4 steps in intensity) of the X-ray surface brightness in the chosen line are
  superposed.  The size of each picture is 2$\times$2 Mpc. The top
  three pictures correspond to the case of no
  motion ($f_\v=0$, $M=0$). The second raw of pictures shows the case
  of simulated gas motions ($f_\v=1$, $M=0$). The third raw shows the
polarization when there are only turbulent motions with Mach number 
$M=0.25$ and the bottom three pictures demonstrate the case of simulated 
gas motions and turbulent motions ($f_\v=1$, $M=0.25$). All results
were adaptively smoothed.    
\label{fig:pol}
}
\end{figure*}

\begin{figure*}
{\centering \leavevmode
\epsfxsize=0.6\columnwidth \epsfbox[40 80 610 670]{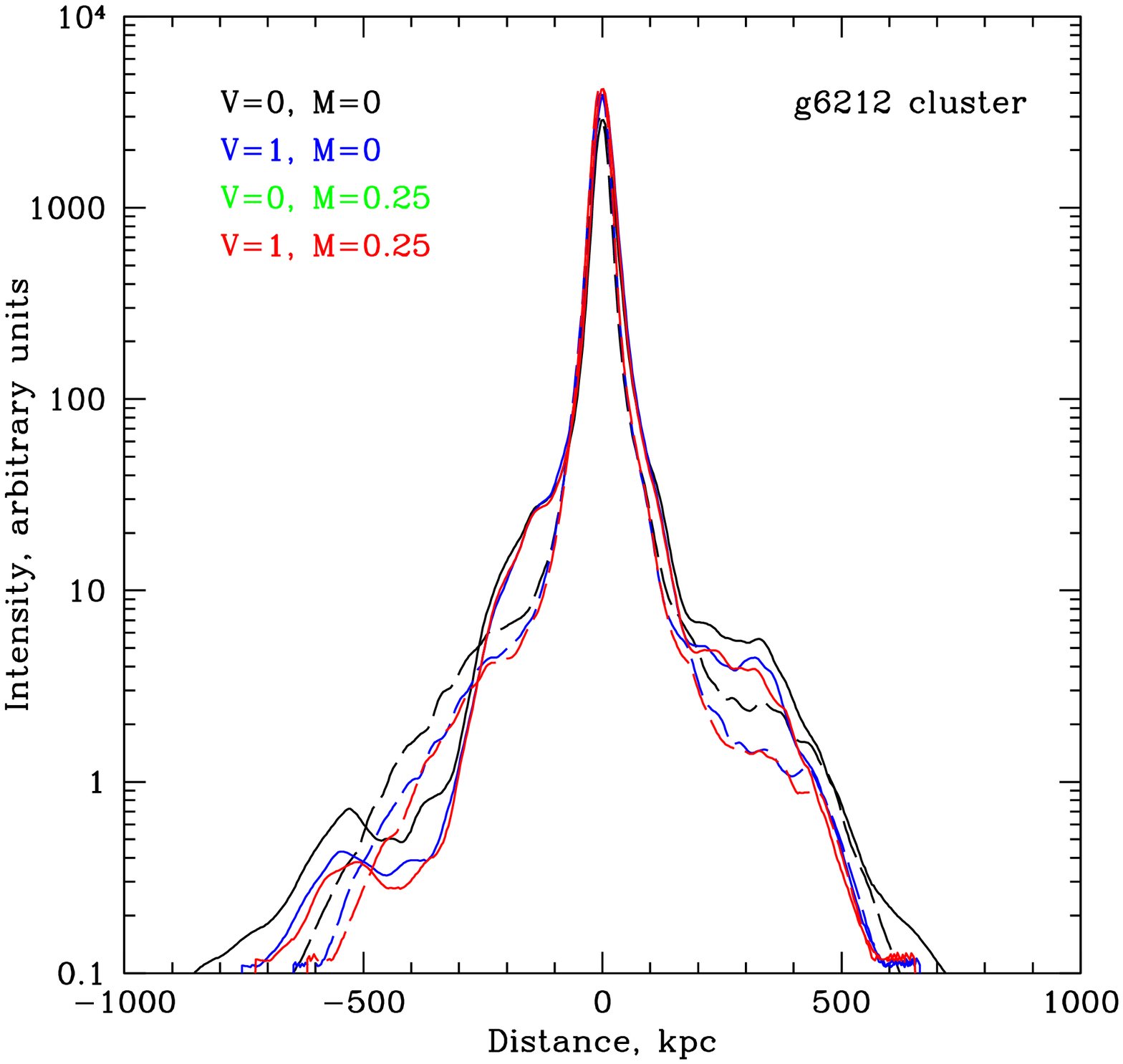}
\epsfxsize=0.6\columnwidth \epsfbox[40 80 610 670]{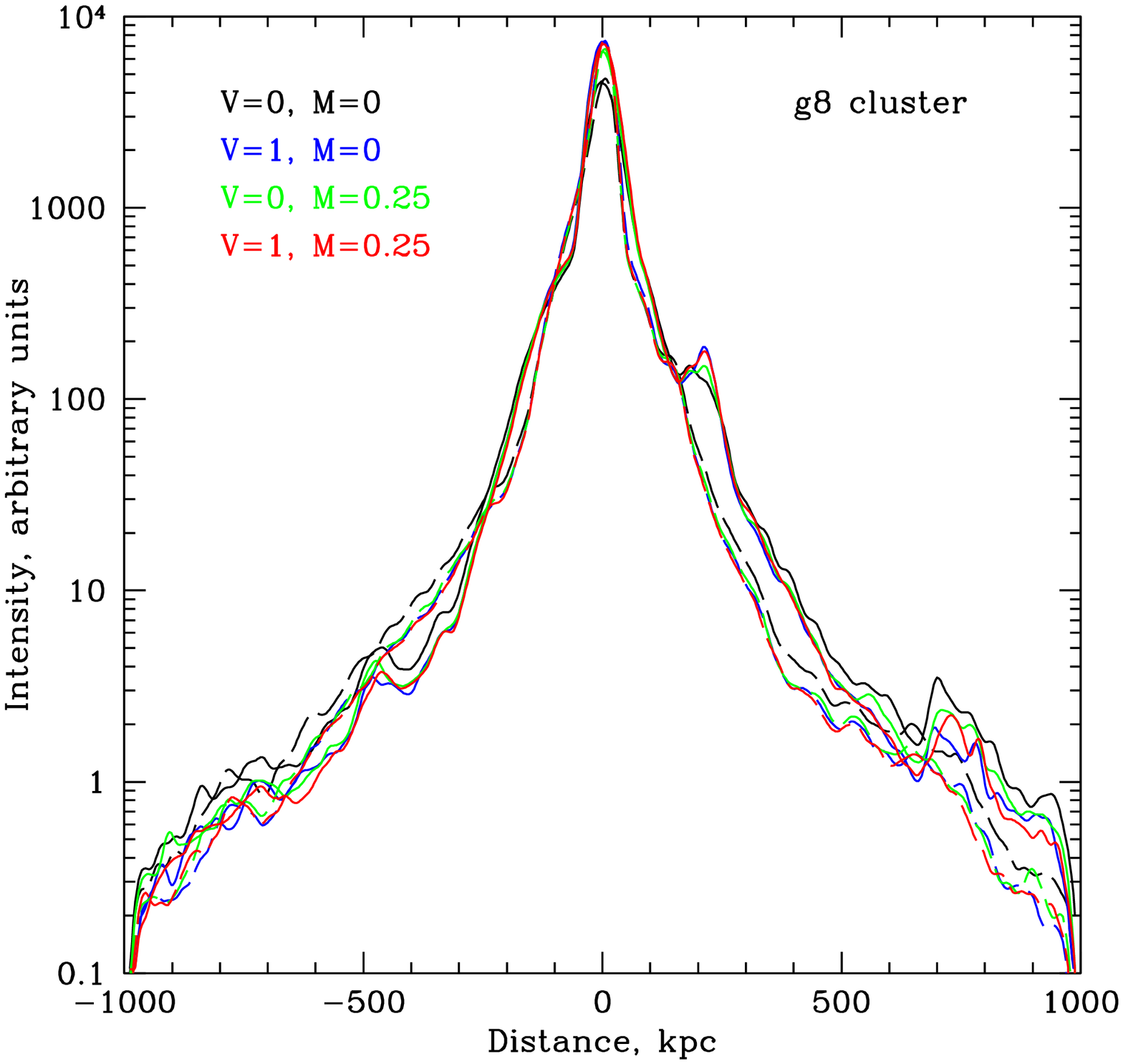}
\epsfxsize=0.6\columnwidth \epsfbox[40 80 610 670]{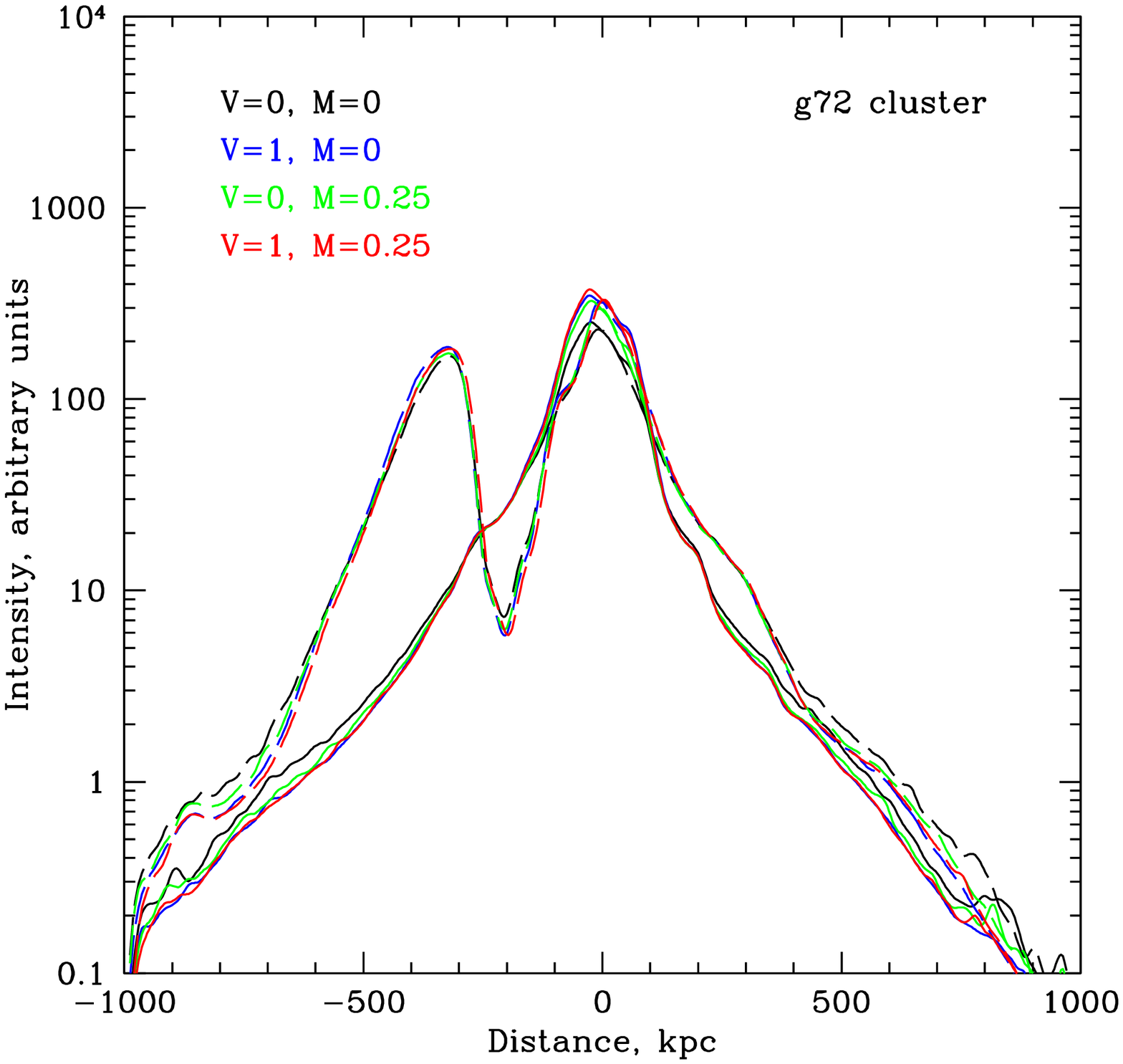}\hfil
\epsfxsize=0.6\columnwidth \epsfbox[40 100 610 690]{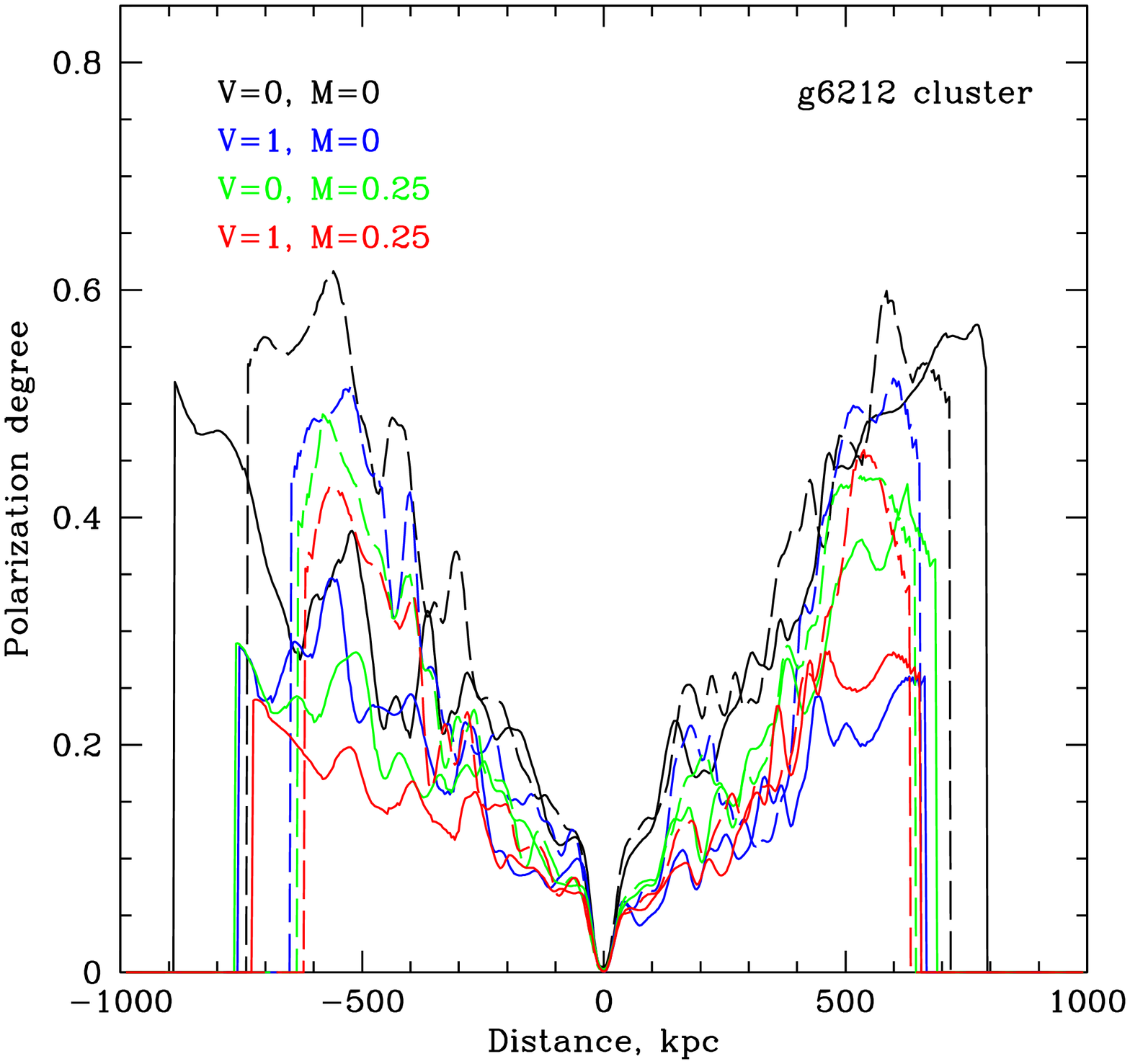}
\epsfxsize=0.6\columnwidth \epsfbox[40 100 610 690]{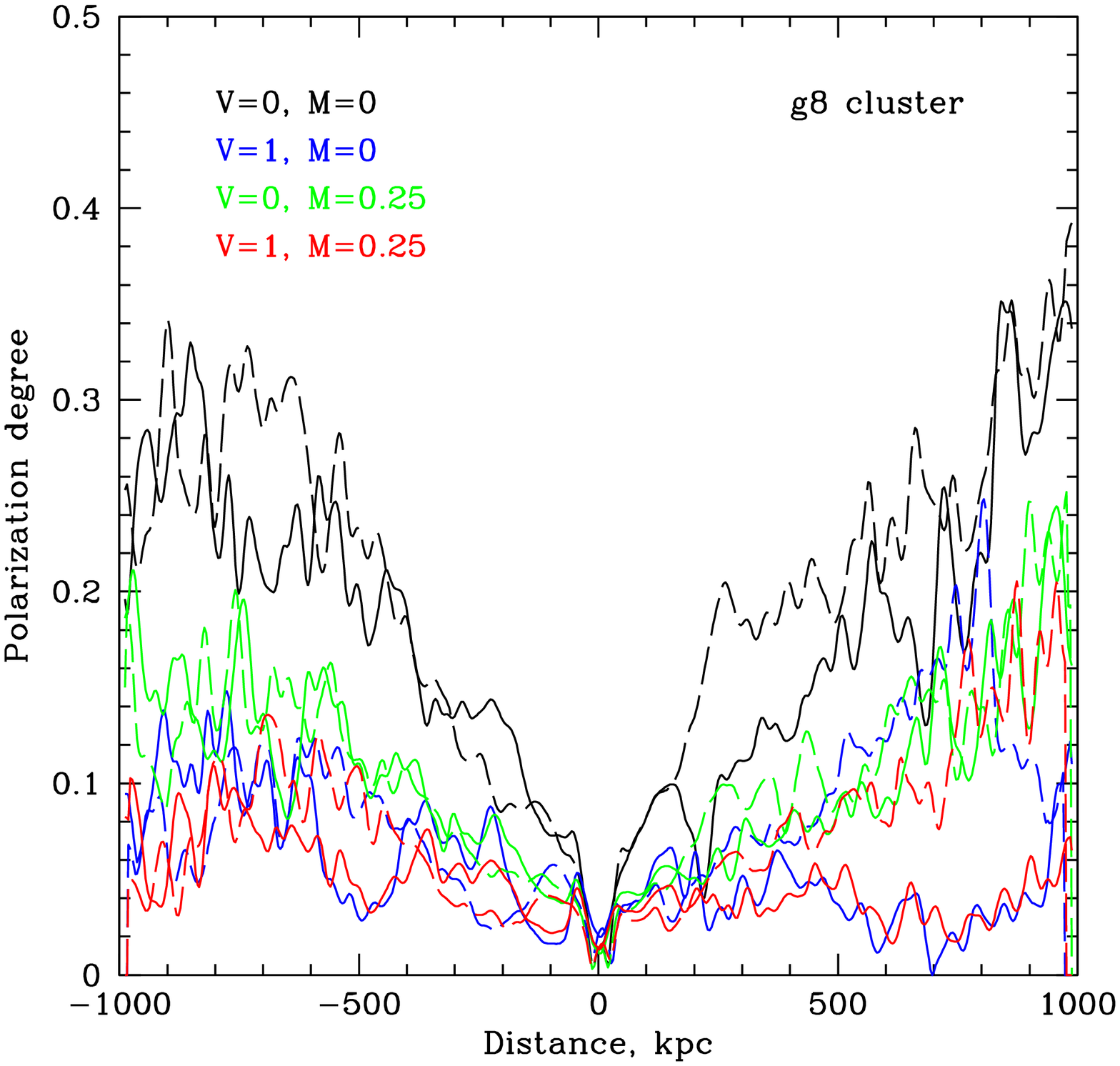}
\epsfxsize=0.6\columnwidth \epsfbox[40 100 610 690]{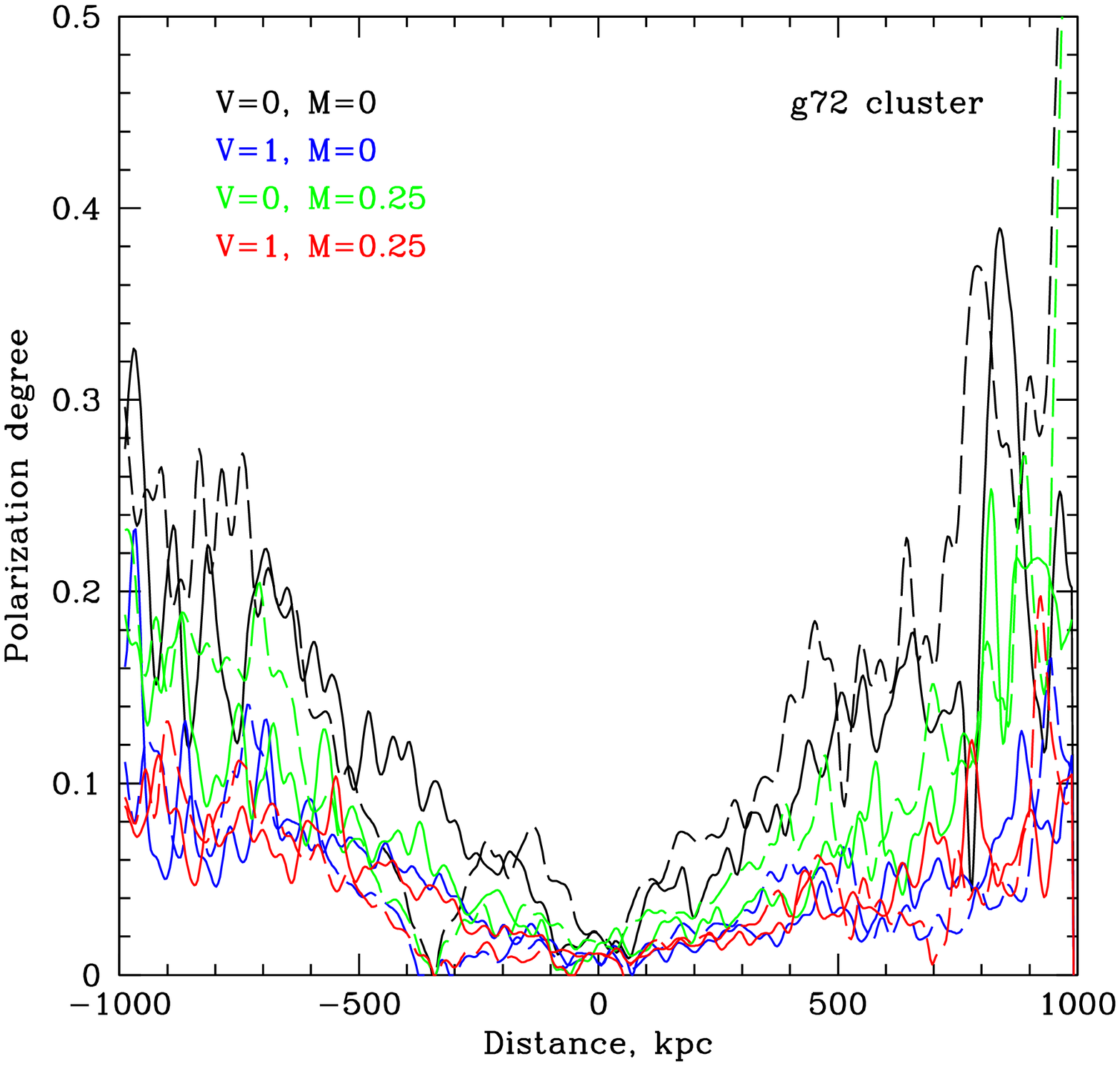}}
\caption{Radial slices of the surface brightness and the polarization degree for
  the g6212 cluster (right column), g8 cluster (middle column) and g72
  cluster (left column), illustrating radial variations of these
  parameters for two position angles. Colors correspond to different
  combinations   of the velocity factor $f_v$ and the Mach number $M$,
  labels are  shown in the top 
  left corner. Solid curves show slices going from the top left conner
  to the bottom right conner of images shown in
  Fig.\ \ref{fig:pol}. The dashed curves show slices going from the top right conner to the left bottom.    
\label{fig:poldiag}
}
\end{figure*}

In Fig.\ref{fig:pol} and Fig.\ref{fig:poldiag} we show the
results of radiative transfer calculations for  the  three clusters discussed
above. Calculations are done for various combinations of the parameters
$f_v$ and $M$, namely ($f_\v=0,M=0$), ($f_\v=1,M=0$),
($f_\v=0,M=0.25$) and ($f_\v=1,M=0.25$). For the g6212 cluster,
calculations are done for the L-shell line of C-like iron at 1.009
keV as discussed above.
For  the g8 and g72 clusters we analyze the Fe XXV line at 6.7 with optical depth
$\sim$3.6 and $\sim$3.2 respectively. For g8 and g6212 clusters we
consider multiple scattering by setting the minimum photon weight to $10^{-8}$ (see section \ref{sec:mc}). For g72 cluster (merger of two subclusters) we take into account only the first scattering. Under the conditions characteristic for galaxy clusters, accounting of multiple scattering does not have strong impact on the degree of polarization.  

As expected in the very center of each cluster the radiation field
is almost isotropic and the polarization degree is accordingly very
low, increasing to the cluster edges.  

We note here that apart
from the real increase of the polarization signal towards the cluster
outskirts, there are two spurious effects which lead to the increase
of the polarization degree close to the edges of the simulated cube.
i) The first effect is similar to the effect of finite $r_\max$
considered in the previous section (see Fig. \ref{fig:polPer}): near
the edges of  the simulated volume,  the  radiation field is strongly anisotropic
and the role of 90\deg scattering is enhanced where  the line of sight is
tangential to the boundary of the simulated volume. From the
experiments with spherically symmetric models and the cubes of
different sizes we concluded that it is safe to use the data at the
projected distance approximately  half of $R_\max$. For our
cubes which are $\sim$2 Mpc on a side, this means that the results within
a 500 kpc circle (radius) are robust. ii) The second effect is caused by limited photon
statistics in the regions close to the image edges, generated by our
Monte-Carlo code. The code is optimized to produce  the smallest
statistical uncertainties in  the more central regions of simulated
clusters. With only few simulated photons in the outskirts of a
cluster, the derived degree of polarization can be spuriously very
high - e.g. it is $\equiv 1$ if there is only one photon in a pixel.
We suppress this effect by making sufficiently large smoothing
windows: $I$, $Q$ and $U$ images are first smoothed and the
polarization degree is calculated using smoothed images.  The net
result of these two effects is that outside the central 0.5 Mpc
(radius) circle the estimated degree of polarization is less robust
than in the inner region. As discussed below the outskirts of
clusters are not a very promising target for measurements of
the polarization.

In the g6212 cluster, the degree of polarization reaches 30-35\% within
a projected distance of $\sim 500$ kpc from the cluster center for the
case ($f_\v=0$, $M=0$). Adding bulk velocity from simulations and
no microturbulence ($f_\v=1$, $M=0$) causes a decrease of the
optical depth from almost 3 to $\sim$2, while the degree of
polarization decreases to $\sim$20-25\%. If only
turbulent motions are included ($f_\v=0$, $M=0.25$), then the
optical depth drops to $\sim$ 1.16. The maximum polarization degree in
this case is about 25$\%$ as shown in Fig.\ref{fig:pol}. The
combination of bulk velocities and micro-turbulence ($f_\v=1$,
$M=0.25$) further decreases the polarization signal and, according to
our simulations, the maximum is about 15$\%$ (Fig.\ref{fig:pol},
the bottom left panel).

In the g8 cluster,  the polarization degree does not exceed
27$\%$ for the case ($f_\v=0$, $M=0$) within 500 kpc from the
cluster center. As in the previous example the
gas motions affect the optical depth, diminishing it and leading to
the decrease of the degree of polarization: for $f_v=1$ the degree of polarization $P$ is less then 10$\%$ (Fig.\ref{fig:pol}).

For the g72 cluster  the situation is similar: when there are no motions,
polarization reaches $\sim$20$\%$ at a distance $r=500$ kpc from the
cluster center
and gas motions reduce the  polarization to 7$\%$. 

The simulations discussed above are
also illustrated in Fig.\ref{fig:poldiag}, where
radial slices of the surface brigtness and the polarization degree are
shown for the same set of simulations as in Fig.\ref{fig:pol}.

\begin{figure*}
\plottwo{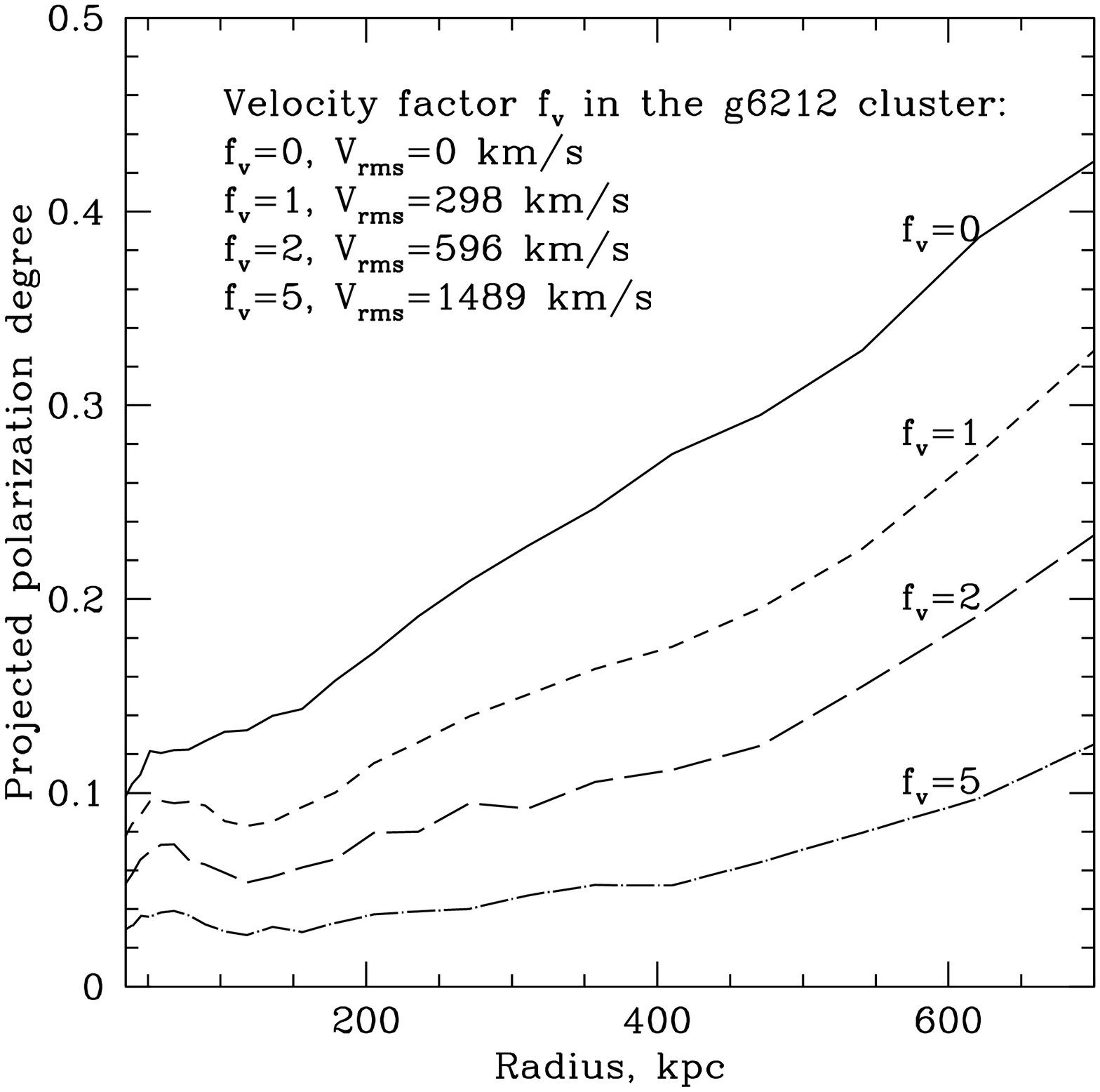}{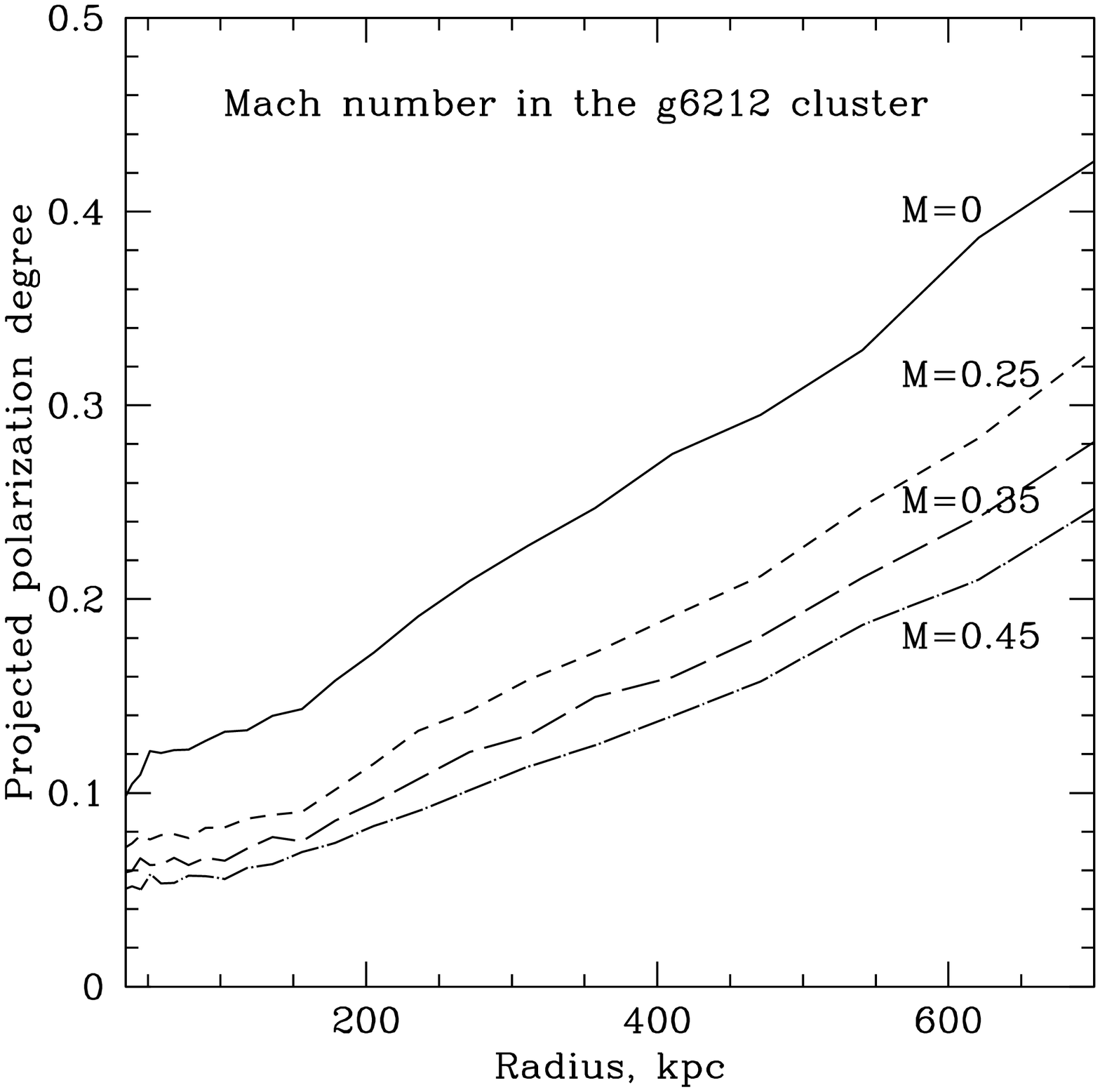}
\caption{Radial profiles of polarization degree in the Fe XXI iron line
  at 1.009 keV for the simulated cluster g6212 for different values of
  velocity factors and Mach numbers. The left panel: the solid line is
  for ($f_\v$=0, $M$=0), the short dashed line for ($f_\v$=1, $M$=0),
  the long dashed line is for ($f_\v$=2, $M$=0) and the dot-dashed
  line is for ($f_\v$=5, $M$=0). The corresponding RMS values of the gas velocities   
  $V_{\rms}$ in km/s are shown in the upper left corner. The right
  panel: the solid line is for ($f_\v=0$, $M=0$), the short dashed
  line for ($f_\v=0$, $M=0.25$), the long dashed line for ($f_\v=0$,
  $M=0.35$) and the dot-dashed line for ($f_\v=0$, $M=0.45$).
\label{fig:polvar}
}
\end{figure*}

Results for various other combinations of $f_\v$ and $M$ 
for the g6212 cluster are shown on Fig.\ref{fig:polvar}. Here, we show
 the projected polarization degree as we did in  the spherically-symmetric
case. The left panel shows profiles
of polarization for velocity factors $f_\v=$0, 1, 2 and 5 and the right
panel for Mach numbers $M=$0, 0.25, 0.35 and 0.45. This results are in
a good agreement with our predictions discussed above.

 In our 3D simulations we assume a flat iron abundance
profile of 0.79 solar. Any changes in the abundance profile
will have an impact on the optical depth in lines and hence can
affect the strength of the line scattering and the degree of
polarization. Obviously, an overall decrease of the abundance will
cause the decrease of the polarization signal.

 An impact of a peaked abundance profile on the polarization signal is
less obvious. Peaked abundance profiles are often observed in cool
core clusters. The abundances vary from $\sim 0.3$ in the outer
regions to $\sim 1.2$ at the center \citep[e.g.][]{Pra07}. To test
this case we consider an averaged metallicity profile for cool core
clusters parametrized by \cite{Gra04} as a function of
$r/r_{200}$, where $r_{200}$ is the radius corresponding to the mean
overdensity of 200$\times$ the critical density of the Universe. We used
the mean temperature of simulated clusters to evaluate $r_{200}$ and assumed
that the metallicity is linearly decreasing with radius for
$r<0.15r_{200}$ and is constant at larger radii. Rescaling the
results of \cite{Gra04} to the iron abundance scale of \cite{Lod03}
,the final iron abundance changes from 0.85 at the
center to 0.44 at large radii. Repeating the radiative transfer
calculations we found only a very slight ($\sim$10\%) decrease of
polarization degree (compared to the flat profile case) in the outer
regions where the abundance has changed by almost a factor of 2: from
0.79 (flat profile) to 0.44 (peaked profile). The reason for this
stability is as follows: polarized flux from outer parts of a cluster
is largely due to the scattered line photons coming from the central
bright region. This flux is obviously
proportional to the product of the flux from the central region
$F_\central$ and the optical depth of the outer region $\tau_\outer$.
The optical depth in turn is proportional to the abundance
$Z_\outer$. The polarization signal is diluted by the flux of "locally"
generated unpolarized line photons $F_\outer$, which is also proportional
to the abundance in the outer regions and to the product of the flux from
the central region and the optical depth of the outer region, i.e.
$\displaystyle
P\approx\frac{a\tau_\outer F_\central}{b\tau_\outer
  F_\central+F_\outer}$, where $a$ and $b$ are some constants, which
depend on the density distribution. Since both numerator and
denominator scales linearly with the abundance $Z_\outer$ this
dependence partly cancels out. In other words in the outer regions of clusters in the limit of low abundances the degree
of the polarization is approximately proportional to the expression
$\displaystyle \frac{\tau_\outer F_\central}{F_\outer}\propto
\frac{Z_\outer F_\central}{Z_\outer}\approx F_\central$ . Thus
these two effects partly cancel each other leading to a relatively
weak dependence of the degree of polarization on the abundance. In other
words, the increase of the flux coming from the central region has
larger impact on the polarization degree than the changes in abundance
in outer regions \citep[see also][for the limiting case of a strong
central source, when a dilution by locally produced line photons is not
important]{Saz02}.

Note, that these calculations assume that the line photons can be
separated from the continuum, i.e. the energy resolution of the
polarimeter is very high (see Section \ref{sec:disc}  for the discussion of the
impact of limited energy resolution on the polarization signal).

\subsection{Major merger}
A particularly  interesting case is when  the polarization degree increases
due to  the anisotropy caused by gas motions. This can happen, for example,
when gas blobs are moving in the cluster center or when two galaxy 
clusters are
merging. The well-known example of merging clusters is the Bullet cluster in which gas
motions reach velocities up to 4500 km/s. Even if the radiation field is originaly isotropic (e.g. in the very center of the
galaxy cluster) the dependence of the scattering cross section on energy (see section \ref{sec:intro}) will produce anisotropy and polarization in the scattered radiation.
\begin{figure*}
\plottwo{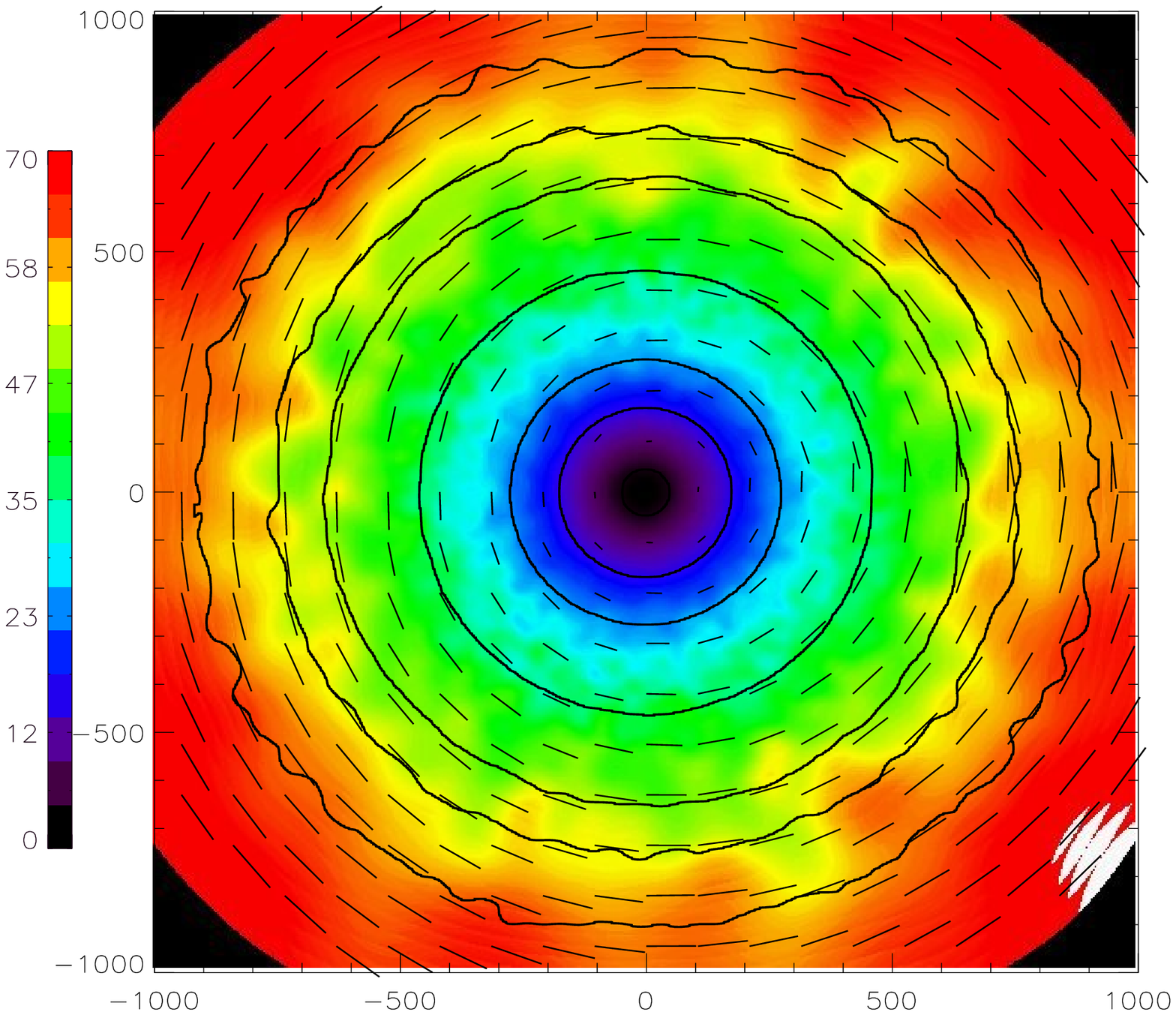}{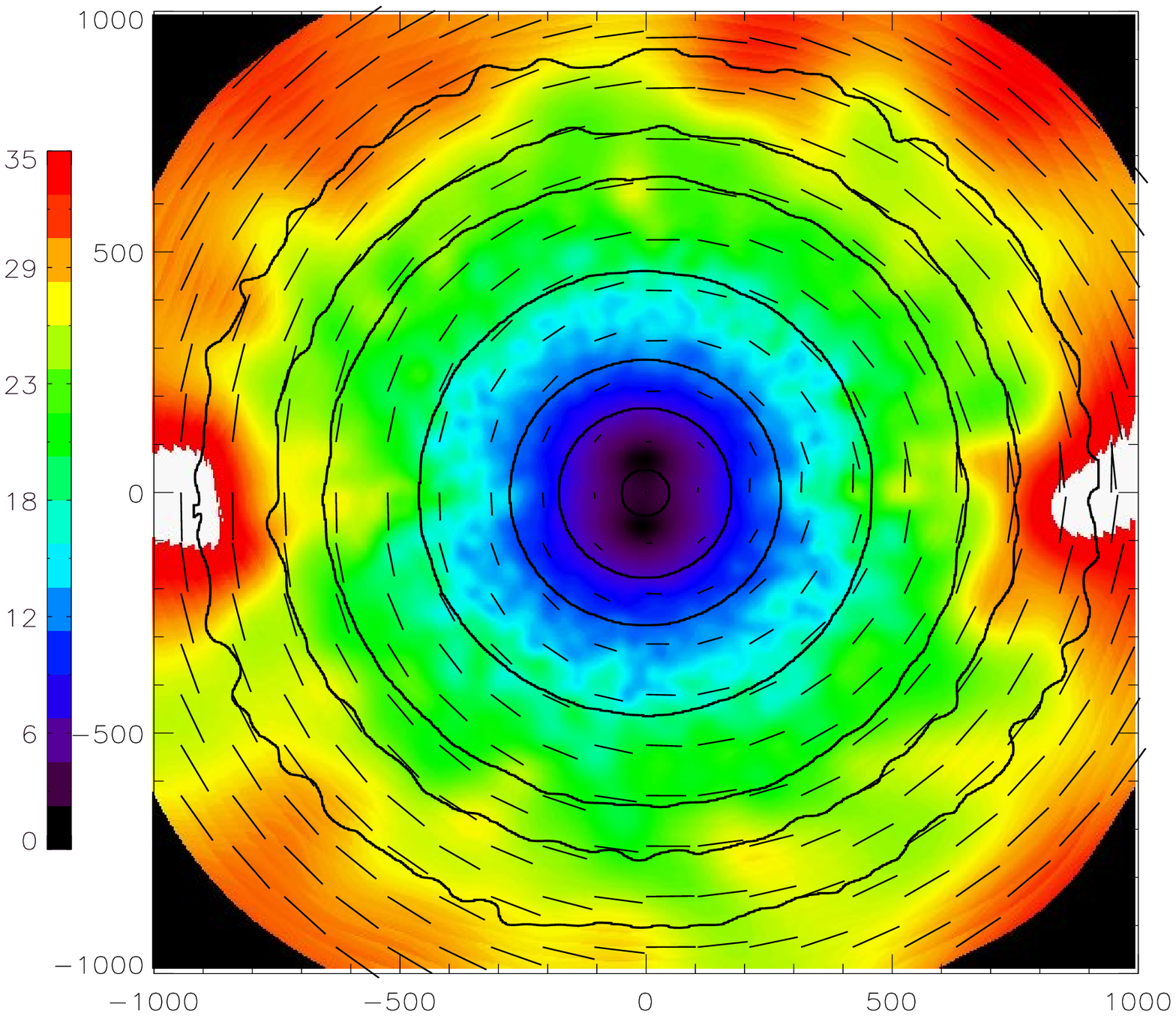}
\caption{Polarization degree in a spherically-symmetric
  $\beta-$model galaxy cluster with $n_0=10^{-2}$ cm$^{-3}$,
  $r_\c=100$ kpc, $\beta=1$.  Temperature is constant $T=5$ keV and
  iron abundance is 0.79 solar.  Calculations were done for the
  He-like iron line at 6.7 keV. The polarization degree was
  evaluated as $P=\sqrt{Q^2+U^2}/I$, the value of which is shown in
  the colorbar in percent. $I$ is intensity of both direct and
  scattered emission.The colors in the images show the
  polarization degree, the short dashed lines show the orientation of the electric
  vector. The contours of the X-ray surface brightness in the chosen
  line are superposed. The size of each picture is 2$\times$2 Mpc. On
  the left panel are shown results when the gas is motionless. On
  the right panel is shown the case when two parts of the cluster
  are moving towards each other with velocities 500 km/s and -500 km/s
  respectively.  In this case the asymmetry in  the radiation field
  appears in the very center of  the galaxy cluster and  the  polarization degree
  reaches about 2$\%$.
\label{fig:mergpol}
}
\end{figure*}

To model the polarization from a major merger, we consider an
illustrative example: two halves of galaxy cluster are moving towards each
other with velocities $\pm$500 km/s. In such a
situation,  an anisotropy in the scattering cross section (see eq. \ref{eq:intro3}) can produce polarized radiation even in the cluster center.
 We used a simple spherically-symmetric
$\beta-$model for the density distribution  $n_0=10^{-2}$ cm$^{-3}$ with
$r_\c=100$ kpc, $\beta=1$.  The temperature of cluster is constant $T=5$ keV and the iron abundance is 0.79 solar.  Calculations were done
for the He-like iron line at 6.7 keV. To show  the sensitivity of
polarization to the tangential component of gas motions, we
assume that only the y-component of velocity is non-zero, while  the
viewing  vector is along  the z-axis (see, e.g. Fig.\ref{fig:intro}).
Fig.{\ref{fig:mergpol}} shows the results of our calculations. We see
that in the case of such motions, the polarization in the center increases
from 0$\%$ to 2$\%$.

\section{Discussion}
\label{sec:disc}
Our results show that, owing to the asymmetry in the radiation
field, the polarization for the brightest X-ray lines from galaxy
clusters can reach  values as high as 30$\%$. In addition, we
have shown that the polarization degree is sensitive to gas
motions. Therefore, polarization measurements provide a method to
estimate velocities of gas motions in the direction perpendicular to
the line of sight.

\begin{figure*}
{\centering \leavevmode
\epsfxsize=1.\columnwidth \epsfbox[70 190 620 680]{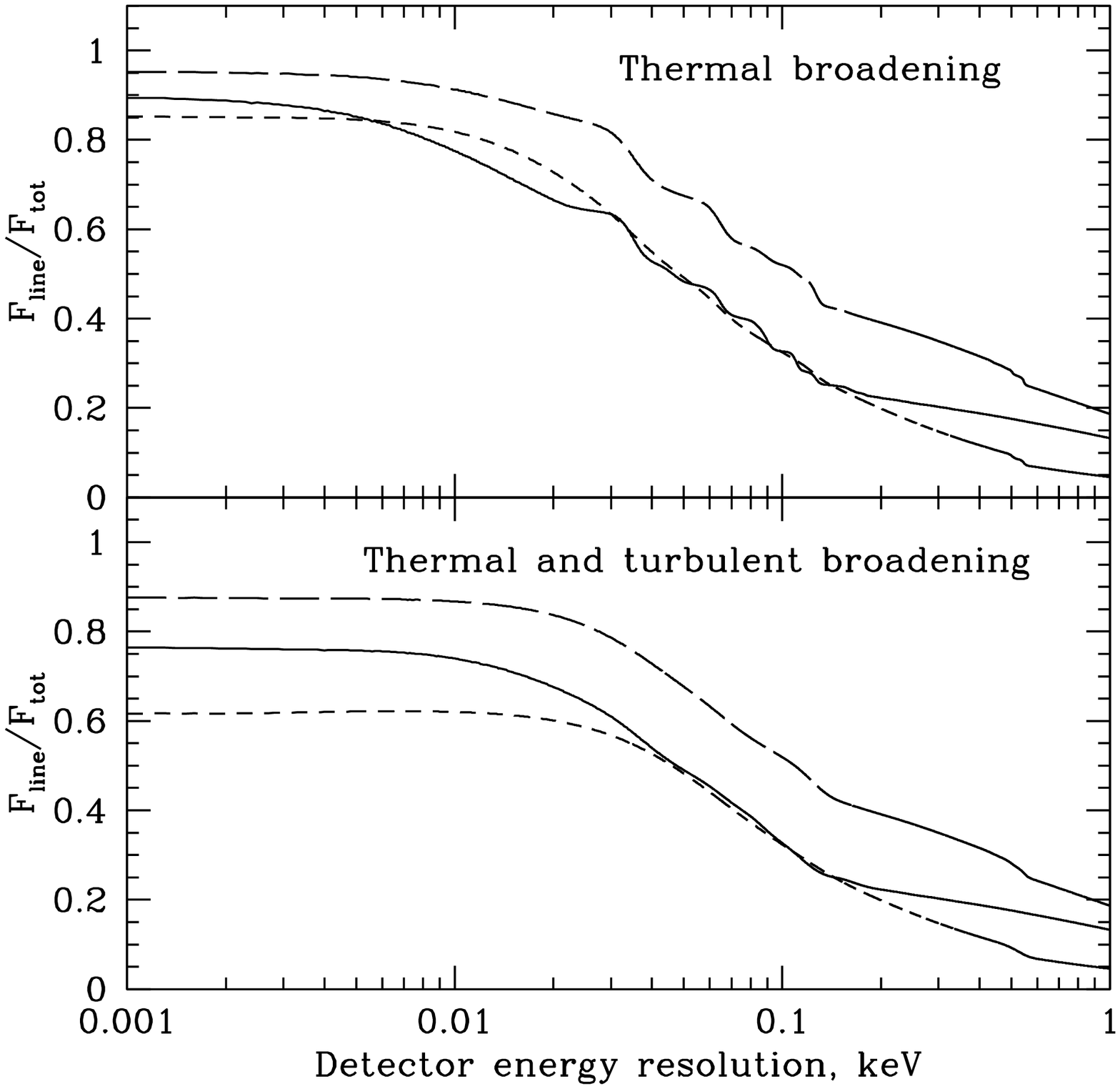}
\epsfxsize=1.\columnwidth \epsfbox[70 190 620 680]{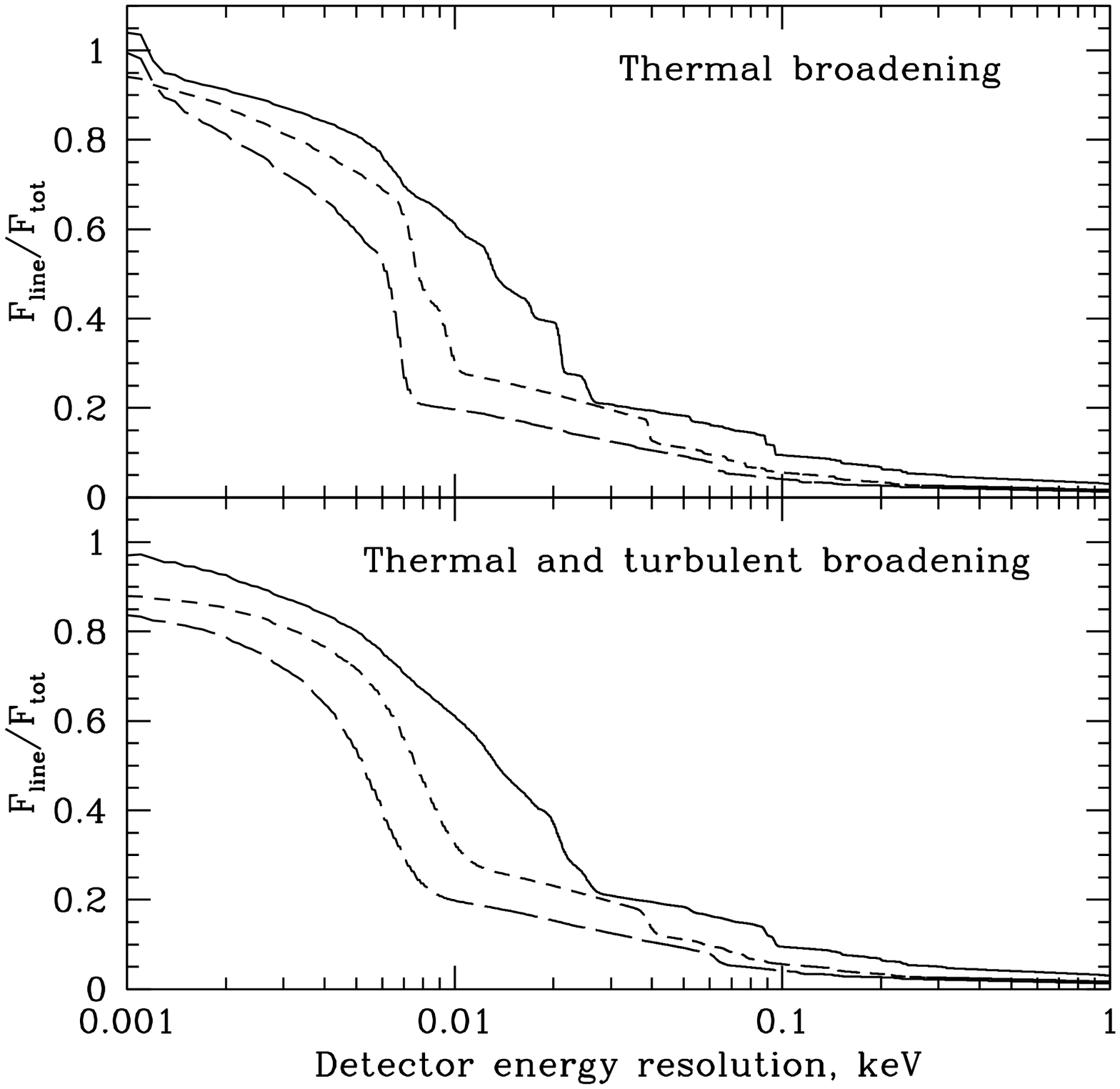 }\hfil
\epsfxsize=1.\columnwidth \epsfbox[100 190 620 730]{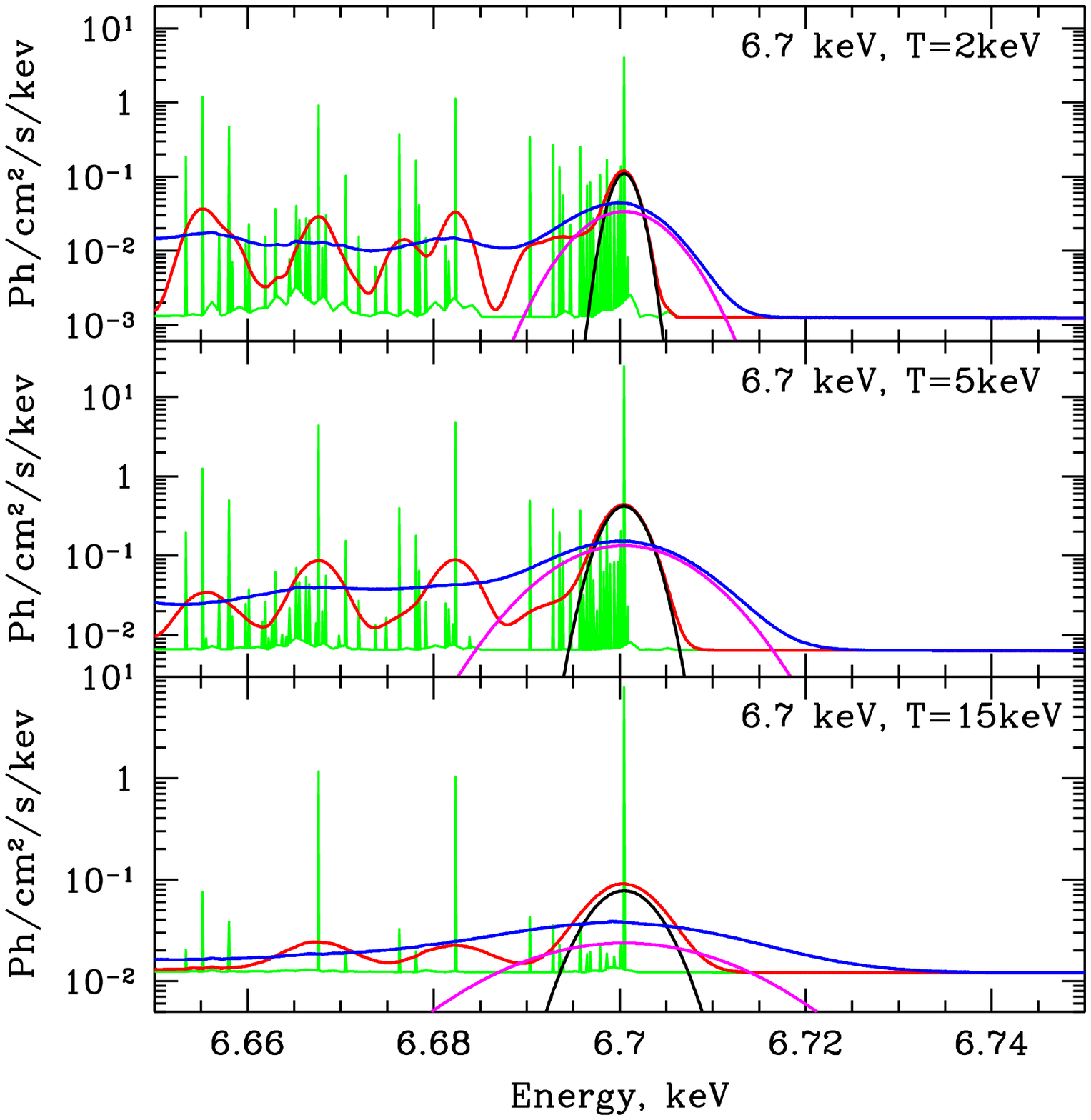}
\epsfxsize=1.\columnwidth \epsfbox[100 190 620 730]{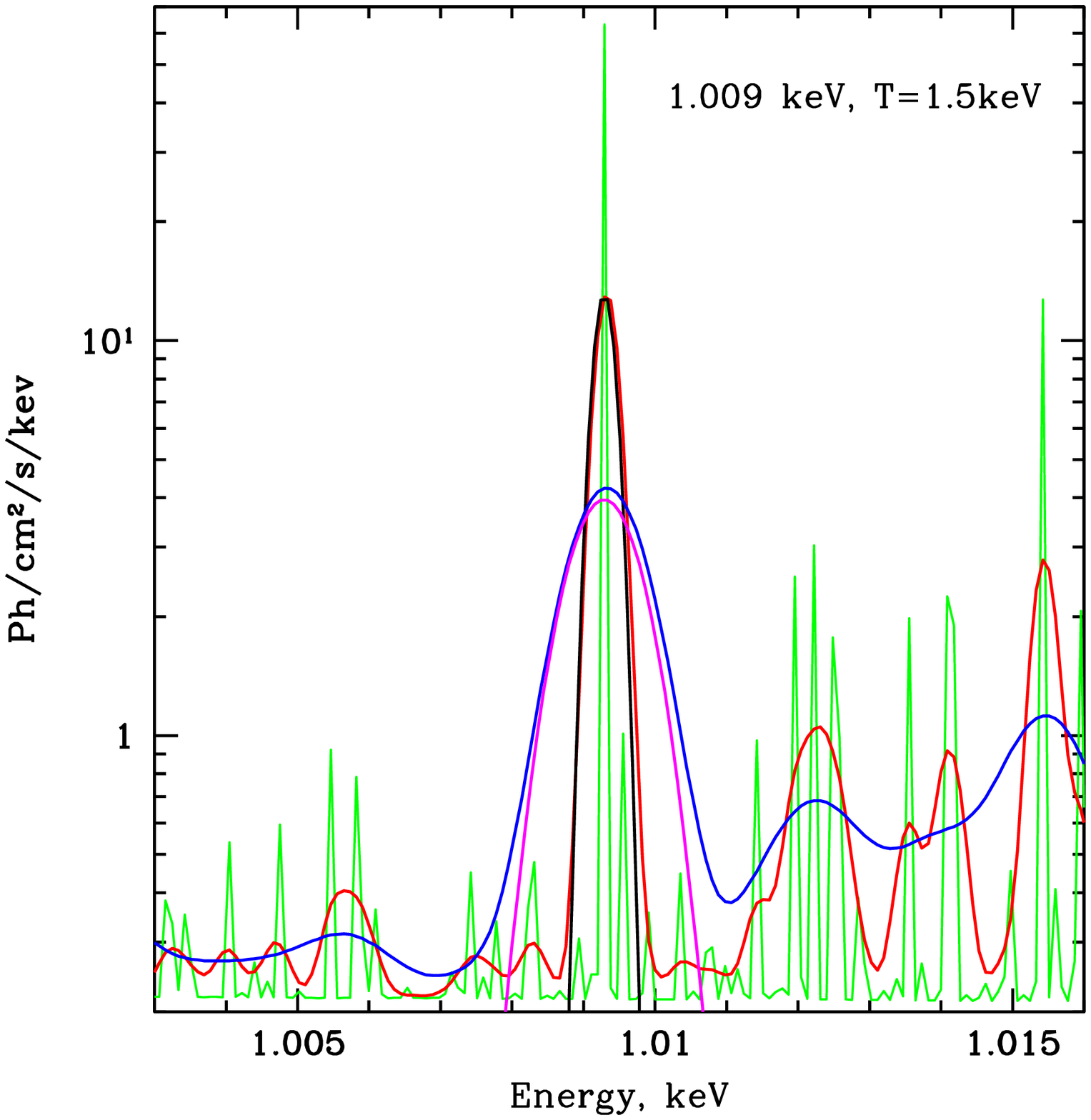}}
\caption{ Top panel: changes of the polarization degree as a function
  of the instrument energy resolution. On the upper panels results
  are shown when only thermal broadening of lines is considered. The
 cases when turbulent broadening is included are shown on the lower
 panels. The lines in the left panel
  show the ratio of the resonant line flux (6.7 keV permitted line) to the
  total (continuum plus all lines) flux within an energy interval of a
  given width centered at the line energy. The solid, long dashed and
  short dashed lines correspond to plasma with the temperature of 2, 5
  and 15 keV respectively. In the right panel similar curves are shown
  for the bright low energy lines: Fe XXI line at 1.009 keV (solid
  line), Fe XXIII line at 1.129 keV (short dashed line) and Fe XXII line at 1.053 keV 
  (long dashed line) for the plasma with the the temperature 1.5 keV.   
The iron abundance is 
assumed to be 0.79
 solar. Only the flux in the line is assumed to be polarized.\newline
Bottom panel: model spectra of the hot optically thin plasma with different
temperatures.\newline
{\bf Color coding:}\newline
{\it Green} - spectrum with no line broadening (apec model in XSPEC).\newline
{\it Red} - spectrum with pure thermal line broadening (bapec model of
XSPEC).\newline
{\it Black} - the Gaussian profile of the main line for pure thermal broadening.\newline
{\it Blue} - spectrum with thermal and turbulent line broadening 
(bapec model of XSPEC). Turbulent velocities are set to 
$V_{\turb}=0.25 c_\s$.\newline
{\it Magenta} - the Gaussian profile of the main line,  accounting
for the thermal and turbulent broadening.\newline 
The left panel: model spectrum of cluster with temperature 2, 5
and 15 keV (the top, middle and bottom panels respectively). The
main line is the Fe XXV line of He-like iron at 6.7 keV. The right
panel: model spectrum for temperature 1.5 keV. The main line is the 
Fe XXI line at 1.009 keV. 
\label{fig:flrat}
}
\end{figure*}

One of the limiting factors which affects the expected degree of
polarization is the contamination of the polarized emission in the
resonant line by the unpolarized emission of the continuum emission of
the ICM and neighboring lines. This factor critically depends on the
energy resolution $\Delta E$ of the polarimeter. If we assume that
only the emission in a given line is polarized, the measurable
polarization will differ from the idealized case by a factor
$f=F_\line/F_\tot$, where $F_\line$ is the flux in the resonant line
and $F_\tot$ is the total flux measured by the detector in an energy
band $\Delta E$ centered on the resonant line energy. In
Fig.\ref{fig:flrat} we plot this factor as a function of the width
of the energy window (detector energy resolution) for different theoretical spectra for the model
clusters discussed above assuming only thermal line broadening and both
thermal and turbulent broadening. 
One can see that the contamination of the 6.7 keV line flux by the unpolarized flux is very minor for an energy resolution better than $\le$ 10 eV. For the energy resolution of 100 eV the fraction of contaminating unpolarized flux is of order 30-60\% depending on the plasma temperature. 
This line is most prominent
in clusters with temperatures $\sim$ 3-6 keV as the line flux at such
temperatures is the largest and neighbouring lines are not as strong
as in a case of $\sim$2 keV plasmas (see the bottom left panel on
Fig.\ref{fig:flrat}). For lines with energies $\sim$1 keV  even better energy
resolution is needed since in the spectrum there is a forest of low energy lines and the
contamination from these lines dramatically decreases the flux of
polarized emission (see the bottom right panel on Fig.\ref{fig:flrat}).

\begin{figure*} 
  {\centering \leavevmode 
\epsfxsize=1.\columnwidth \epsfbox[35 160 610 650]{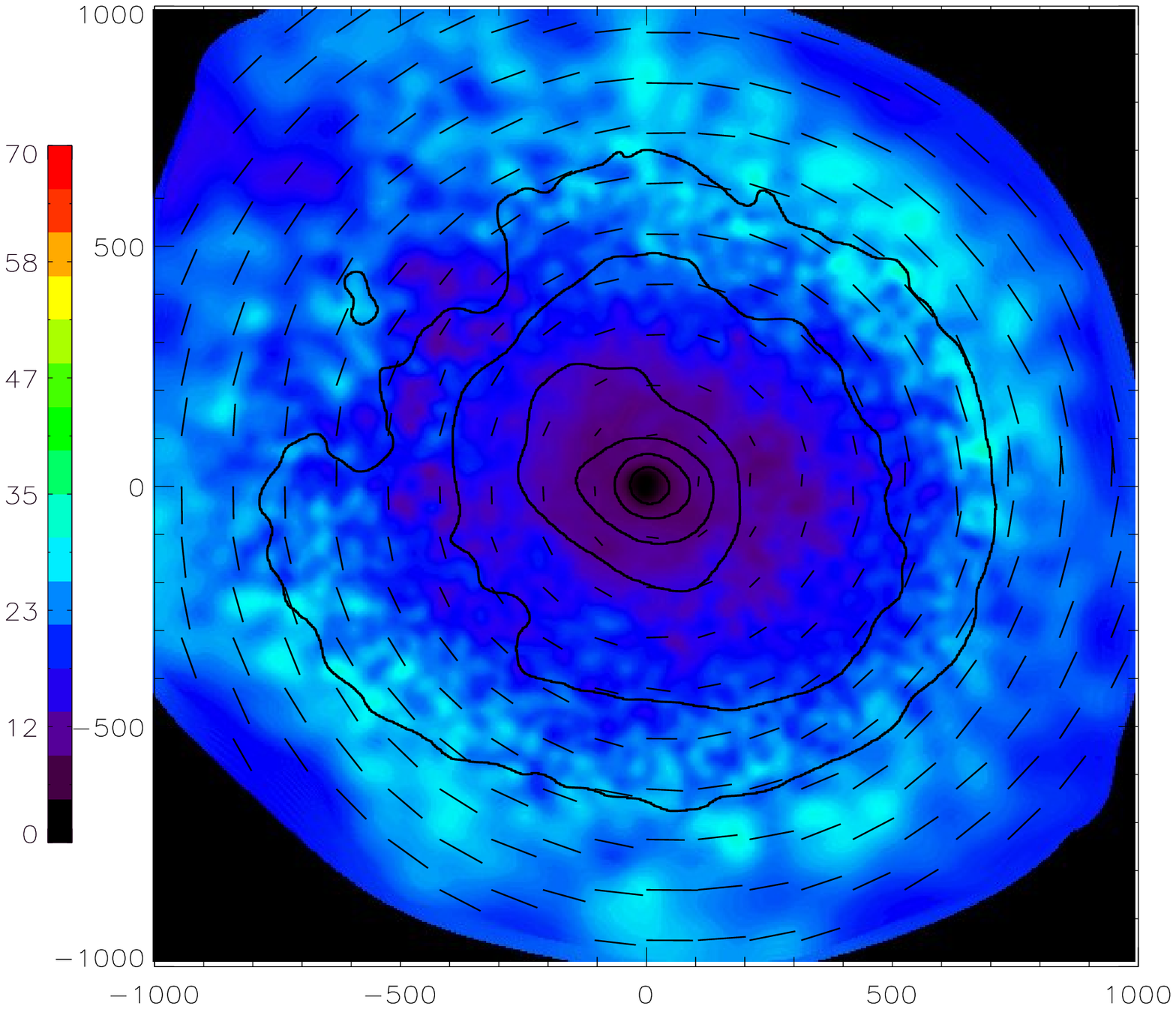} 
\epsfxsize=1.\columnwidth \epsfbox[35 160 610 650]{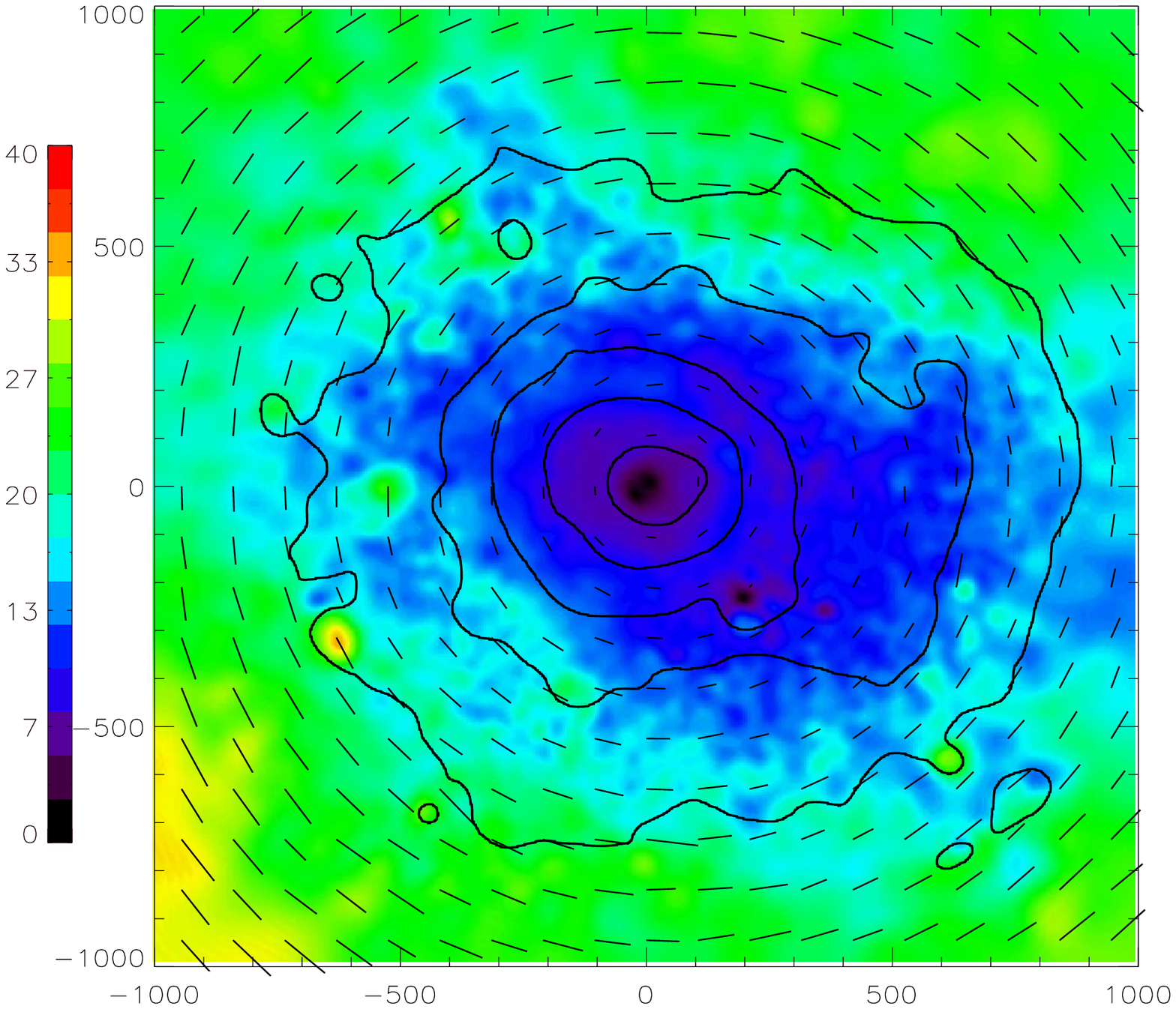}\hfil
  \epsfxsize=1.\columnwidth \epsfbox[35 160 610 650]{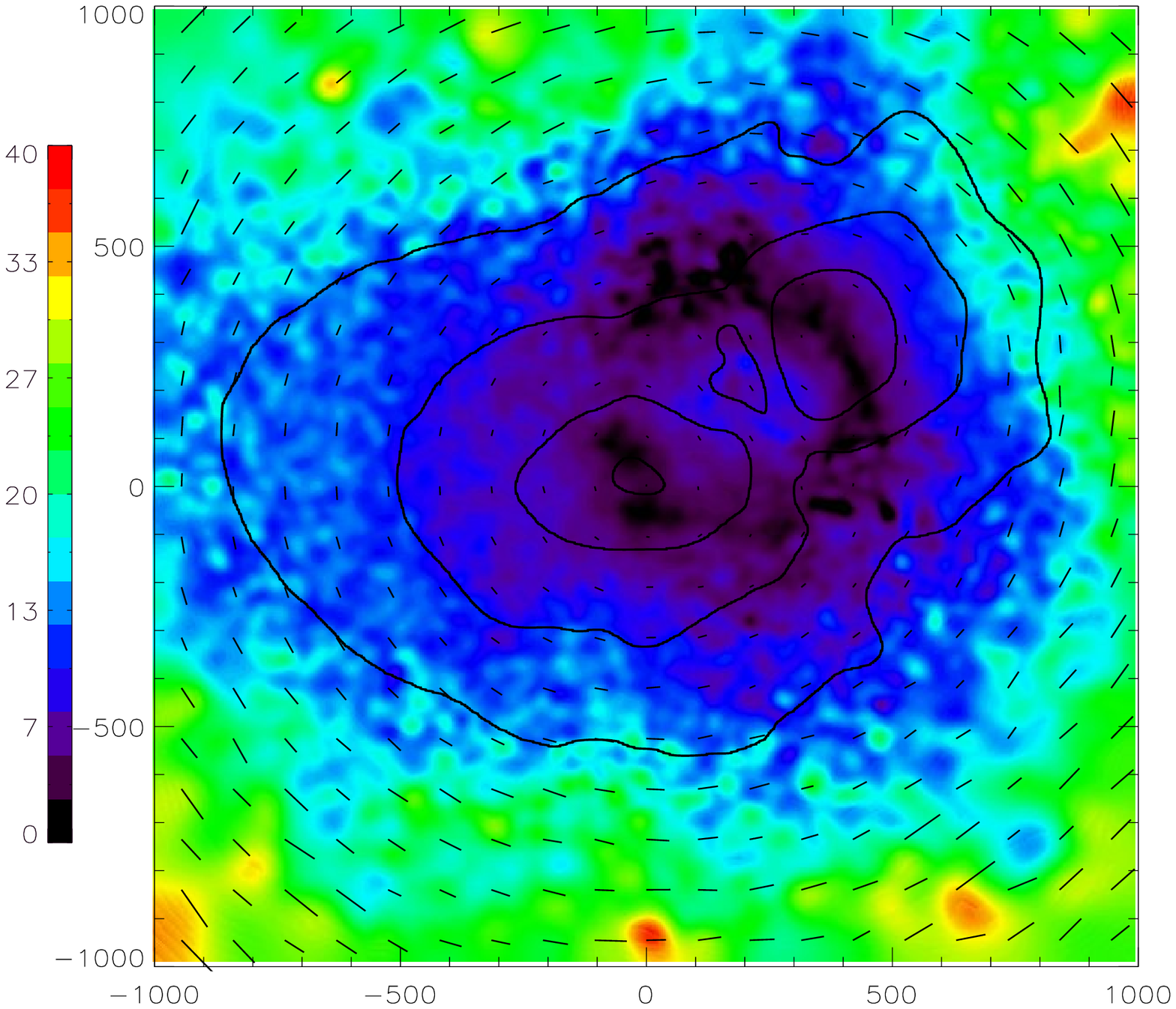}}
  \caption{Polarization degree for simulated clusters g6212 (the top
    left panel), g8 (the top right panel) in the lines of Fe XXI at
    1.0092 keV and Fe XXV at 6.7 keV, respectively, and for the g72
    cluster (the middle bottom panel) in the 6.7 keV line. The polarization degree was evaluated as
  $P=\sqrt{Q^2+U^2}/I$, the value of which is shown in colorbar in
  percent. $I$ in a total intensity, i.e. intensity of scattered and
  direct emission. The colors in the images show polarization
  degree, the short dashed lines show the orientation of the electric vector. The
  contours of the X-ray surface brightness  (factor of 4 steps) in the
  chosen line are   superposed. The size of each picture is 2$\times$2 Mpc. The gas is
    motionless ($f_\v=0$, $M=0$). The CXB radiation is included and
    the energy resolution is 0.01 keV. Also the decrease of
    polarization degree due to the continuum and neighboring
    non-polarized lines is taken into
    account. According to Fig.\ref{fig:flrat} the limiting factor for
    the 1.0092 keV line in g6212 cluster is 0.6, for the 6.7 keV
    line in g8 cluster is 0.85 and for the same line in the g72
    cluster is 0.95.
\label{fig:polcxbicm}
}
\end{figure*}

Another effect which also depends on the energy resolution of the
polarimeter is the contamination of line flux by the contribution
of the cosmic X-ray background (CXB) which starts to dominate at some
distance from the galaxy cluster center. Assuming that the CXB is
not resolved into individual sources by the polarimeter the
contribution of the CXB starts to be the dominant source of
contamination at a distance from the cluster center where the
surface brightness of the cluster continuum emission matches the CXB
surface brightness (at the line energy). As an illustrative example we
use the results of simulations for the g6212 cluster in the Fe XXI line
at 1.0092 keV, setting $f_v=0$ and $M=0$, and similarly for the g8
cluster in the Fe XXV line at 6.7 keV and for the g72 cluster in the
same line.  All clusters are placed at a distance of 100 Mpc and the
energy resolution of the instrument is set to 10 eV. Using the CXB
intensity from \cite{Gru99} \citep[see also][]{Chur07} we calculated the expected degree of
polarization (Fig.\ref{fig:polcxbicm})  including a CXB
``contamination'' flux $I_{CXB}\times \Delta E$. Accounting for
the finite energy
resolution and contamination by the ICM emission affects the
polarization signal from any region of a cluster, while the CXB
contribution (for an instrument with a modest energy resolution)
effectively overwhelms the polarization signal in the cluster outskirts.
In Fig.\ref{fig:polcxbicm} we show results for  the  g6212, g8 and g72
clusters, where  the limiting factors discussed above are taken into
account. According to Fig.\ref{fig:flrat} the limiting factors due to
the 
ICM  are 0.6, 0.85 and 0.95 for g6212, g8 and g72 clusters if we assume
an energy resolution of 10 eV. First we see that inclusion of the CXB reduces
the very high polarization degree in the outer parts of clusters. Second,
 the polarization degree is reduced everywhere  by
 ``contamination'' from 
unpolarized radiation. Within  the central region ($r\approx$500 kpc),
 the polarization degree decreases from 30-35$\%$ to 20$\%$ in
 the g6212
cluster, from 27$\%$ to 20$\%$ in  the g8 cluster. For the g72 cluster,
the degree of polarization does not change significantly, since the
contamination by the unpolarized radiation is small in this case.

It is also interesting to note, that when gas motions with velocities
up to a few 1000 km/s exhibit, the Doppler shifts corresponding to such
velocities are comparable or larger than some doublet separations (see
e.g. Fig.\ref{fig:flrat}). And flux from one line can be scattered in
another neighbouring line. This can easily happen, for example, in
doublet with the components at 6.97 keV and 6.93 keV.

\section{Requirements for future X-ray polarimeters}

Galaxy clusters are promising, but challenging targets for future
polarimetric observations which pose requirements to essentially all
basic characteristics of the telescopes, such as angular and energy
resolutions, effective area and the size of the field of view.  (i)
First of all, in order to detect polarization in lines 
polarimeters should have good energy resolution to avoid
contamination of the polarized line flux by the unpolarized
continuum and nearby lines (Fig.\ref{fig:flrat}).  (ii) An angular
resolution has to be better than an arcminute even for nearby
clusters. Indeed, the polarization largely vanishes when integrated
over the whole cluster. For instance, the polarization of the 6.7
keV line flux, integrated over the whole cluster, is 1.2\%, 0.55\%
and 1.2\% for the g6212, g8 and g72 clusters respectively. As is clear
from e.g. Fig.\ref{fig:pol} the polarization direction is predominantly
perpendicular to the radius and the polarization degree 
varies with radius (see Fig.\ref{fig:polPer}). Therefore, to get
most significant detection it is necessary to resolve individual
wedges and to combine the polarization signal by aligning the reference
frame, used for the calculations of Stokes parameters, with the
radial direction. For nearby clusters (e.g. Perseus) this requires an
arcminute angular resolution. (iii) Since the polarization is small in
the central bright part of the cluster (e.g. Fig.\ref{fig:polPer}) the
instrument should have sufficiently large effective area to collect
enough photons from the cluster outskirts. (iv) For the same reason a
large Field-of-View is important for collecting a large number of
photons from the brightest nearby clusters like Perseus.

Currently, several polarimeters based on the photoelectric effect are actively
discussed (see e.g. \citet{Cos01, Mul09, Bel07, Jah07, Sof01}). These
are basically gas detectors with fine 2-D position resolution to
exploit the photoelectric effect. When a photon is absorbed in the
gas, a photoelectron is emitted preferentially along the direction of
polarization. As the photoelectron propagates its path is traced by
the generated electron-ion pairs, which are amplified by a Gas
Electron Multiplier (GEM) and are collected by a pixelized
detector. Hence the detector sees the projection of the track of the
photoelectron.

One of the polarimeters based on photoelectric effect is
GEMS
(Gravity and Extreme Magnetism SMEX) \footnote{http://projects.iasf-roma.inaf.it/xraypol/Presentations/\\090429/morning/JSwank$\_$GEMS.pdf}of the NASA SMEX program. This
polarimeter will provide polarization measurements in the standard
2-10 keV band with energy resolution $\Delta E/E=0.2$, a modulation
factor $\mu=$0.55 and an effective area of $\sim$600 cm$^2$ at 6 keV.

Another  proposed instrument is EXP (Efficient X-ray Photoelectric
Polarimeter) polarimeter on-board HXMT (Hard X-ray Modulation
Telescope). This telescope has a large field of view (FoV) $22'\times 22'$,
mirrors effective area $\sim$280 cm$^2$ at 6.7 keV and the detector
quantum efficiency (QE) of about 3$\%$. The energy resolution is
$\Delta E/E=0.2$ at 6.7 keV \citep{Cos07, Sof08}.

Polarimeters have been also
proposed as focal plane instruments\footnote{E.g. http://projects.iasf-roma.inaf.it/xraypol/Presentations/\\090428/afternoon/roma$\_$09$\_$04$\_$sandro$\_$Brez.pdf} for the International X-ray observatory (IXO) \citep{Cos08}.

Most important characteristics of discussed polarimeters are
summarized in Table \ref{tab:polar}. We also show in this table the
characteristics of a hypothetical polarimeter, which would be
suitable for the study of polarized emission lines in galaxy
clusters. The key difference of this hypothetical polarimeter with the
currently discussed GEM polarimeters is the energy resolution of 100
eV and larger efficiency.

Obviously only the brightest clusters have a chance of being detected
with the current generation of X-ray polarimeters. Discussed below is
the case of the Perseus cluster, which is the brightest cluster in the
sky and as such is the most promising target. The degree of polarization in
the 6.7 keV resonant line flux is shown in Fig.\ref{fig:polPer}. While
the surface brightness declines with radius, the degree of polarization
is instead steadily increasing with radius. It is possible therefore that
the optimal signal-to-noise ratio (S/N) is achieved at some
intermediate radius. The expected modulation pattern in a given pixel
of the cluster image is proportional to
\be 
c+\mu pc\cos(2\phi+\phi_0),
\label{eq:polsig}
\ee 
where $c$ is a number of counts in this pixel, $\mu$ is a
modulation factor, $p$ and $\phi_0$ are the degree of polarization
and the angle of the polarization plane with respect to the reference
coordinate system. One can fit the observed modulation pattern with a function
\be
x_1+x_2\cos(2\phi),
\label{eq:polsig2}
\ee 
where $x_1$ and $x_2$ are the free parameters. In the above
expression we set $\phi_0$ to zero, since 
the polarization plane is perpendicular to the radial direction (in the first approximation) and
therefore the angle is known. We assume below that for any pixel of
the cluster image the reference system is aligned with the radial
direction. This way the Stokes Q and U parameters obtained in
individual pixel can be averaged over any region of interest. The final S/N ratio, characterizing the significance of the polarization detection from the region is 
\be
\disp\left(\frac{S}{N}\right)_0=\disp\frac{< x_2 >}{\sigma_{< x_2 >}}=\disp\mu\frac{\sum\limits_{i}p_{i}c_{i}}{\sqrt{2\sum\limits_{i}c_{i}}},
\label{eq:sn0}
\ee where the summation is over all pixels in the region. If an
unpolarized flux of the continuum and neighbouring lines is
contributing to the total flux seen by the
detector (see \S\ref{sec:disc}) then the S/N ratio will scale as the square
root of the line and total fluxes: 
\be
\disp\frac{S}{N}=\left(\frac{S}{N}\right)_0\sqrt{\frac{F_\line}{F_\tot}},
\label{eq:sn}
\ee
where $\left(\frac{S}{N}\right)_0$ is the S/N ratio for the pure line
flux. 

\begin{table*}
 \centering
  \caption{The main characteristics of HXMT\citep{Cos07,Sof08}, IXO\citep{Cos08}, GEMS\citep{Jah07} and
    hypothetical polarimeters.} 
  \begin{tabular}{@{}rcccc@{}}
  \hline
  & HXMT & IXO & GEMS & Hypothetical polarimeter\\
\hline
 Mirrors Effective Area @ 6.7 keV & 280 cm$^2$ & 5600 cm$^2$ & 600
 cm$^2$  & 1000 cm$^2$\\
 Field of View &$22'\times22'$& $2.6'\times 2.6'$& $14'$ & $20'\times20'$\\
Angular Resolution & $1'$ & $5''$& $1.8'$[HPD] & $1'$\\
Energy Resolution & $\Delta E/E=0.2$   & 0.2 & 0.2  & 100 eV\\
Modulation Factor @ 6 keV $\mu$ & 0.65 & 0.65  & 0.55 & 0.65\\
Detector Quantum Efficiency @ 6 keV & 0.023 & 0.023& 0.1 & 0.5\\
\hline
\label{tab:polar}
\end{tabular}
\end{table*}

\begin{figure}
\plotone{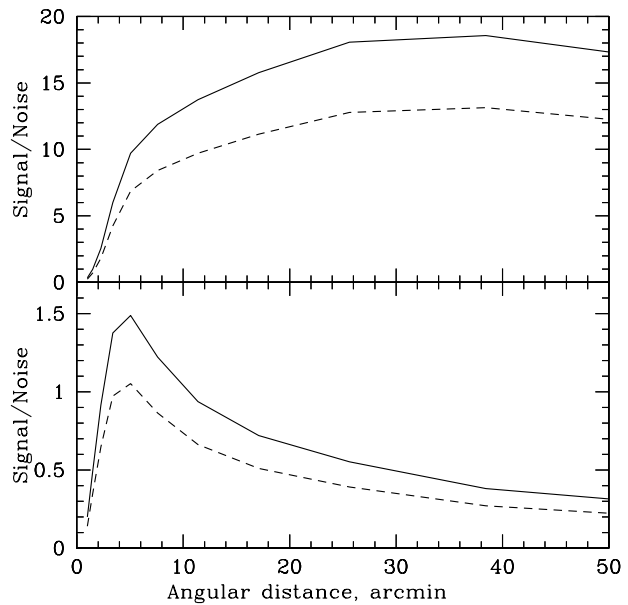}
\caption{ Significance of the 6.7 keV line polarization detection in
  Perseus cluster. For this plot we assume that a hypothetical
  polarimeter with  characteristics listed in Table \ref{tab:polar} is
  observing the cluster for 1 Megasecond. The solid line shows the
  detection significance if only line photons are detected, while the
  dashed line shows the S/N ratio for an instrument with the energy
  resolution of 100 eV (and therefore the continuum and nearby lines
  decrease the degree of polarization). {\bf Top panel:} the S/N ratio is
  calculated using eq.(\ref{eq:sn0}) and eq.(\ref{eq:sn}) for a set of annuli with inner and
  outer radii $R_1=R$ and $R_2=1.5\times R$ and plotted as a function
  of $R$. {\bf Bottom panel:} the same as in the top panel, but for a $1'\times 1'$ square region.   
\label{fig:sn}
}
\end{figure}

The S/N
ratio for the 6.7 keV line in Perseus is shown in
Fig.\ref{fig:sn}. For this plot we assume that a hypothetical
polarimeter with the characteristics given in Table \ref{tab:polar} is
observing the cluster for 1 Megasecond. In the bottom panel the S/N ratio was
calculated for  $1'\times 1'$ square regions located at a given
distance from the center of the Perseus cluster. The solid line is
calculated assuming that only 6.7 keV line photons are detected (i.e. $\left(\frac{S}{N}\right)_0$), while the
dashed line shows the S/N ratio for an instrument with the energy
resolution of 100 eV (and therefore the continuum and nearby lines
decrease the degree of polarization). In the top panel of
Fig.\ref{fig:sn} the S/N ratio is calculated for a set of annuli with inner and
outer radii $R_1=R$ and $R_2=1.5\times R$ and plotted as a function of $R$.  In
our model of the Perseus cluster its surface brightness in the outer regions is 
described by a $\beta$ model with $\beta=0.5$. Thus for large radii
the surface brightness declines as $1/R^2$. If the solid angle of the region
scales as $R^2$ (as is the case for the annuli used in the top panel
of Fig.\ref{fig:sn}) these two factors cancel each other and the S/N ratio as a function of radius simply
follows the variations in the degree of polarization with radius (see Fig.\ref{fig:polPer}). The
solid line in the top panel of Fig.\ref{fig:sn} again corresponds to
the line photons only, while the dashed line assumes 100 eV energy
resolution. Setting the energy resolution to 1 keV (realistic number
for a gas detector) would decrease the S/N ratio from $\sim$18 to $\sim$8
in an annulus near 30$'$.

The above analysis suggests that for an instrument with a very small
FoV (few arcminutes) the optimal S/N ratio will be achieved if a
region at a distance of $4'-5'$ from the center of the Perseus cluster
is observed. For a large FoV (few tens of arcminutes) the optimal
strategy is to point directly at the center of the cluster and to
combine the signal from the cluster outskirts.

\begin{figure}
\plotone{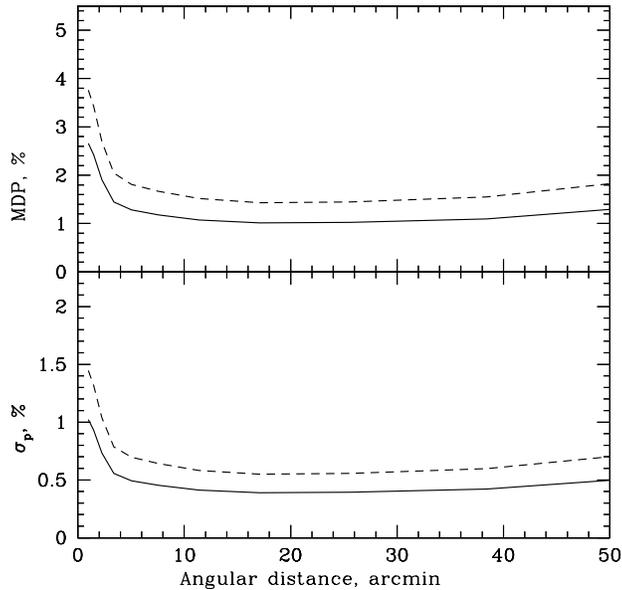}
\caption{Top panel: MDP calculated using eq.\ref{eq:mdp} in the Perseus
  cluster for a set of annuli with inner and outer radii $R_1=R$ and
  $R_2=1.5\times R$ as a function of $R$. The solid curve shows
  MDP when only flux from the line is included, the dashed curve
  shows the case when the unpolarized continuum is added. Characteristics of the hypothetical
  polarimeter were assumed (see Table \ref{tab:polar}) \newline
The bottom panel: statistical uncertainty in $P$ in the Perseus cluster. 
 Notations are the same as in the top panel. 
\label{fig:mdp}
}
\end{figure}

We can recast these results  in a form of Minimum Detectable
Polarization (MDP) in the line at the 90$\%$ confidence level (S/N=2.6). From
eq.\ref{eq:sn} we get 
\be
{\rm
  MDP}_\line=\frac{2.6}{\mu}\sqrt{\frac{2}{\sum\limits_{i}c_{i}}}\frac{F_\tot}{F_\line}={\rm
  MPD}_0\sqrt{\frac{F_\tot}{F_\line}},
\label{eq:mdp}
\ee
where ${\rm MPD}_0$ corresponds to the case when only line photons are
detected. Resulting MDP is shown on the top panel of Fig.\ref{fig:mdp}.  We see that
outside the central 4$'$ the MDP is $\sim$1.5\% for any annuli
where the ratio of outer to inner radii is a factor of 1.5. Shown in the bottom panel of
Fig.\ref{fig:mdp} is the 1~$\sigma$ statistical error in the measured
polarization signal in the line
$\disp\sigma_\p=\frac{\sqrt{2}}{\mu\sqrt{\sum\limits_{i}c_{i}}}\frac{F_\tot}{F_\line}=\sigma_{p,0}\sqrt{\frac{F_\tot}{F_\line}}$,
where $\sigma_{p,0}$ corresponds to the case when only line photons are
detected.

Assuming parameters of the hypothetical polarimeter listed in Table
\ref{tab:polar} we can explicitly estimate the expected polarization signal
 from the Perseus cluster. The FoV
of the polarimeter is $20'\times20'$ and from Fig. \ref{fig:sn} we see
that the maximal S/N ratio is expected for the largest annulus which still
fits in the FoV, that is with $R\approx 10'$. From Fig. \ref{fig:polPer} we see
that at this distance (10$'$ corresponds to $\sim$200 kpc for the Perseus cluster) the degree of polarization in the 6.7
keV line flux is about 6\%. For the annuli with $R=10'$ the
expected statistical error is 0.6\% (Fig. \ref{fig:mdp}; bottom panel,
dashed line,
assuming 100 eV energy resolution). Thus the
expected polarization signal in the direction perpendicular to
the center of the Perseus cluster is $6\% \pm 0.6\%$\footnote{We
  emphasize again that this estimate refers to the polarization
  perpendicular to the direction towards the center of the cluster.}

From Table \ref{tab:polar} we see that two key requirements should be
improved to detect polarization from galaxy clusters: detector quantum
efficiency and energy resolution. One possibility to achieve good energy resolution is to use
narrow-band dichroic
filters\footnote{http://projects.iasf-roma.inaf.it/xraypol/Presentations/\\090427/Morning/XpolRoma$\_$Fraser.pdf}
\citep{Mar07,Ban06}. In such instruments the polarization sensitivity
arises from the difference in transmission of a family of materials
which exhibit dichroism in narrow ($\sim$ 10 eV) energy bands close to
atomic absorption edges, where the electron is excited into a bound,
molecular orbital. The simplicity of such a device makes it an
interesting option for future X-ray observatories as it decouples
the polarization sensitivity from the intrinsic efficiency of the
detector. Effective area is defined by mirrors of the telescope,
energy resolution by a bolometer. Hence, the resulting polarimeter
can achieve
quantum efficiency close to $\sim 100 \%$ and thus such filters may
offer a simple and compact method of measuring X-ray polarization. The
only problem is that narrow-band polarimeters operate  only at a
certain energy and we have to select the appropriate filter material  and
redshift of the cluster in order to study a certain line. For example,
if a filter based on the K-edge of Mn at 6539 eV is feasible then  it would be possible to study
the polarization degree in the 6.7 keV line in nearby clusters at a
redshift of $z\approx 0.024$.

A Bragg polarimeter could be another possibility. In the soft band
(below 1 keV) there are strong resonant lines emitted by relatively
cool gas of elliptical galaxies. For a plasma with temperature below
0.4-0.5 keV the lines of oxygen and nitrogen are extremely
strong. However elliptical galaxies usually contain somewhat hotter
gas ($T\ge$0.6 keV).  At these temperatures, the lines of Fe XVII and Fe
XVIII, rather than N and O lines, become very strong. We performed calculations for a set of elliptical
galaxies and found a polarization degree of $\sim 20\%$ in the line of
Fe XVIII at 0.8 keV in the galaxy NGC1404. The degree of polarization
increases rapidly with radius and reached a maximum at $r\sim 10$
kpc. If the Bragg angle can be changed (by tilting the
scattering surface) to probe several lines in the 0.8-1 keV region
then such instrument could be useful for studies of elliptical
galaxies.

\section{Acknowledgements} 

IZ and EC are grateful to Prof. Dmitry Nagirner for numerous useful
discussions.  EC and WF thank Eric Silver for helpful comments and
discussions about polarimeters. RS thanks Kevin Black for demonstration
of Bragg polarimeter. IZ thanks Fabio Muleri for important comments and corrections about
    future polarimeters. This work was supported by the DFG grant CH389/3-2, NASA
contracts NAS8-38248, NAS8-01130, NAS8-03060, the program ``Extended
objects in the Universe'' of the Division of Physical Sciences of the
RAS, the Chandra Science Center, the Smithsonian Institution, MPI
f\"{u}r Astrophysik, and MPI f\"{u}r Extraterrestrische Physik.

\label{lastpage}
\end{document}